\documentclass[longauth]{aa}  

\usepackage{graphicx}
\usepackage{txfonts}
\usepackage[dvipsnames]{xcolor}
\usepackage[colorlinks=true, linkcolor=ForestGreen, citecolor=ForestGreen, urlcolor=ForestGreen]{hyperref}
\newcommand{\cii}{\mbox{[\ion{C}{ii}]}}
\newcommand{\hii}{\mbox{\ion{H}{ii}}}

\newcommand{\oi}{\mbox{[\ion{O}{i}]}}
\DeclareUnicodeCharacter{0308}{\"{}}

\begin{document}

   \title{The ALMA-CRISTAL survey: Gas, dust, and stars in star-forming galaxies when the Universe was $\sim1$ Gyr old}

   \subtitle{I. Survey overview and case studies}

   \author{R. Herrera-Camus \inst{1,2}, J. González-López \inst{3,2,34}, N. Förster Schreiber \inst{4}, M. Aravena \inst{5,2}, I. de Looze \inst{6}, J. Spilker \inst{7}, K. Tadaki \inst{8,9}, L. Barcos-Muñoz \inst{10,33}, R. J. Assef \inst{5}, J. E. Birkin \inst{7}, A. D. Bolatto \inst{11}, R. Bouwens \inst{12}, S. Bovino \inst{13}, R. A. A. Bowler \inst{14}, G. Calistro Rivera \inst{7}, E. da Cunha \inst{14},  R. I. Davies \inst{4}, R. L. Davies \inst{15, 16}, T. Díaz-Santos \inst{17,18}, A. Ferrara \inst{19}, D. Fisher \inst{15, 16}, R. Genzel \inst{4}, J. Hodge  \inst{12}, R. Ikeda \inst{20,10}, M. Killi \inst{5}, L. Lee  \inst{4}, Y. Li \inst{7}, J. Li \inst{14}, D. Liu \inst{21}, D. Lutz \inst{4}, I. Mitsuhashi \inst{22,23,9}, D. Narayanan \inst{31,32}, T. Naab \inst{24}, M. Palla \inst{25,26}, S. H. Price \inst{27}, A. Posses \inst{7}, M. Relaño \inst{30}, R. Smit \inst{28}, M. Solimano \inst{5}, A. Sternberg \inst{29,4}, L. Tacconi \inst{4}, K. Telikova \inst{5}, H. Übler \inst{4}, S. A. van der Giessen \inst{6,30}, S. Veilleux \inst{11}, V. Villanueva \inst{1}, \and M. Baeza-Garay \inst{1}
          }

   \institute{Departamento de Astronomía, Universidad de Concepción, Barrio Universitario, Concepción, Chile\\
              \email{rhc@astro-udec.cl} 
        \and
            Millenium Nucleus for Galaxies (MINGAL) 
         \and
         Instituto de Astrof\'isica, Facultad de F\'isica, Pontiﬁcia Universidad Cat\'olica de Chile, Santiago 7820436, Chile 
         \and
         Max-Planck-Institut für extraterrestrische Physik, Giessenbachstrasse 1, 85748 Garching, Germany 
         \and
         Instituto de Estudios Astrofísicos, Facultad de Ingeniería y Ciencias, Universidad Diego Portales, Av. Ejército 441, Santiago 8370191, Chile 
          \and
         Sterrenkundig Observatorium, Ghent University, Krijgslaan 281 - S9, B9000 Ghent, Belgium 
         \and
         Department of Physics and Astronomy and George P. and Cynthia Woods Mitchell Institute for Fundamental Physics and Astronomy, Texas A\&M University, 4242 TAMU, College Station, TX 77843-
4242, USA 
         \and
         Faculty of Engineering, Hokkai-Gakuen University, Toyohira-ku, Sapporo 062-8605, Japan 
         \and 
         National Astronomical Observatory of Japan, 2-21-1 Osawa, Mitaka, Tokyo 181-8588, Japan 
         \and
         National Radio Astronomy Observatory, 520 Edgemont Road, Charlottesville, VA 22903, USA 
         \and
         Department of Astronomy and Joint Space-Science Institute, University of Maryland, College Park, MD 20742, USA 
         \and
         Leiden Observatory, Leiden University, NL-2300 RA Leiden, The Netherlands 
         \and
         Chemistry Department, Sapienza University of Rome, P.le A. Moro,
00185 Rome, Italy 
	\and
	International Centre for Radio Astronomy Research (ICRAR), The University of Western Australia, M468, 35 Stirling Highway, Crawley, WA 6009, Australia 
	\and
	Centre for Astrophysics and Supercomputing, Swinburne University of Technology, Hawthorn, VIC 3122, Australia 
	\and
	ARC Centre of Excellence for All Sky Astrophysics in 3 Dimensions (ASTRO 3D) 
	\and
	Institute of Astrophysics, Foundation for Research and Technology - Hellas (FORTH), Voutes, 70013 Heraklion, Greece 
	\and
	School of Sciences, European University Cyprus, Diogenes street, Engomi, 1516 Nicosia, Cyprus 
	\and
	Scuola Normale Superiore, Piazza dei Cavalieri 7, 50126 Pisa, Italy 
	\and
	Department of Astronomical Science, SOKENDAI (The Graduate University for Advanced Studies), Mitaka, Tokyo 181-8588, Japan 
	\and
	Purple Mountain Observatory, Chinese Academy of Sciences, 10 Yuanhua Road, Nanjing 210023, China 
	\and
	Department for Astrophysical \& Planetary Science, University of Colorado, Boulder, CO 80309, USA 
	\and
	Waseda Research Institute for Science and Engineering, Faculty of Science and Engineering, Waseda University, 3-4-1 Okubo, Shinjuku, Tokyo 169-8555, Japan 
	\and
	Max-Planck-Institut für Astrophysik, Karl-Schwarzschild-Str. 1, D-85748, Garching, Germany 
	\and
	Dipartimento di Fisica e Astronomia “Augusto Righi”, Alma Mater Studiorum, Università di Bologna, Via Gobetti 93/2, 40129 Bologna, Italy 
	\and
	INAF – Osservatorio di Astrofisica e Scienza dello Spazio di Bologna, Via Gobetti 93/3, 40129 Bologna, Italy 
	\and
	Department of Physics and Astronomy and PITT PACC, University of Pittsburgh, Pittsburgh, PA 15260, USA 
	\and
	Astrophysics Research Institute, Liverpool John Moores University, 146 Brownlow Hill, Liverpool L3 5RF, UK 
	\and
	School of Physics and Astronomy, Tel Aviv University, Tel Aviv 69978, Israel 
	\and
	Dept. Fisica Teorica y del Cosmos, Universidad de Granada, Spain 
	\and
	Department of Astronomy, University of Florida, 211 Bryant Space Sciences Center, Gainesville, FL 32611 USA 
	\and
	Cosmic Dawn Center at the Niels Bohr Institute, University of Copenhagen and DTU-Space, Technical University of Denmark 
	\and
	Department of Astronomy, University of Virginia, 530 McCormick Road, Charlottesville, VA 22903, USA 
	\and
	Las Campanas Observatory, Carnegie Institution of Washington,  Ra\'ul Bitr\'an 1200, La Serena, Chile 
             }


 
  \abstract
   {We present the ALMA-CRISTAL survey, an ALMA Cycle 8 Large Program designed to investigate the physical properties of star-forming galaxies at \(4 \lesssim z \lesssim 6\) through spatially resolved, multi-wavelength observations. This survey targets 19 star-forming main-sequence galaxies selected from the ALPINE survey, using ALMA Band 7 observations to study \cii~158~$\mu$m line emission and dust continuum, complemented by JWST/NIRCam and HST imaging to map stellar and UV emission. The CRISTAL sample expanded to 39 after including newly detected galaxies in the CRISTAL fields, archival data, and pilot study targets. The resulting dataset provides a detailed view of gas, dust, and stellar structures on kiloparsec scales at the end of the era of reionization. The survey reveals diverse morphologies and kinematics, including rotating disks, merging systems, \cii\ emission tails from potential interactions, and clumpy star formation. Notably, the \cii\ emission in many cases extends beyond the stellar light seen in HST and JWST imaging. Scientific highlights include CRISTAL-10, exhibiting an extreme \cii\ deficit similar to Arp 220; and CRISTAL-13, where feedback from young star-forming clumps likely causes an offset between the stellar clumps and the peaks of \cii\ emission. CRISTAL galaxies exhibit global \cii/FIR ratios that decrease with increasing FIR luminosity, similar to trends seen in local galaxies but shifted to higher luminosities, likely due to their higher molecular gas content. CRISTAL galaxies also span a previously unexplored range of global FIR surface brightness at high-redshift, showing that high-redshift galaxies can have elevated \cii/FIR ratios. These elevated ratios are likely influenced by factors such as lower metallicity gas, the presence of significant extraplanar gas, and contributions from shock-excited gas.}
    \titlerunning{The ALMA-CRISTAL Survey}
    \authorrunning{Herrera-Camus et al.}
    
   \maketitle
%

\section{Introduction}

The study of galaxy formation and evolution is key to understanding the diversity of galaxies in the Universe. Investigating the first billion years after the Big Bang is especially important, as it marks a critical period when galaxies transitioned from primordial gas clouds into organized systems, rapidly assembling their stellar mass, forming stars, growing their gas, metal and dust content, and playing an important role in reionizing the Universe \cite[e.g.,][]{rhc_bromm11,rhc_stark16}.

Multiwavelength observations are crucial for studying these distant galaxies, as they allow us to probe different components of the interstellar (ISM) and circumgalactic (CGM) media. In particular, observations with the Atacama Large Millimeter/submillimeter Array (ALMA) are key for tracing the cold gas and dust reservoirs in galaxies, which fuel star formation and are closely linked to galaxy evolution \cite[e.g.,][]{rhc_tacconi20, rhc_hodge20}. When combined with observations with the {\it James Webb Space Telescope} (JWST) of the stellar and nebular components \citep[e.g.,][]{rhc_robertson22}, these provide a comprehensive view of how galaxies evolved during the first billion years of cosmic history.

Among the diverse population of galaxies in the Universe, this paper focuses on those that follow the well-established correlation between stellar mass ($M_{\star}$) and star formation rate (SFR), commonly referred to as the main sequence of star-forming galaxies \citep[e.g.,][]{rhc_brinchmann04,rhc_rodighiero11,rhc_speagle14}. These galaxies represent the majority of star-forming systems across cosmic time, and this relationship has been observed to persist at least up to $z\sim6$ \citep[e.g.,][]{rhc_rinaldi25}.  Their location on the main-sequence reflects a balance between gas accretion, star formation, and feedback processes, making them ideal laboratories to study the interplay between these phenomena, commonly referred to as the baryon cycle of galaxies \citep[e.g., ][]{rhc_tumlinson17}.

In the early days of submillimeter astronomy, studies of high-redshift galaxies focused on the brightest systems due to the limited sensitivity of the available facilities. These included extreme objects such as luminous starbursts, quasars (QSOs), and highly magnified galaxies observed through gravitational lensing \cite[e.g.,][]{rhc_maiolino05,rhc_iono06,rhc_stacey10,rhc_wagg10,rhc_cox11,rhc_deBreuck11,rhc_venemans12,rhc_walter12,rhc_vieira13,rhc_gullberg15,rhc_decarli18}. Initially, these studies were confined to global properties of galaxies, but the advent of facilities like ALMA and the Northern Extended Millimeter Array (NOEMA) enabled spatially resolved investigations on kiloparsec scales \cite[e.g.,][]{rhc_riechers13,rhc_gullberg18,rhc_neeleman19,rhc_neeleman20,rhc_rizzo21,rhc_rizzo22,rhc_fraternali21,rhc_rhc21,rhc_spilker22,rhc_fujimoto24}. While these studies provided valuable insights into the most extreme cases of galaxy evolution, they did not fully capture the broader, more representative population of main-sequence galaxies that dominate cosmic star formation at high redshifts.

The first comprehensive efforts to study the cold gas and dust in main-sequence galaxies at $4\lesssim z \lesssim6$ began with small ALMA surveys, observing the \cii~158~$\mu$m line and dust continuum emission in roughly ten galaxies in the COSMOS field \citep{rhc_capak15}. These were followed by larger surveys, such as the ALMA Large Program ALPINE \citep{rhc_lefevre20, rhc_bethermin20, rhc_faisst20}, which expanded the sample to over one hundred galaxies. Additionally, the REBELS survey \citep{rhc_bouwens22} extended the study of massive, star-forming galaxies to even higher redshifts ($z\sim7-8$), offering insights into the cosmic dawn.

While these surveys focused on main-sequence star-forming galaxies provided valuable global measurements of cold gas and dust, they lacked the angular resolution needed to resolve their detailed internal structures. This also complicates the comparison between the cold gas and dust traced by ALMA and the stellar and ionized gas components traced by HST and JWST. To address this limitation, the “[{\bf C}II] {\bf R}esolved {\bf I}SM in {\bf ST}ar-forming galaxies with {\bf AL}MA” (CRISTAL) survey was designed. The primary scientific goal for the CRISTAL survey is to build a detailed census of gas, dust, and stars on $\sim$kiloparsec scales in typical star-forming galaxies at $4 < z < 6$. 

The goal of this survey paper is to detail the selection process of the CRISTAL galaxies, describe the observations and data reduction methods, summarize the available multi-wavelength data, present the results regarding detections and the main properties of the \cii\ and dust continuum emission, and highlight key scientific findings from both individual cases and the overall sample. 

This paper is organized as follows: In Section \ref{science} we summarize the contributions of the CRISTAL survey to four key areas of galaxy evolution: kinematics, outflows, morphologies, and the properties of the interstellar medium (ISM) and star formation. In Section \ref{sample}, we describe the selection criteria and main properties of the CRISTAL galaxy sample. Section \ref{sec:obs} provides an overview of the ALMA observations, while Section \ref{sec:pipeline} outlines the data reduction and processing steps for the ALMA data. Section \ref{sec:ancillary} presents the ancillary data available for CRISTAL galaxies, including observations from the Hubble Space Telescope (HST) and JWST. In Section \ref{sec:results}, we report the results, featuring a multi-wavelength view of the CRISTAL galaxies. Section \ref{case-studies} explores two case studies focused on CRISTAL-10 and CRISTAL-13. In Section \ref{CII2FIR}, we analyze the \cii/FIR ratio across the CRISTAL sample. Finally, Section \ref{conclusions} presents the summary and conclusions. Throughout this paper, we assume a flat universe with cosmological parameters of $\Omega_{\rm M}= 0.3$, $\Omega_{\Lambda} = 0.7$, and $H_0 = 70$ km s$^{-1}$ Mpc$^{-1}$. All stellar masses and SFRs are normalized to a \cite{rhc_chabrier03} initial mass function (IMF).

\section{Overview of first science results of the CRISTAL survey}\label{science}

\subsection{Kinematics} The kinematic properties of galaxies offer valuable insights into their formation history and their current state of evolution. For example, the intrinsic velocity dispersion ($\sigma_0$) of the gas is a measure of the level of turbulence, with high $\sigma_0$ values resulting from star formation feedback and radial gas transport \citep[e.g.,][]{rhc_krumholz18}. Moreover, the ratio between the rotation velocity ($V_{\rm rot}$) and $\sigma_0$ serves as an indicator of disk stability. Galaxies with ordered rotation and $V_{\rm rot}/\sigma_0 \gtrsim 2-3$ are generally classified as dynamically cold, stable disks \citep[e.g., ][]{rhc_nfs20}. 

For main-sequence galaxies in the redshift range $4 < z < 6$, early morpho-kinematic analyses using low angular resolution \cii\ observations suggest that only a small fraction ($\lesssim 40$\%) exhibit evidence of ordered rotation \citep{rhc_lefevre20}, including a more detailed analysis by \cite{rhc_jones21} using the tilted ring model fitting code $^{\rm 3D}$Barolo \citep{rhc_diteodoro15}.  More recent high-resolution \cii\ observations at $\sim$kiloparsec scales have provided a clearer view of the kinematics of these galaxies. For instance, there is evidence of rotationally supported cold disks in star-forming main-sequence galaxies at $z=4.3$ \cite[$V_{\rm rot}/\sigma_0=3.4$;][]{rhc_neeleman20} and $z=5.5$ \cite[HZ4 or CRISTAL-20: $V_{\rm rot}/\sigma_0=2.2$;][]{rhc_rhc22}, although JWST/NIRSpec data for the latter system suggests a more complex scenario \citep{rhc_parlanti25}. The same is true for dusty star-forming galaxies above the main-sequence \cite[e.g., ][]{rhc_rizzo21}. Similar conclusions are drawn from NIR rest-frame morphological studies, which reveal the existence of massive, unperturbed disk-like galaxies at $z\sim4-6$ \citep[e.g., ][]{rhc_huertas-company24}.

More recently, based on ALMA observations from the CRISTAL survey, in \citet{rhc_posses24} we analyze the kinematics of the main-sequence galaxy CRISTAL-05 (also known as DEIMOS\_COSMOS\_683613 and HZ3). In contrast to earlier classifications of CRISTAL-05 as a single source, these new observations reveal it to be a close interacting pair surrounded by an extended region of carbon-enriched gas. The overall velocity is irregular across the full system, although one of the pair member is consistent with disk rotation; the velocity dispersion is roughly constant and around 80~km~s$^{-1}$. Another notable example is the HZ10 system (CRISTAL-22), where CRISTAL and JWST/NIRSpec IFU observations show that this massive galaxy consists of at least three closely projected components, with complex kinematics suggesting a possible close merger or disturbed disk \citep[][]{rhc_jones24,rhc_telikova24}. 

The findings from individual CRISTAL systems highlight the intricate kinematics of star-forming galaxies during the first $\sim1$~Gyr. In Lee et al. (in prep.), we present a detailed morpho-kinematic analysis of 35 CRISTAL galaxies, reporting that nearly $\sim50$\% of them are classified as disk-dominated. Additionally, CRISTAL galaxies exhibit high intrinsic velocity dispersions ($\sim70$~km~s$^{-1}$), likely driven by gravitational instabilities.

\subsection{Outflows}

Outflows, driven by stellar and active galactic nuclei feedback, are expected to play a key role in regulating star formation and shaping galaxy evolution \citep[e.g., ][]{rhc_veilleux20}. Despite recent progress, we still lack a comprehensive characterization of the outflow properties in individual galaxies during the early cosmic epochs. Recently, JWST/NIRSpec observations have found potential signature of outflows in low-mass galaxies at $z\gtrsim4$, probing the ionized gas phase \cite[e.g., ][]{rhc_zhang24,rhc_carniani24}. However, low-ionization or colder ($T\sim300-400$~K) gas in the outflow, traced for example by \cii\ emission \citep[e.g.,][]{rhc_contursi13}, remains largely unexplored. Understanding this cold component is critical, as it likely dominates the mass of outflows \citep[e.g., ][]{rhc_fluetsch19,rhc_rhc20a}. Simulations \citep[e.g., ][]{rhc_pizzati20,rhc_pizzati23} suggest that these outflows are also connected to the diffuse \cii\ emission observed around galaxies, extending into the circumgalactic medium \citep{rhc_fujimoto19, rhc_fujimoto20}.

For main-sequence galaxies at \(z \gtrsim 4\), \cite{rhc_gallerani18} conducted one of the earliest efforts to characterize outflows using stacked global \cii\ spectra from galaxies in the \cite{rhc_capak15} sample, finding tentative evidence for outflows. Building on this work, \cite{rhc_ginolfi20}, using the larger ALPINE sample, detected broad \cii\ emission in stacked profiles, with velocities of \(\sim 500\) km s\(^{-1}\). This broad component was particularly prominent in galaxies with star formation rates exceeding \(\approx 25\) \(M_{\odot}\) yr\(^{-1}\). 

For individual galaxies, \cite{rhc_rhc21} focused on the neutral outflow phase in HZ4 (CRISTAL-20, DEIMOS\_COSMOS\_494057), reporting evidence of outflowing gas from the central star-forming region. They found projected velocities of \(\sim 400\) km s\(^{-1}\) and a neutral gas mass outflow rate \(\sim 3-6\) times higher than the star formation rate in the central region. Recent JWST/NIRSpec observations have also identified the ionized phase of the HZ4 outflow, confirming that the mass outflow rate is predominantly driven by neutral gas traced by the \cii\ transition \citep{rhc_parlanti25}. 

Among the CRISTAL sample, we identify additional evidence for outflowing gas traced by \cii\ line emission. One key advantage relative to previous programs such as ALPINE is the higher angular resolution that allow searching for outflows in the nuclear regions of galaxies, or centered at star-forming clumps identified with HST or JWST, at comparable or better sensitivity. For example, CRISTAL-02 exhibits the strongest \cii\ outflow, with the ionized counterpart also detected in recent JWST/NIRSpec observations \citep{rhc_davies25}. CRISTAL-08 exhibits evidence of outflowing gas associated with the giant star-forming clumps abundant in its disk, which will be discussed in detail in a forthcoming letter by Herrera-Camus et al. (in prep.). Additionally, in Birkin et al. (in prep.) we report tentative evidence for modest outflows with velocities of $\sim300$~km~s$^{-1}$ based on the stacking of \cii\ spectra for CRISTAL galaxies, excluding systems with kinematic evidence for gravitational interactions from high-angular-resolution observations. Overall, we find that mass loading rates in CRISTAL galaxies are generally modest but can reach values comparable to those observed in nearby starbursts.

\subsection{Morphology}

During the early cosmic epochs covered by the CRISTAL survey, galaxies experienced rapid growth and intense star formation. Observing their sizes and morphologies—such as disks, clumps, and spatial offsets between various tracers—reveals how these galaxies assembled under the influence of gravity, mergers, and feedback. Ultimately, morphological studies are important to characterize the turbulent conditions of the early Universe, providing important tests for models of galaxy growth and the role of physical processes like feedback. 

For main-sequence galaxies at $4 \lesssim z \lesssim 6$, one of the key discoveries by ALMA was detecting \cii\ line emission extending well beyond the star-forming regions traced by rest-frame UV or dust continuum emission. \cite{rhc_fujimoto19} found evidence for a 10 kpc-scale extended gas component through stacked \cii\ emission from ALPINE galaxies, with further analysis by \cite{rhc_fujimoto20} revealing \cii\ emission sizes typically $2-3$ times larger than the UV or dust continuum. Several scenarios could explain these extended \cii\ emissions, such as outflows depositing gas in the surrounding medium \citep[e.g., ][]{rhc_pizzati20,rhc_pizzati23}, unresolved satellite galaxies, or circumgalactic photodissociation regions (PDRs). Disentangling these origins requires deep, multi-wavelength, spatially resolved observations. Such data have started to emerge; for example, \cite{rhc_lambert23} found that \cii\ emission in HZ7 (CRISTAL-21) is about twice the size of its rest-frame UV emission, likely due to a merger resulting in a non-rotating disk. For HZ4 (CRISTAL-20, DEIMOS\_COSMOS\_494057), the extended \cii\ emission may be linked to the outflowing gas from the central region \citep{rhc_rhc21}.  
 
The CRISTAL survey offers an advantage over previous studies by providing deep, spatially resolved \cii\ observations that allow for testing various scenarios to explain the extended nature of \cii\ emission. Based on CRISTAL observations, we report in \cite{rhc_ikeda25} that, on average, \cii\ emission extends approximately three times farther than rest-frame UV emission and twice as far as the dust continuum. Notably, this extended emission can be well-modeled by an exponential disk, without requiring an additional halo-like component. Interestingly, \cite{rhc_ikeda25} detects extended \cii\ emission in CRISTAL galaxies both with and without companions, suggesting that the contributions from PDRs, in addition to diffuse neutral medium (atomic gas) at large radius, may further enhance the spatial distribution of \cii\ emission.

An excellent illustration of the complex morphology and spatial distribution of the \cii\ emission in $z\sim5$ galaxies is the analysis of the CRISTAL-01 field (DEIMOS\_COSMOS\_842313) by \cite{rhc_solimano24a,rhc_solimano25_jwst}. This system features a close interaction between the disturbed CRISTAL-01 and the submillimeter galaxy (SMG\footnote{Here we use the definition in \cite{rhc_hodge20} of a submillimeter galaxy: a system with a high submillimeter flux density of $S_{\rm 850 mm} \gtrsim~1{\rm mJy}$.}) J1000+0234 \citep{rhc_g-j18, rhc_fraternali21}, surrounded by two companion systems (CRISTAL-01b and -01c; see Section \ref{multi-view}) and embedded within a giant Lyman-$\alpha$ blob \citep{rhc_jimenez-andrade23}. Most notably, a $\sim$15 kpc plume of \cii\ gas emerges from the center of the SMG, potentially driven by AGN activity, as revealed by JWST/NIRSpec observations \citep{rhc_solimano25_jwst}.

Regarding the dust morphology of star-forming galaxies at $z\sim5$, the depth of the CRISTAL survey enables probing FIR surface densities an order of magnitude lower than those typical of SMGs. In \cite{rhc_mitsuhashi24} we find that the effective radius of the dust continuum is, on average, twice as large as that of the UV continuum—opposite to trends predicted by the TNG50 simulations \citep{rhc_popping22}. One possible interpretation for this intriguing result is that the dust continuum emission traces more extended, global (obscured) star formation, while the rest-frame UV continuum is associated with concentrated star-forming clumps within galaxies and/or less obscured lines-of-sights.

\subsection{Interstellar medium and star formation}

By observing both the \cii\ line and dust continuum emission, the CRISTAL survey provides valuable information about the ISM of $z\sim4-6$ star-forming galaxies. The \cii\ line, the primary coolant of the ISM, arises from multiple phases—including dense PDRs, diffuse atomic gas, and ionized gas—making it an effective tracer of the ISM complex structure \citep[e.g., ][]{rhc_pineda13}.  Observations of the dust continuum, on the other hand, provide information on the heating mechanisms within the ISM. Dust grains and polycyclic aromatic hydrocarbons (PAHs) absorb UV photons and release electrons, heating the gas through the photoelectric effect. This balance of \cii\ cooling and dust-driven heating, often measured by the \cii/FIR ratio, is important for studying star formation and galaxy evolution in the early universe \cite[e.g., ][]{rhc_wolfire22}.

In an ISM in thermal equilibrium, cooling via the \cii\ line can also trace heating from star formation activity. A strong relationship, both globally and on spatially resolved scales, exists between \cii\ line emission and SFR in nearby galaxies \citep[e.g.,][]{rhc_pineda14,rhc_delooze14,rhc_rhc15,rhc_rhc18b}, with little evolution observed in this relation across the redshift range of CRISTAL galaxies \citep{rhc_schaerer20}. For CRISTAL galaxies, \cite{rhc_li24} report a strong correlation between \cii\ and SFR. However, they also identify significant scatter at larger radii and variable slopes in the \cii-SFR relation across different CRISTAL systems. A more detailed analysis of the sources of this scatter will be presented in Palla et al. (in prep.).

In spatially resolved observations of CRISTAL galaxies, \cite{rhc_rhc21} find that the \cii/FIR ratio in HZ4 (CRISTAL-20) spans $\sim2-4\times10^{-3}$, similar to values seen in star-forming regions of nearby starbursts like M82 and M83, which exhibit comparable levels of star-formation rate surface density. In CRISTAL-05, \cite{rhc_posses24} report \cii/FIR ratios in the main galaxy component of $\sim2-8\times10^{-3}$, while in outer regions where \cii\ line emission is extended, the high \cii/FIR lower limits ($\gtrsim10^{-2}$) are comparable to values observed in local merging systems where \cii\ emission is enhanced by shocks \citep[e.g.,][]{rhc_appleton13,rhc_peterson18}. In the system HZ10 (CRISTAL-22), \cite{rhc_villanueva24} measure \cii/FIR ratios between $\sim1-3\times10^{-3}$. This analysis is particularly robust, benefiting from ALMA Band 9 observations at the peak of the dust spectral energy distribution (SED), enabling a robust determination of the dust temperature of $T_{\rm dust}=46.7\pm6.8$ K, and thus a more accurate FIR luminosity. 

Spatially resolved dust continuum observations in CRISTAL galaxies are also important to improve the SED modeling and obtain more accurate mapping of physical properties (e.g., age, SFR, $M_{\star}$, dust extinction $A_{\rm V}$). One key finding we present in \cite{rhc_li24} and \cite{rhc_lines24} is that, on $\sim0.5$–1 kpc scales, the stellar mass is not underestimated despite the fact that light from younger stars can outshine and mask the contribution from older, fainter stars—a process typically referred to as outshining. 

Additionally, as shown by \cite{rhc_li24}, combining JWST/NIRCam stellar light data with ALMA dust continuum measurements helps resolve the age-dust degeneracy, leading to more robust estimates of dust obscuration, which are often overestimated when ALMA data is excluded. These refined obscuration estimates impact the inferred spatial distribution of specific star formation rates, offering a clearer view of trends in galaxy growth and quenching, such as inside-out evolution. There results are further supported by the fact that CRISTAL galaxies show a high fraction of dust-obscured star formation—averaging around 50\% but varying widely from 20\% to 90\%—indicating substantial diversity in dust properties and morphology, as we discuss in \cite{rhc_mitsuhashi24}. This level of dust obscuration is comparable to what has been observed in star-forming galaxies at $4\lesssim z \lesssim 6$ in the ALPINE sample \citep{rhc_khusanova21,rhc_fudamoto20}, and massive star-forming galaxies at $z\gtrsim6.5$ \citep{rhc_inami22}.

\subsection{The emerging picture}

The CRISTAL survey provides a detailed view on $\sim$kiloparsec scales of galaxies during the first billion years of cosmic history, highlighting their complexity and dynamic evolution. The kinematic analyses suggest that star-forming galaxies in this epoch exhibit a wide range of behaviors, from rotationally supported cold disks ($\sim50\%$) to complex, irregular velocity fields indicative of interactions or mergers (Lee et al. in prep.). Such findings challenge simplified models of early disk formation and point to a dynamic interplay of turbulence, feedback, and environmental interactions shaping galaxy evolution at $z=4-6$.

Beyond kinematics, the CRISTAL survey has significantly advanced our understanding of outflows, morphologies, and the interstellar medium (ISM) properties in young galaxies. Outflows traced by \cii\ emission, observed in systems such as CRISTAL-20 \citep[][]{rhc_rhc21,rhc_parlanti25}, CRISTAL-02 \citep[also detected in the ionized phase with JWST/NIRSpec; ][]{rhc_davies25}, and through the stacking of non-interacting CRISTAL systems (Birkin et al., in prep.), highlight the role that the neutral gas ejection can have regulating star formation and enriching the circumgalactic medium. 

Our morphological studies reveal extended \cii\ emission components, often spanning several times the extent of UV or dust emission regions, which suggest contributions from diffuse gas or feedback-driven processes \citep{rhc_ikeda25}. Moreover, our analysis of the dust morphology highlights the presence of extended obscured star formation, with dust emission typically reaching approximately twice the size of the rest-frame UV emission \citep{rhc_mitsuhashi24}. 

Collectively, these findings emphasize the importance of multi-phase, multi-wavelength observations in unraveling the complex assembly and evolutionary processes of galaxies during the early universe.

\begin{figure*}
\centering
   \includegraphics[width=\hsize]{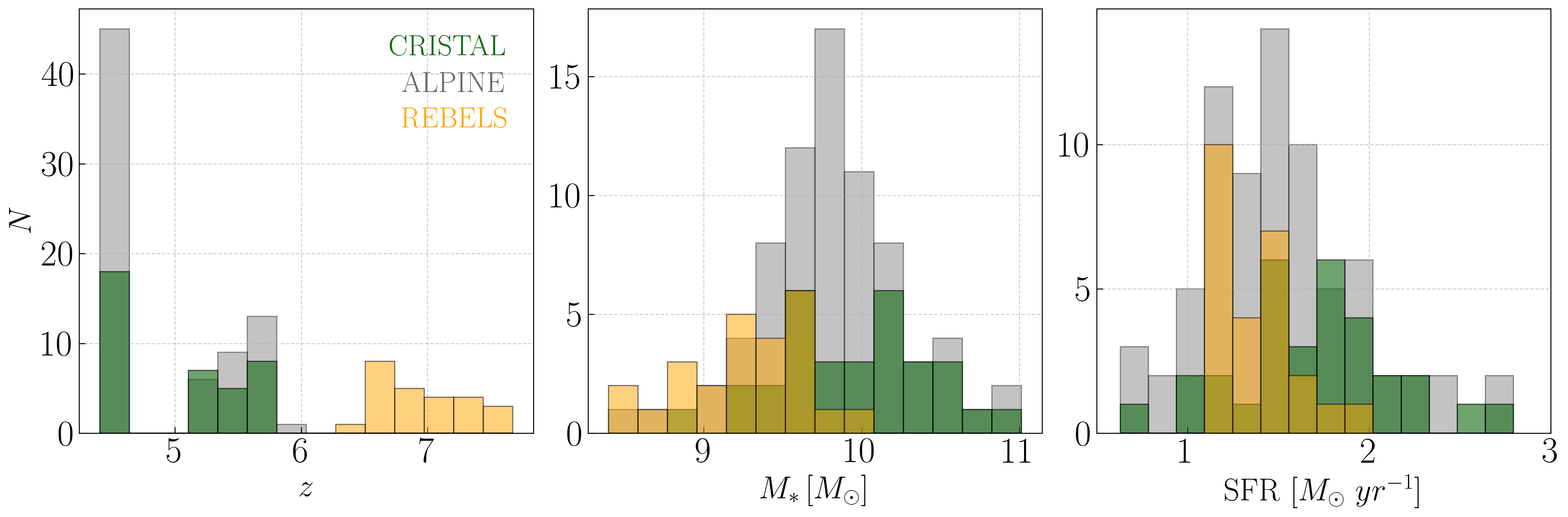}
      \caption{Histograms showing the distribution of redshift (left), stellar mass (center), and star formation rate (right) for the CRISTAL (green), ALPINE \citep[gray;][]{rhc_lefevre20, rhc_bethermin20, rhc_faisst20}, and REBELS samples \citep[gold;][]{rhc_bouwens22}.} \label{Hist_surveys}
\end{figure*}

\begin{figure}
\centering
   \includegraphics[width=\hsize]{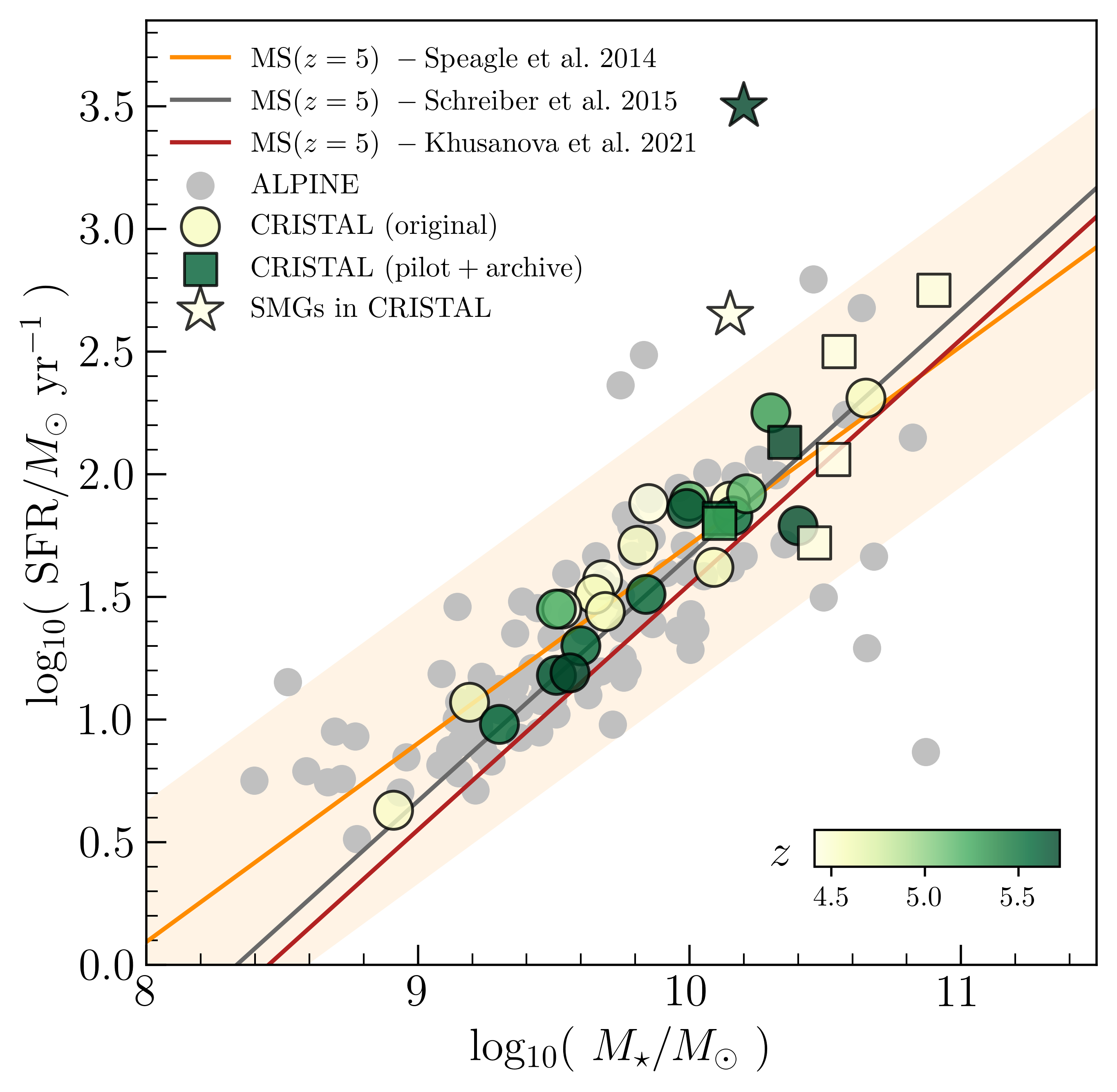}
      \caption{The CRISTAL sample consists of 39 star-forming galaxies between $4\lesssim z \lesssim6$ that are representative of the population of massive ($M_{\star}\gtrsim10^{9}$~$M_{\odot}$) galaxies at this redshift range. The figure shows the stellar mass-star formation rate plane for galaxies between $4\lesssim z \lesssim6$. CRISTAL galaxies are shown as green circles, color-coded according to their redshift. The main-sequence of star-forming galaxies at $z=5$ is shown following the calibrations by \cite{rhc_speagle14} (with $\pm0.5$ dex width represented by the orange shaded region), \cite{rhc_schreiber15}, and \cite{rhc_khusanova21}. Galaxies from the pilot programs and drawn from the archive are shown as squares. SMGs in the CRISTAL fields are shown as stars. Star-forming galaxies detected in \cii\ line emission observed as part of the ALPINE survey \citep{rhc_lefevre20, rhc_bethermin20, rhc_faisst20} are shown as gray circles.} \label{Fig1}
\end{figure}

\begin{figure*}
\centering
   \includegraphics[width=\hsize]{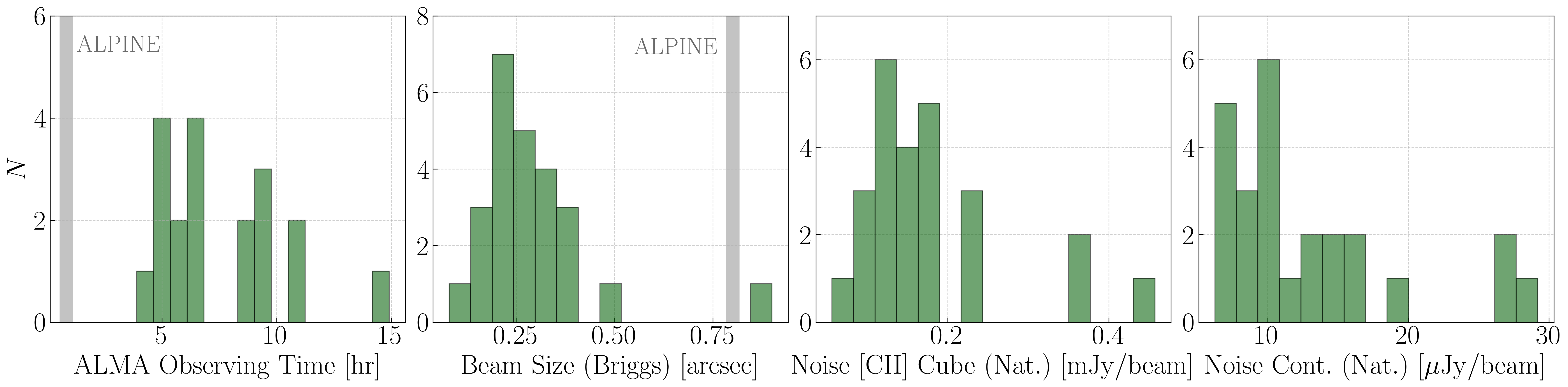}
      \caption{Histogram showing the distribution of observing time (first panel), synthesized beam size (second panel), noise measured in the cubes for 20~km~s\(^{-1}\) channels (third panel) and Band 7 continuum (fourth panel). The average values for ALPINE galaxies \citep[][]{rhc_lefevre20, rhc_bethermin20, rhc_faisst20} are indicated by a gray line.} \label{Obs_cristal}
\end{figure*}

\section{The CRISTAL sample} \label{sample}

\subsection{Sample selection}

The first step in the CRISTAL target selection was to search in the ALMA archive for galaxies in the redshift range $4\leq z \leq6$ detected in \cii\ line emission with signal-to-noise $S/N\geq3$. This was motivated by the need of pre-selecting galaxies with a reliable redshift determination and a measured \cii\ line flux that allows us to plan follow-up spatially-resolved observations. The choice of the redshift range, in addition to the scientific motivation described before, was set by the requirement to observe the \cii\ transition in Band 7. 

About 95\% of the pre-selected galaxies from the ALMA archive at the time were part of the \cite{rhc_capak15} sample and the ALPINE survey \citep{rhc_lefevre20, rhc_bethermin20, rhc_faisst20}. In these two programs, galaxies were typically observed for $\lesssim1$~hr and with an angular resolution $\gtrsim0\farcs8$ ($\gtrsim5$~kpc), so most of the systems were barely resolved or spatially unresolved.

The next step was to select galaxies based on the following criteria:

\begin{enumerate}

    \item {\it Main-sequence}: With the goal of selecting typical or representative systems of the galaxy population between $4\leq z \leq6$, we only considered galaxies within $\pm0.5$~dex of the main-sequence (MS) of star-forming galaxies as defined by \cite{rhc_speagle14}. As shown in Figure~\ref{Fig1}, this calibration is is in good agreement with those of \cite{rhc_schreiber15} and \cite{rhc_khusanova21} for a similar redshift range. The star formation rates and stellar masses for these systems were drawn from the literature \citep[e.g.,][]{rhc_capak15,rhc_faisst20}.
    
    \item {\it Stellar mass}: We selected systems with stellar masses greater than $M_{\star}\geq10^{9.5}~M_{\odot}$ according to the values from the ALPINE catalog presented in \cite{rhc_faisst20}. Given that the pre-selected systems are part of the main-sequence, a cut in stellar mass also correspond to a cut in SFR of approximately $\geq20~M_{\odot}~yr^{-1}$, which in turn is also closely associated with the \cii\ luminosity \citep[e.g.,][]{rhc_sargsyan12,rhc_delooze14,rhc_rhc15,rhc_schaerer20}. For galaxies with $M_{\star}\lesssim10^{9.5}~M_{\odot}$ we estimated that we would need observing times larger than $\sim15$~hr to achieve our required sensitivity. Therefore, applying a stellar mass threshold at $M_{\star}=10^{9.5}~M_{\odot}$ was important to keep the program within the observing time available for Large Programs in ALMA Cycle 8.
    
    \item {\it Rest-frame UV emission}: To achieve our scientific goals regarding the study of kinematics, morphology, and star formation, we selected galaxies detected in rest-frame UV emission with the HST Wide Field Camera 3 (WFC3). Given that most of the pre-selected galaxies that fulfill this criteria are in the COSMOS field \citep{rhc_scoville07}, observations in at least three WFC3 filters were available for all systems, except for CRISTAL-14 and CRISTAL-19 for which only F160W data exist.  
        
\end{enumerate}

Having JWST data available, which trace the stellar and ionized gas light, was not part of the selection criteria (as the telescope had not yet been launched at the time of the proposal). However, because a large fraction of the pre-selected systems are in the COSMOS field, there are programs such as COSMOS-Web \citep[PID 1727; co-PIs: Kartaltepe \& Casey; ][]{rhc_casey23} and PRIMER (PI: Dunlop, PID 1837) that provide NIRCam and/or MIRI observations in multiple filters for a large fraction of systems. There is also NIRSpec observations of the main nebular lines from programs such as GA-NIFS (PI: Luetzgendorf, PID 1217), ORCHIDS (PI: Aravena, PID 5974) and PID 3073 (PI Faisst). We discuss this in more detail in Section~\ref{sec:ancillary}. 

After applying the initial selection criteria based on redshift ($4\leq z \leq 6$), \cii\ signal-to-noise ratio ($S/N\geq3$), and main-sequence membership ($|\log_{10}(\Delta{\rm MS})| \leq 0.5$~dex), we selected 81 out of 124 systems. Subsequently, applying the additional criteria of stellar mass ($M_{\rm star}\geq10^{9.5}$~M$_{\odot}$), observability, and the availability of ancillary HST data reduced the sample to 25 galaxies. Seven of them, however, already had spatially-resolved \cii\ line observations in the ALMA archive. These systems are: HZ4 \citep[PI R. Herrera-Camus;][]{rhc_rhc21,rhc_rhc22}, HZ3 \citep[PI M. Aravena][]{rhc_posses24}, HZ7 \citep[PI M. Aravena;][]{rhc_lambert23}, HZ10 \citep[PI M. Aravena][]{rhc_telikova24} (all of them designed as pilot programs for the CRISTAL ALMA Large Program), and DEIMOS\_COSMOS\_818760, DEIMOS\_COSMOS\_873756, and vuds\_cosmos\_5101218326 \citep[PI E. Ibar; ][]{rhc_devereaux24}. Except for HZ3, we did not request additional Band 7 observations as the sensitivity and angular resolution achieved in the programs was good enough to pursue the main scientific goals of CRISTAL.

In summary, the sample that was observed with ALMA as part of the CRISTAL Large Program consisted of 19 targets. The details of the ALMA observations are presented in Section~\ref{sec:obs}.

\subsection{Final CRISTAL sample}

After the ALMA observations were completed, the CRISTAL sample increased in size for two main reasons: 

\begin{enumerate}
    \item {\it Additional galaxies in CRISTAL fields:} The deeper, higher angular resolution observations revealed additional companion galaxies in the field-of-view of the main 19 CRISTAL targets. In total we detected 7 additional new galaxies in the field (CRISTAL-01b,c, -07c,d, -09b, -10b, -13b) and we spatially-resolved 4 interacting systems into pairs (CRISTAL-04b, -06b, -07b, -16b). A thorough search and characterization of all systems in the CRISTAL \cii\ cubes will be presented in van Leeuwen et al. (in prep).
    \item {\it Galaxies in the ALMA archive:} In total we included 6 systems from the ALMA archive with angular resolution and sensitivity comparable to that achieved by the CRISTAL program. Four of them are from pilot programs (HZ4, HZ7, and HZ10), and the other three are from an ALPINE follow-up program \citep[PI E. Ibar; DEIMOS\_COSMOS\_818760, DEIMOS\_COSMOS\_873756, vuds\_cosmos\_5101218326;][]{rhc_devereaux24}. The high-angular resolution observations of HZ10 and DEIMOS\_COSMOS\_818760 allow us to spatially resolve the systems into two (CRISTAL-22a, b) and three galaxies (CRISTAL-23a, b, c), respectively.
    \end{enumerate}

\begin{table*}
\caption{The ALMA-CRISTAL Sample}             
\label{table:sample}      
\centering                          
\begin{tabular}{l c c c c c c}        
\hline\hline                
Name & Other names  & $\alpha_{\rm J2000}$ & $\delta_{\rm J2000}$ & $z_{\cii}$~$^{a}$ & $\log_{10}(M_{\star})$ & $\log_{10}(\rm SFR)$\\    
 & & & & & $M_{\odot}$ & $M_{\odot}~{\rm yr}^{-1}$ \\ 
\hline                        
 & & Large Program & & & & \\
\hline
CRISTAL-01a	&	DC\_842313	&	10:00:54.509	&	+02:34:34.407	&	4.554	&	$	10.65	\pm	0.50	$	&	$	2.31	\pm	0.74	$	\\
CRISTAL-01b	&		&	10:00:54.774	&	+02:34:28.330	&	4.530	&	$	9.81	\pm	0.34	$	&	$	1.71	\pm	0.29	$	\\
CRISTAL-01c	&		&	10:00:54.185	&	+02:34:37.294	&	4.540	&	$				$	&	$				$	\\
CRISTAL-02	&	DC\_848185, HZ6, LBG-1	&	10:00:21.503	&	+02:35:11.054	&	5.294	&	$	10.30	\pm	0.28	$	&	$	2.25	\pm	0.42	$	\\
CRISTAL-03	&	DC\_536534, HZ1	&	09:59:53.255	&	+02:07:05.358	&	5.689	&	$	10.40	\pm	0.29	$	&	$	1.79	\pm	0.31	$	\\
CRISTAL-04a	&	vuds\_5100822662	&	09:58:57.907	&	+02:04:51.407	&	4.520	&	$	10.15	\pm	0.29	$	&	$	1.89	\pm	0.21	$	\\
CRISTAL-04b	&		&	09:58:57.944	&	+02:04:53.003	&	4.520	&	$	8.91	\pm	0.55	$	&	$	0.63	\pm	0.37	$	\\
CRISTAL-05	&	DC\_683613, HZ3	&	10:00:09.431	&	+02:20:13.905	&	5.541	&	$	10.16	\pm	0.35	$	&	$	1.83	\pm	0.30	$	\\
CRISTAL-06a	&	vuds\_5100541407	&	10:01:00.910	&	+01:48:33.706	&	4.562	&	$	10.09	\pm	0.30	$	&	$	1.62	\pm	0.34	$	\\
CRISTAL-06b	&		&	10:01:01.002	&	+01:48:34.987	&	4.562	&	$	9.19	\pm	0.46	$	&	$	1.07	\pm	0.33	$	\\
CRISTAL-07a	&	DC\_873321, HZ8	&	10:00:04.059	&	+02:37:35.841	&	5.154	&	$	10.00	\pm	0.33	$	&	$	1.89	\pm	0.26	$	\\
CRISTAL-07b	&		&	10:00:03.973	&	+02:37:36.326	&	5.154	&	$	9.90	\pm	0.66	$	&	$	1.45	\pm	0.49	$	\\
CRISTAL-07c	&		&	10:00:03.222	&	+02:37:37.732	&	5.155	&	$	10.21	\pm	0.35	$	&	$	1.92	\pm	0.41	$	\\
CRISTAL-07d	&		&	10:00:03.191	&	+02:37:35.502	&	5.155	&	$				$	&	$				$	\\
CRISTAL-08	&	vuds\_efdcs\_530029038	&	03:32:19.046	&	-27:52:38.256	&	4.430	&	$	9.85	\pm	0.36	$	&	$	1.88	\pm	0.23	$	\\
CRISTAL-09a	&	DC\_519281	&	09:59:00.892	&	+02:05:27.553	&	5.575	&	$	9.84	\pm	0.39	$	&	$	1.51	\pm	0.32	$	\\
CRISTAL-09b	&		&	09:59:00.775	&	+02:05:26.904	&	5.575	&	$				$	&	$				$	\\
CRISTAL-10a	&	DC\_417567, HZ2	&	10:02:04.124	&	+01:55:44.279	&	5.671	&	$	9.99	\pm	0.31	$	&	$	1.86	\pm	0.20	$	\\
CRISTAL-10b	&		&	10:02:04.498	&	+01:55:49.963	&	5.671	&	$				$	&	$				$	\\
CRISTAL-11	&	DC\_630594	&	10:00:32.596	&	+02:15:28.515	&	4.439	&	$	9.68	\pm	0.33	$	&	$	1.57	\pm	0.31	$	\\
CRISTAL-12	&	CG\_21	&	03:32:11.936	&	-27:41:57.514	&	5.572	&	$	9.30	\pm	0.47	$	&	$	0.98	\pm	0.40	$	\\
CRISTAL-13a	&	vuds\_5100994794	&	10:00:41.169	&	+02:17:14.283	&	4.579	&	$	9.65	\pm	0.34	$	&	$	1.51	\pm	0.41	$	\\
CRISTAL-13b	&		&	10:00:41.151	&	+02:17:15.875	&	4.579	&	$				$	&	$				$	\\
CRISTAL-14	&	DC\_709575	&	09:59:47.072	&	+02:22:32.894	&	4.411	&	$	9.53	\pm	0.38	$	&	$	1.45	\pm	0.38	$	\\
CRISTAL-15	&	vuds\_5101244930	&	10:00:47.660	&	+02:18:02.116	&	4.580	&	$	9.69	\pm	0.33	$	&	$	1.44	\pm	0.24	$	\\
CRISTAL-16a	&	CG\_38	&	03:32:15.900	&	-27:41:24.353	&	5.571	&	$	9.60	\pm	0.39	$	&	$	1.30	\pm	0.35	$	\\
CRISTAL-16b	&		&	03:32:15.817	&	-27:41:24.723	&	5.571	&	$				$	&	$				$	\\
CRISTAL-17	&	DC\_742174	&	10:00:39.133	&	+02:25:32.335	&	5.635	&	$	9.51	\pm	0.40	$	&	$	1.18	\pm	0.31	$	\\
CRISTAL-18	&	vuds\_5101288969	&	09:59:30.648	&	+02:19:53.760	&		&	$	9.56	\pm	0.15	$	&	$	1.19	\pm	0.17	$	\\
CRISTAL-19	&	DC\_494763	&	10:00:05.105	&	+02:03:12.101	&	5.233	&	$	9.51	\pm	0.36	$	&	$	1.45	\pm	0.36	$	\\
\hline   																					
	&		&	Pilot Programs	&		&		&	$				$	&	$				$	\\
\hline   																					
CRISTAL-20	&	DC\_494057, HZ4	&	09:58:28.504	&	+02:03:06.593	&	5.545	&	$	10.11	\pm	0.35	$	&	$	1.82	\pm	0.28	$	\\
CRISTAL-21	&	HZ7	&	09:59:30.467	&	+02:08:02.626	&	5.255	&	$	10.11	\pm	0.32	$	&	$	1.80	\pm	0.32	$	\\
CRISTAL-22a	&	HZ10	&	10:00:59.297	&	+01:33:19.458	&	5.653	&	$	10.35	\pm	0.37	$	&	$	2.13	\pm	0.34	$	\\
CRISTAL-22b	&	HZ10	&	10:00:59.250	&	+01:33:19.382	&	5.653	&	$				$	&	$				$	\\
\hline   																					
	&		&	ALMA Archive	&		&		&	$				$	&	$				$	\\
\hline   																					
CRISTAL-23a	&	DC\_818760	&	10:01:54.865	&	+02:32:31.512	&	4.560	&	$	10.55	\pm	0.29	$	&	$	2.50	\pm	0.34	$	\\
CRISTAL-23b	&	DC\_818760	&	10:01:54.966	&	+02:32:31.534	&	4.562	&	$	10.46	\pm	0.24	$	&	$	1.72	\pm	1.33	$	\\
CRISTAL-23c	&	DC\_818760	&	10:01:54.684	&	+02:32:31.438	&	4.565	&	$				$	&	$				$	\\
CRISTAL-24	&	DC\_873756	&	10:00:02.716	&	+02:37:39.971	&	4.546	&	$	10.53	\pm	0.08	$	&	$	2.06	\pm	0.22	$	\\
CRISTAL-25	&	vuds\_5101218326	&	10:01:12.495	&	+02:18:52.546	&	4.573	&	$	10.90	\pm	0.32	$	&	$	2.75	\pm	0.29	$	\\
\hline  										$				$		$				$	
	&		&	SMGs in the FOV$^{b}$	&		&		&	$				$	&	$				$	\\
\hline   																					
J1000+0234	&		&	10:00:54.480	&	+02:34:36.120	&	4.539	&	$	10.15-10.90			$	&	$	2.65			$	\\
CRLE	&		&	10:00:59.184	&	+01:33:06.840	&	5.667	&	$	10.20			$	&	$	3.50			$	\\
\hline 
\end{tabular}
\tablefoot{$^{a}$ Redshifts based on the \cii\ observations from the CRISTAL survey.}
\tablefoot{$^{b}$ The stellar masses and star formation rates for J1000+0234 and CRLE are drawn from \cite{rhc_fraternali21} and \cite{rhc_pavesi18}, respectively.}
\tablefoot{When applicable, the prefix DEIMOS\_COSMOS\_ has been abbreviated as DC\_}
\end{table*}

In total, the final CRISTAL sample includes 39 main-sequence star-forming galaxies, including companion galaxies detected around the originally targeted main galaxies. Among the systems added to the original sample, those with measurable stellar masses meet the criteria for being on the main sequence. However, their stellar masses can extend below the initial selection threshold of $M_{\star}=10^{9.5}~M_{\odot}$.

Table~\ref{table:sample} summarizes the main properties of all CRISTAL systems.  The stellar masses and star formation rates were derived using the CIGALE code \citep{rhc_boquien19}, and the details for these calculations are described in \cite{rhc_mitsuhashi24}.\footnote{\cite{rhc_mitsuhashi24} use all available broad and medium bands in optical to near-infrared for the SED fitting. From the COSMOS2015 catalog, these include 10 broad bands ($u^{*},B,V,r^{+},i^{+},z^{++},Y,J,H,Ks$), 12 medium bands on Ground-based telescopes, and 4 Spitzer bands. \cite{rhc_mitsuhashi24} also take 4 broad bands ($U, B, R, Ks$) and 23 medium bands on Ground-based telescopes, and 10 HST and 4 Spitzer bands on space telescopes from the ASTRODEEP catalog. The CRISTAL galaxies for which the ALMA Band 7 continuum was included in the SED fitting are listed in Table~1 of \cite{rhc_mitsuhashi24}.}
In general, there is a good agreement between the stellar masses and SFRs derived by \cite{rhc_mitsuhashi24} using CIGALE, by \cite{rhc_li24} using MAGPHYS (including JWST/NIRCam data), and those in the ALPINE database \citep{rhc_faisst20} based on the LePhare SED fitting code \citep{rhc_arnouts99, rhc_ilbert06}, with a mean difference of only $\sim0.1$~dex between the estimates. The coordinates and redshifts in Table~\ref{table:sample} are derived from the \cii\ data, with the exception of CRISTAL-18, which is undetected in the \cii\ transition despite deep observations (see Section~\ref{sec:results} for further details).

Together with the 39 star-forming galaxies in CRISTAL, our ALMA observations also included two SMGs, J1000+0234 \citep{rhc_fraternali21} and CRLE \citep{rhc_pavesi18}, which are within the field of view of CRISTAL-01 and CRISTAL-22 observations, respectively. Their properties are also listed in Table~\ref{table:sample}.  All the ALMA Band 7 observations --new and archival-- for the CRISTAL galaxies and the SMGs were processed together through a common and optimized pipeline described in Section~\ref{sec:pipeline}.

Figure~\ref{Hist_surveys} presents the distribution of redshift, stellar mass, and star formation rate for the CRISTAL galaxies in comparison to the ALPINE \citep{rhc_lefevre20, rhc_bethermin20, rhc_faisst20} and REBELS \citep{rhc_bouwens22} samples. The median properties of the CRISTAL galaxies are: redshift \(z = 5.1\), stellar mass \(M_{\star} = 10^{10.1}~M_{\odot}\), and star formation rate \({\rm SFR} = 58~M_{\odot}~{\rm yr}^{-1}\). 

Figure~\ref{Fig1} shows the CRISTAL galaxies in the context of the stellar mass-SFR plane for star-forming galaxies between $4\leq z \leq6$. The scaling relations between SFR and $M_{\star}$ at $z \approx 5$ from \cite{rhc_speagle14}, \cite{rhc_schreiber15}, and \cite{rhc_khusanova21}, which are displayed in the figure, are generally consistent with one another. The colorscale indicates the redshift of the systems. Galaxies in the original CRISTAL sample and detected in the CRISTAL fields are shown as circles. There are three systems with stellar masses below the $M_{\star}\geq10^{9.5}$ $M_{\odot}$ selection cut. One of these is CRISTAL-12, for which our CIGALE-based stellar mass resulted in $M_{\star}\geq10^{9.3}$ $M_{\odot}$, a factor of $\sim3$ lower than the value in the ALPINE database \citep{rhc_faisst20}. The other two galaxies are the minor companions in the interacting systems CRISTAL-04 and CRISTAL-06, with stellar masses of $10^{8.91}$ $M_{\odot}$ and $10^{9.19}$ $M_{\odot}$, respectively. Finally, the SMGs in the CRISTAL fields are shown as stars. 

\section{Observations} \label{sec:obs}

ALMA observations for the CRISTAL galaxies were carried out during Cycle 8 and Cycle 9, between December 2021 and April 2023. Each galaxy was observed using a combination of a compact (typically C43-1 or -2) and a more extended (typically C43-4 or -5) array configuration. The goal was to spatially resolve the \cii\ line and dust continuum emission on $\sim$kiloparsec scales, while at the same time continue to be sensitive to large-scale ($\sim5\arcsec$ or $\sim15$~kpc), more diffuse \cii\ line and dust emission that has been found in high-$z$ galaxies \citep[e.g.,][]{rhc_fujimoto19,rhc_fujimoto20,rhc_lambert23,rhc_ikeda25,rhc_mitsuhashi24}.

For each CRISTAL source, we determined the required integration time and angular resolution based on the global \cii\ flux from ALPINE \citep{rhc_bethermin20} and the expected \cii\ size. To estimate the latter, we scaled the rest-frame UV size measured from the HST data by a factor of $\times1.5$ following the size analysis of the ALPINE galaxies by \cite{rhc_fujimoto20}. Then, we determined the observing time and angular resolution required to detect ($\gtrsim4\sigma$) and spatially resolve in \cii\ line emission each CRISTAL galaxy with at least $\sim$4 independent beams within one \cii\ effective radius. We chose the best combination of array configurations to achieve the desired angular resolution with the help of the \texttt{simobserve} task in CASA and the recommendations made by the ALMA Observing Tool. The final list of array configurations used and angular resolutions achieved for each CRISTAL target can be found in Table~\ref{table:obs}. 

The number of antennas used for the observations varied from track to track ranging from 41 to 50, with an average of 45. The flux calibrator for most of the CRISTAL sources was the quasar J1058+0133. The detail of the mean number of antennas and flux calibrator used for each CRISTAL target can be found in Table~\ref{table:obs}. As can be seen from the first panel of Figure~\ref{Obs_cristal}, the observing times typically ranged from $\sim4-5$~hr for the most massive sources to $\sim8-15$~hr for the less massive systems.

\begin{figure*}
\centering
   \includegraphics[scale=0.43]{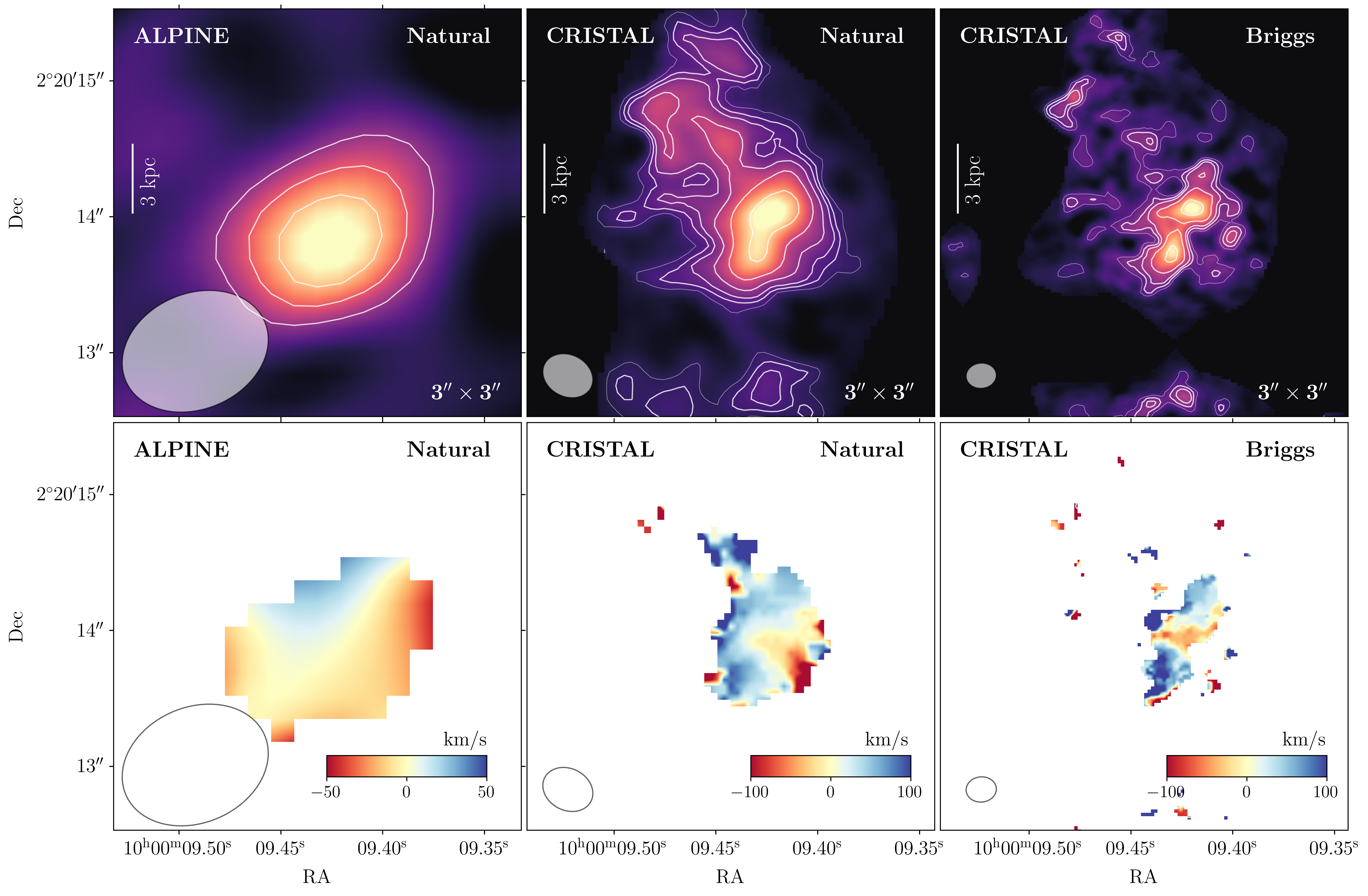}
      \caption{Comparison between the ALPINE and CRISTAL \cii\ observations of the CRISTAL-05 galaxy at $z=5.5$ \citep{rhc_posses24}. {\it (Top)} \cii\ integrated intensity map based on the ALPINE Natural weighting observations (left), and the CRISTAL higher-angular observations using the Natural (without applying a $uv$-taper; center) and Briggs  (right) weighting. {\it (Bottom)} Similar to the top panels, but this time showing the \cii\ velocity field.} \label{C05}
\end{figure*}

The observations were performed in Frequency Division Mode (FDM). In order to fully characterize the \cii\ line profile of each CRISTAL target, in particular of systems with broad line profiles like J1000+0234 in the CRISTAL-01 field, we placed two of the four spectral windows next to each other with an overlap of 0.12~GHz centered at the frequency of the line as detected in the ALPINE lower-angular resolution data \citep{rhc_lefevre20,rhc_bethermin20}. This way we provided contiguous frequency coverage of about 3.6~GHz around the line. The spectral resolution for the two spectral windows assigned for the line detection was set to 3.9~MHz, which corresponds to about $\sim$4~km~s$^{-1}$ for the representative frequency of the CRISTAL galaxies. The remaining two spectral windows were placed in the opposite sideband with the goal of detecting the continuum emission. We chose, however, to use a spectral window with a relatively high resolution of 7.8~MHz ($\sim$8~km~s$^{-1}$) in order to use these data to search for serendipitous line detection in the field given the depth of the CRISTAL data (van Leeuwen et al. in prep.)

\begin{figure*}
\centering
   \includegraphics[width=\hsize]{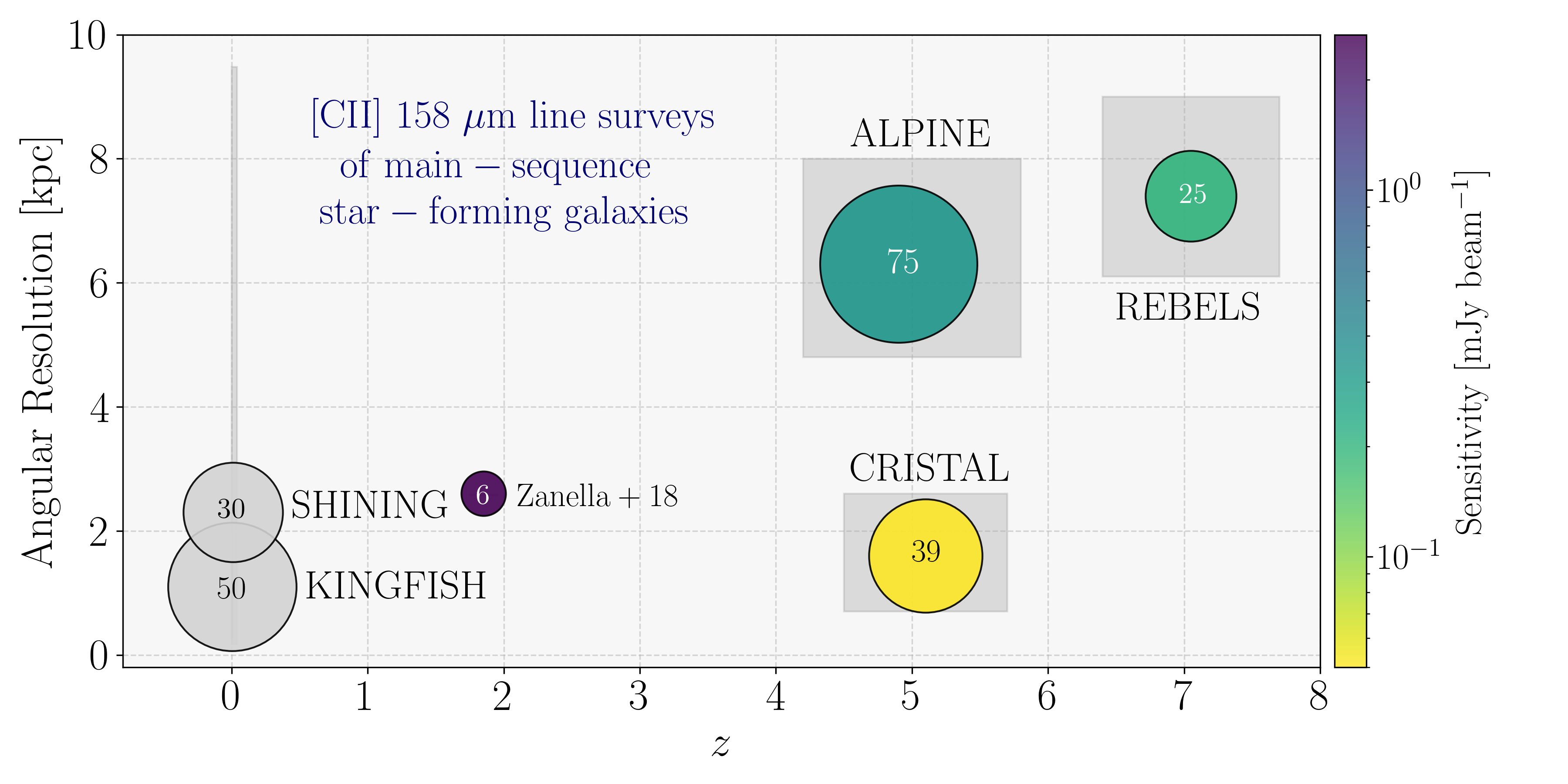}
      \caption{\cii\ 158~$\mu$m line surveys of typical or main-sequence star-forming galaxies as a function of redshift. The surveys included are: Herschel Space Observatory-based KINGFISH \citep{rhc_kennicutt11} and SHINING \citep{rhc_gracia-carpio11,rhc_rhc18a,rhc_rhc18b}, as well as ALMA-based ALPINE \citep{rhc_lefevre20, rhc_bethermin20, rhc_faisst20}, REBELS \cite{rhc_bouwens22}, CRISTAL, and \cite{rhc_zanella18}. The circles represent the average redshift and physical angular resolution (in kiloparsecs) achieved by each survey. The circle sizes indicate the number of galaxies detected in \cii\ line emission in these surveys (also listed inside each circle). The gray boxes indicate the full range of redshift and angular resolution covered by each survey. The color bar illustrates the sensitivity achieved in 200 km s$^{-1}$ channels, except for the {\it Herschel}-based surveys.} \label{survey_comparison}
\end{figure*}

\section{Data Processing} \label{sec:pipeline}

\subsection{ALPINE-ALMA} \label{sec:alpine}

All CRISTAL galaxies were first observed with ALMA Band 7 as part of the ALPINE Large Program \citep{rhc_lefevre20,rhc_bethermin20,rhc_faisst20}, and some also as part of the \cite{rhc_capak15} program (CRISTAL-02, 03, 05, 07, 10, 20, 21, 22). The main differences between the ALPINE and the CRISTAL observations are two:

\begin{enumerate}
    \item ALPINE was a program with the aim of building a large sample of star-forming galaxies at $4\lesssim z\lesssim6$ detected in \cii\ line emission. For this reason ALMA observations were carried out in the compact array configuration, achieving angular resolutions typically of $\theta_{\rm beam}\sim0\farcs8-1\farcs2$ \citep{rhc_bethermin20}, which corresponds to physical scales of $\sim5-7$~kpc at this redshift. With CRISTAL, our goal was to spatially resolve our sources, achieving as close as possible to $\sim$kiloparsec scale resolution. 
    \item ALPINE observations were carried out in Time Division Mode (TDM), which implies a spectral resolution of 31.2~MHz. For a typical Band 7 representative frequency of 300~GHz, this corresponds to a velocity resolution of 31.2~km~s$^{-1}$. In contrast, CRISTAL observations were carried out in FDM mode, achieving a spectral resolution $8\times$ higher.
\end{enumerate}

In the next section (\ref{sec:combination}) we describe how we combined the ALPINE and CRISTAL data and address the differences in the spectral observing mode (TDM versus FDM).

\subsection{Combination, calibration, and imaging} \label{sec:combination}

Data calibration and a combination of the different observations were performed using the CASA software \citep{rhc_casa22}. Table~\ref{table:ALMA_ID} lists the IDs of the ALMA observing programs that were used to produce the CRISTAL data products. 

We processed the datasets from the different programs using the corresponding pipeline versions: 5.6.1 for the ALPINE program, 5.6.1 for the CRISTAL pilot program, and finally, version 6.5.2 for the CRISTAL program. No extra manual flagging was needed beyond what was already identified by the observatory and automatically flagged by each pipeline. We combined the calibrated datasets into a single measurement set ({\it ms}) used to create the images and data cubes, using the task \texttt{concat}. This task and the following procedure are performed using version 6.5.2 of CASA software. For the concatenation process, we opted for the default values for the frequency (\texttt{freqtol}) and direction (\texttt{dirtol}) tolerances. This decision was deliberate, as it ensures that all individual field and spectral windows (SPWs) are kept separately in the concatenated {\it ms}. This approach effectively handles any combination of the observations by \texttt{tclean}, not \texttt{concat}. 

During the whole process of image and data cube creation, we used clean and the \texttt{auto-multithresh} algorithm. This mode automatically masks regions based on the signal-to-noise of the emission in the image. The creation of these regions is determined by the following parameters: \texttt{Sidelobethreshold} = 2, \texttt{Noisethreshold} = 4.5, \texttt{Lownoisethreshold} = 2, and \texttt{Minbeamfrac} = 0.0. This value choice is intended to match similar results when using manual cleaning. In all the cases, the cleaning was performed down to 1$\sigma$ by selecting \texttt{nsigma} = 1, where $\sigma$ is estimated automatically by \texttt{tclean} using robust statistics ($\sigma = 1.4826\times {\rm MAD}$ with MAD being the median absolute deviation). The critical parameter is \texttt{Noisethreshold} = 4.5, which puts cleaning regions around pixels with signal-to-noise higher than 4.5. This value works well for continuum images, where the combination of the synthesized beam and image size results in 4.5 being a reasonable limit for significant positive emission. We kept the same value for the cubes, even where noise peaks above 4.5 are more common than in the multi-frequency synthesis ({\it mfs})  images. We checked that the number of 4.5 noise peaks in the cubes was low and that the background noise distribution did not change by cleaning them. 

\begin{figure}[h!]
\centering
   \includegraphics[width=\hsize]{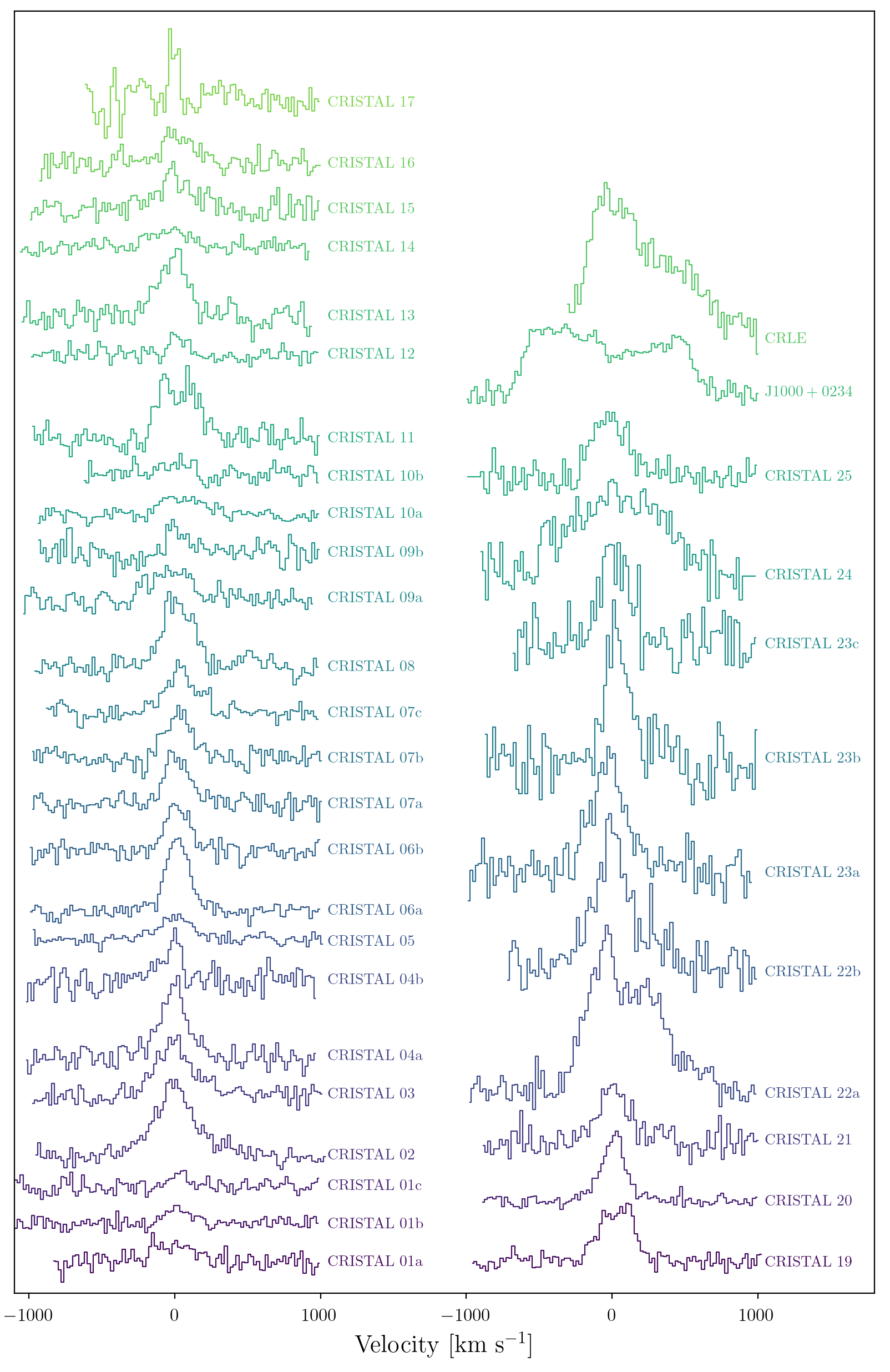}
      \caption{\cii\ 158~$\mu$m line spectra for all the star-forming galaxies in the CRISTAL sample listed in Table~\ref{table:sample}, except for CRISTAL-18 which is undetected. \cii\ line intensities are arbitrary.} \label{spec_all}
\end{figure}

The first step to create the data cubes and continuum maps was to create an initial {\it mfs} using natural weighting and a $uv$-taper to 1 arcsec to identify the presence of detections manually. Then, we created data cubes using natural weighting and a channel width of 40~km~s$^{-1}$ for each of the sidebands. These cubes are used to identify the frequency ranges where the \cii\ line is detected and define the spectral range to avoid when creating the continuum images. Pure continuum images were then created by avoiding the frequency ranges where the line was identified. In the case of continuum emission detected at the position of the central CRISTAL galaxy, we then subtract the continuum using \texttt{uvcontsub} and \texttt{fitorder} = 0. 

From the {\it ms} file with the continuum subtracted, we created data cubes with different spectral resolutions. First, we measured the full-width at half maximum (FWHM) in the same cube where the frequency range for the \cii\ detection was selected. Subsequently, we generated spectral cubes with resolutions of FWHM/5, 20~km~s\(^{-1}\), and 10~km~s\(^{-1}\). All these cubes cover a velocity range from \(-1000\)~km~s\(^{-1}\) to \(+1000\)~km~s\(^{-1}\), relative to the reference frequency of the \cii\ emission as defined by the source in the ALPINE catalog \citep{rhc_lefevre20}. The cubes were created using three different weighting schemes: $(i)$ natural weighting, $(ii)$ Briggs weighting with \texttt{robust} = 0.5, and $(iii)$ natural weighting combined with a \(uv\)-taper of 1\arcsec.

The final step was to inspect the synthesized beams for the continuum and cubes created with Natural and Briggs weighting and retrieve their BMAJ value ranges. We then selected the maximum value of BMAJ in those products and used it to circularize the beams for the continuum and spectral cubes. We used the same parameters for creating the continuum and \cii\ cubes as stated above, but now we are setting the  \texttt{restoringbeam} parameter into a circular beam of size with the maximum value of BMAJ. These cubes should be a good reference for comparing the properties of the continuum and \cii\ emission in beam-by-beam or pixel-to-pixel basis. 
 
One final note on the ALMA-CRISTAL data reduction is that the products presented here are not corrected for the ``JvM effect'' \citep{rhc_jorsater95,rhc_czekala21}. However, the impact of this effect is expected to be minimal because the data were cleaned deeply, using $\texttt{nsigma} = 1$. The ``JvM effect'' occurs in datasets combining multiple array configurations of an interferometer, where the TCLEAN algorithm produces final images that mix dirty and clean beam units. A detailed analysis of the JvM correction for the individual source CRISTAL-05 is provided in \cite{rhc_posses24}, while a broader assessment of its impact on the full CRISTAL sample will be presented in González-López et al. (in prep.).

\begin{figure}
\centering
   \includegraphics[width=\hsize]{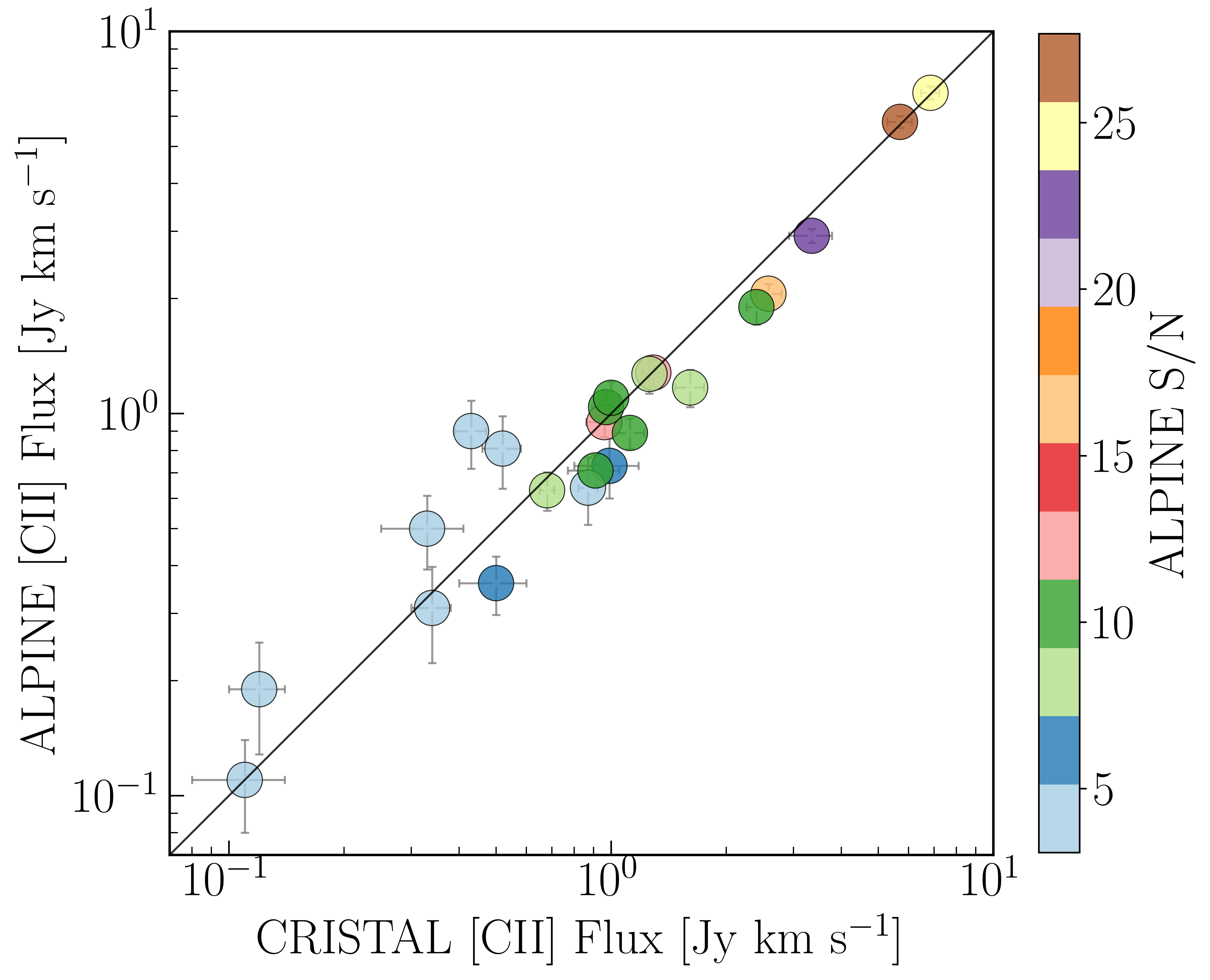}
      \caption{\cii\ integrated flux measured from the CRISTAL and ALPINE data. The colorscale represents the $S/N$ of the integrated line detection in the ALPINE data \citep{rhc_bethermin20}. In general there is good agreement between the CRISTAL and ALPINE fluxes, except for four systems that were detected with $S/N\lesssim5$ in the ALPINE data.} \label{flux_comp}
\end{figure}

\subsection{Moment maps} \label{sec:moments}

We generated moment maps for integrated intensity, intensity-weighted velocity, and velocity dispersion (moments 0, 1, and 2). The maps were produced in Python, with two versions created. In the first version we integrated the line emission over a frequency range defined by the full width at tenth maximum (FWTM). We refer to the resulting moment maps as the “FWTM-mask” moment maps. The second version involved applying a blanking mask to suppress noise that could otherwise dominate faint line emission. Starting from the naturally-weighted cube binned in 20 km~s$^{-1}$ channels, we convolved with a $\sigma=100$ km s$^{-1}$ Gaussian kernel along the velocity axis, and a $\sigma = 10$ pix -  2D Gaussian kernel in the spatial axes (the pixel size in arcseconds depends on the beam size of the cube, and it was determined using the CASA extension tool \texttt{pickCellSize} with \texttt{npix}=7). We then measured the rms in the signal-free regions of the convolved cube. Finally, we split cells above and below a $2\times {\rm rms}$ threshold into a 3D mask, which we then fed to casa task \texttt{immoments} to obtain the intensity, velocity, and velocity dispersion maps from the original cube. We refer to the resulting moment maps as the “dilated-mask” moment maps.

The dilated-mask maps are particularly valuable for identifying systems within the \cii\ cubes that exhibit line emission outside the frequency range used to generate the FWTM moment maps of the main CRISTAL targets. They are also effective in detecting companion systems with a central frequency within the FWTM frequency range but with significantly narrower \cii\ line profiles. This is the case for example of the detection of CRISTAL-01c, located approximately 32 kpc away from CRISTAL-01a, which we confirmed as a real detection after identifying its stellar counterpart in NIRCam imaging of the field. A full analysis of the detection of serendipitous sources in the \cii\ cubes of the CRISTAL fields will be presented in van Leeuwen et al. (in prep). Another demonstration of the utility of the dilated-mask approach is the detection of extended emission around CRISTAL galaxies where the gas traced by the \cii\ line emission extends beyond the stellar light traced by JWST and HST images, such as the case of CRISTAL-09 or CRISTAL-13 (see Section~\ref{multi-view}).

\subsection{Achieved beam sizes and sensitivities}

The achieved synthesized beam sizes for the Natural- and Briggs-weighted \cii\ cubes are listed in Table~\ref{table:obs}. For the natural weighting, the median size of the minor axis is 0\farcs43 (ranging from 0\farcs11 to 0\farcs68) and for the major axis is 0\farcs53 (ranging from 0\farcs12 to 0\farcs82). For the Briggs weighting, the median size of the minor axis is 0\farcs26 (ranging from 0\farcs08 to 0\farcs44) and for the major axis is 0\farcs30 (ranging from 0\farcs08 to 0\farcs54). The second panel of Figure~\ref{Obs_cristal} displays the distribution of beam sizes (calculated as the geometric average between the minor (BMIN) and major (BMAJ) beam axis size) for the Briggs weighting applied to the CRISTAL galaxies.

To illustrate the improvement in angular resolution achieved by CRISTAL relative to ALPINE, Figure~\ref{C05} shows the \cii\ integrated (top) and velocity field (bottom) maps of CRISTAL-05 (vuds\_cosmos\_5100822662, HZ3). From left to right, we show the ALPINE and the CRISTAL data using Natural and Briggs weighting. The increase in the angular resolution going from the ALPINE  ($\sim1$\arcsec, $\sim6$~kpc) to the CRISTAL ($\sim0\farcs2$, $\sim1.2$~kpc) observations reveal that this system is not a single source, but in fact an interacting system with at least two major components, plus extended emission for about $\sim10$~kpc in the north-east direction. The kinematics go from what could be considered a smooth velocity gradient in the ALPINE data, to complex, disturbed kinematics consistent with a multi-component system. A detailed morpho-kinematic analysis of this source is presented in \cite{rhc_posses24}.

Table~\ref{table:obs} also list the noise measured in the \cii\ cubes in channels of 20 km~s$^{-1}$ width and the continuum maps. For the cubes, and for the Natural and Briggs weighting, the median noise levels are 0.15 mJy~beam$^{-1}$ and  0.18 mJy~beam$^{-1}$, respectively. For the continuum maps,  the median noise levels are 10.3 $\mu$Jy~beam$^{-1}$ and 11.2 $\mu$Jy~beam$^{-1}$ for the Natural and Briggs weighting, respectively. The third and fourth panels of Figure~\ref{Obs_cristal} show the distribution of noise measured in the Natural cubes (20 km~s$^{-1}$ channels) and continuum images, respectively.

All the relevant ALMA data products discussed here can be downloaded from the CRISTAL data repository: \href{http://www.cristal.udec.cl}{www.cristal.udec.cl}.

\begin{figure*}
\centering
   \includegraphics[width=\hsize]{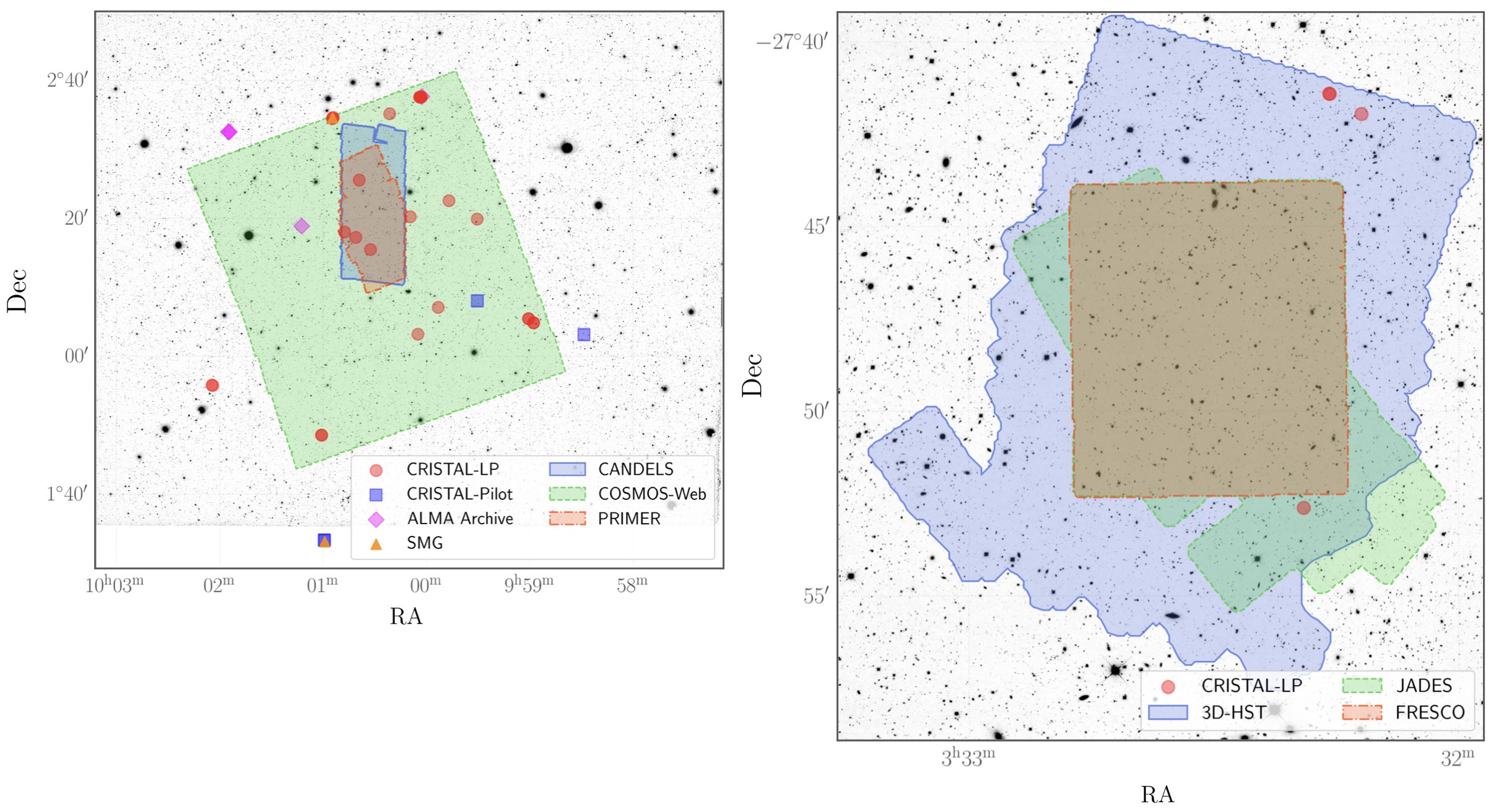}
      \caption{Position in the sky of the CRISTAL galaxies including the JWST pointings footprints from programs COSMOS-Web, PRIMER, JADES, and FRESCO, and HST pointings footprints from programs COSMOS-DASH, CANDELS, and 3D-HST. CRISTAL galaxies from the original sample, pilot programs, and extracted from the ALMA archive are shown as red circles, blue squares, and magenta diamonds, respectively. SMGs in the field of CRISTAL galaxies are shown as orange triangles. CRISTAL galaxies that are not part of large HST surveys are covered by individual HST programs.} \label{coverage}
\end{figure*}

\subsection{Comparison to other surveys of \cii\ 158~$\mu$m line emission in normal, star-forming galaxies}

To contextualize the contribution of the CRISTAL survey to the study of normal, main-sequence galaxies, Figure~\ref{survey_comparison} summarizes \cii\ 158~$\mu$m line surveys of star-forming galaxies at different cosmic epochs.

At $z\sim0$, we include the {\it Herschel} Space Observatory PACS \citep{rhc_pilbratt10,rhc_poglitsch10} KINGFISH \citep{rhc_kennicutt11} and SHINING \citep{rhc_gracia-carpio11,rhc_rhc18a,rhc_rhc18b} surveys. For consistency, we exclude galaxies that significantly deviate from the main sequence, such as elliptical galaxies, dwarf systems, and luminous infrared galaxies. At $z\sim2$, we include the sample of six main-sequence star-forming galaxies studied by \cite{rhc_zanella18}. At $4\lesssim z \lesssim6$, we include the ALPINE and CRISTAL surveys. For reference, at $6\lesssim z \lesssim8$, we include the REBELS survey \citep{rhc_bouwens22} as the closest match at higher redshift to the ALPINE and CRISTAL surveys.

Figure~\ref{survey_comparison} illustrates the physical resolution achieved by each survey (y-axis). The circles are centered at the average resolution, while the gray boxes indicate the range covered by each survey. The circle size represents the number of galaxies detected in \cii\ 158~$\mu$m line emission, and the color scale indicates the sensitivity of each survey (measured in 200~km~s$^{-1}$ channels), excluding the {\it Herschel}-based surveys. While sensitivity measurements for the KINGFISH and SHINING surveys are unavailable, the lower end of \cii\ surface brightness for KINGFISH regions, around $\sim10^5$~$L_{\odot}$~kpc$^{-2}$ \citep[e.g.,][]{rhc_rhc15}, is comparable to the sensitivity achieved by the CRISTAL survey.

Building on the pioneering work of surveys like ALPINE and REBELS, the CRISTAL survey extends the study of star-forming galaxies to $\sim$kiloparsec scales for a significant sample when the Universe was $\sim$1 Gyr old. Notably, CRISTAL galaxies also have rest-frame UV and optical observations (see Section~\ref{sec:ancillary}), providing a comprehensive view of the gas, dust, and stars in early galaxies. This approach parallels the detailed understanding of nearby galaxies achieved over the past decade through a combination of {\it Herschel}/PACS observations, ground-based optical data, and GALEX UV imaging.

\section{Ancillary data} \label{sec:ancillary}

Most of the CRISTAL galaxies are located in the COSMOS field \citep{rhc_scoville07,rhc_weaver22}, which implies that there is a wealth of optical and near-infrared data available \citep[e.g.,][]{rhc_faisst20}. Figure~\ref{coverage} shows the position in the sky of the CRISTAL galaxies and the JWST and HST footprints pointings from large surveys such as COSMOS-Web \citep[PID 1727; co-PIs: Kartaltepe \& Casey; ][]{rhc_casey23}, PRIMER (PI: Dunlop, PID 1837), COSMOS-CANDELS \citep{rhc_grogin11}, COSMOS-DASH \citep{rhc_mowla19} and 3D-HST \citep{rhc_bramer12}. Table~\ref{table:ancillary} summarizes the HST/WFC3 and the JWST Near Infrared Camera \citep[NIRCam; ][]{rhc_rieke23} data available for each CRISTAL source. In this section we describe how we reduced and processed the HST and JWST data. 

\subsection{HST}

We employed the \texttt{grizli} pipeline \citep{rhc_brammer23} to retrieve and process archival HST Advanced Camera for Surveys (ACS) and HST/WFC3 data for all selected targets. The pipeline automatically retrieves, calibrates, and re-samples the individual raw exposures that overlap with the target coordinates for each filter. The calibrated frames were then precisely aligned using several astrometric reference catalogs, including the DESI Legacy Imaging Survey DR9, PANSTARRS (PS1), and Gaia, leading to an uncertainty of about 0\farcs1. Finally, the aligned frames were combined into the final mosaics.

\subsection{JWST/NIRCam}

The extensive JWST/NIRCam data available for the CRISTAL galaxies enable us to probe the rest-frame optical light, complementing the HST data, which is limited to the rest-frame UV and sensitive to dust obscuration. Of the 25 main systems in our sample, 19 have been observed with NIRCam, with the longest wavelength covered by the F444W filter, as part of several programs: PRIMER (PI: Dunlop, PID 1837), COSMOS-Web \citep[PID 1727; co-PIs: Kartaltepe \& Casey; ][]{rhc_casey23}, GOODS-S (PID 1286, PI Eisenstein), and various GO programs (PIDs 3215, 2198, 3990). 

For the CRISTAL galaxies in the PRIMER-COSMOS field, we used the data products from the Dawn JWST Archive (DJA) Mosaic release v7 \citep{rhc_heintz25}, which used the \texttt{grizli}  pipeline \citep{rhc_brammer23}. For the rest of the CRISTAL systems, we processed the data using a modified version of the official JWST pipeline (version 1.10.0, pmap 1075). The processing followed the methods outlined in \citet{rhc_bagley23} to remove wisps, snowballs, and $1/f$ noise, applying flat field corrections and wisp templates from the NIRCam team. We also removed stripes using de-striping techniques aligned with the diffraction spikes of the PSF, specifically addressing tilted stripes affecting PRIMER data. We then applied constant background subtraction following prescription from \citet{rhc_bagley23}. The final mosaic was drizzled to a pixel scale of $0\farcs03$ with the default square kernel and pixel fraction of 1.0. For the astrometry, we first use the 3D-HST catalog and align it to Gaia DR3, then use the astrometry-corrected 3D-HST catalog as the absolute reference catalog for the JWST pipeline, so that the final mosaic has an astrometry corrected. 

\begin{figure*}
\centering
   \includegraphics[scale=0.41]{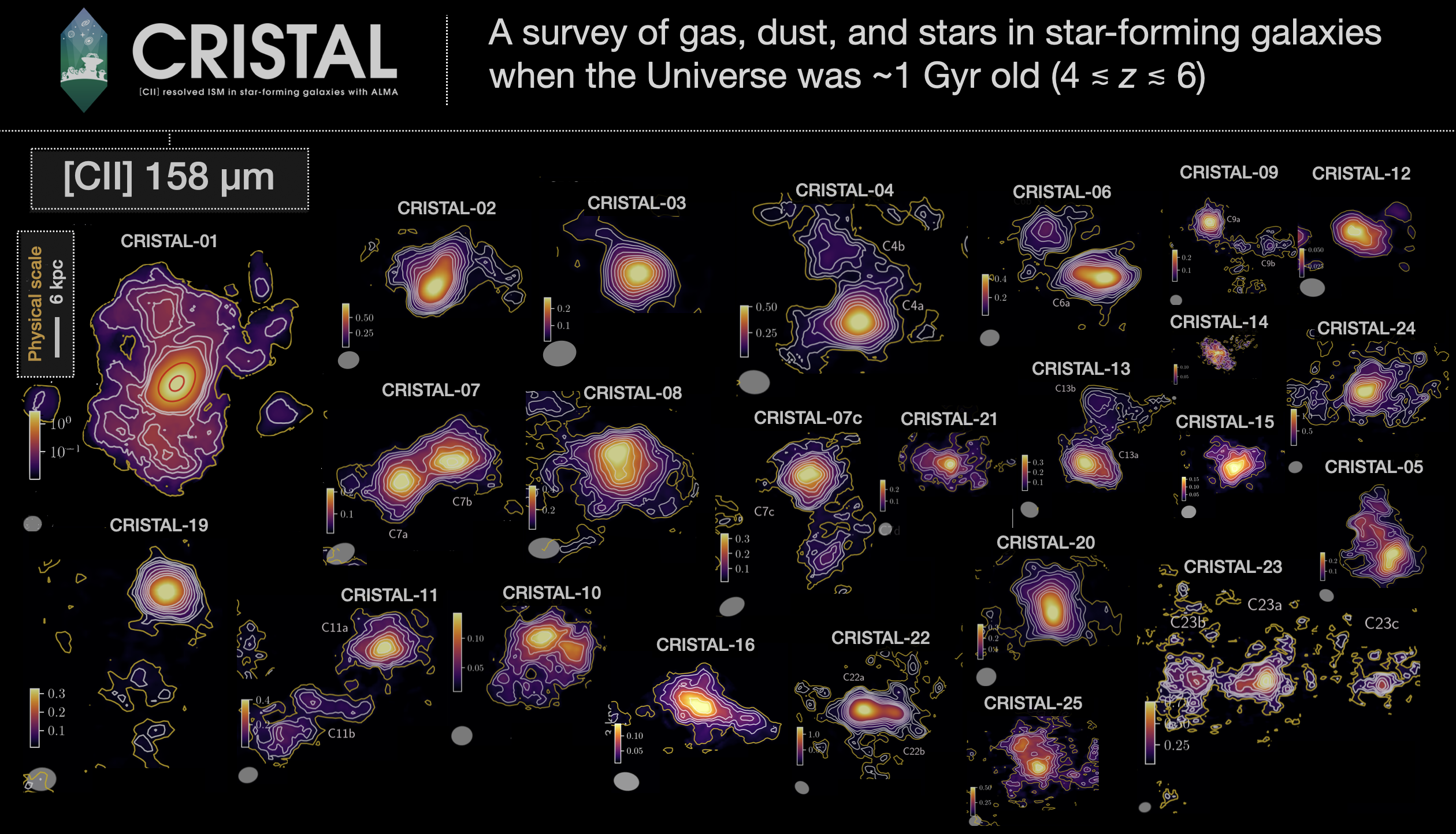}
      \caption{\cii\ integrated line emission maps for the CRISTAL galaxies, constructed from the naturally weighted cubes. All maps have been scaled to the same physical size, indicated by a white line on the left side representing 6 kpc. The colorscale represents the integrated flux emission, and the contours correspond to [3, 4, 5, 7, 9, 11, 13, 15]$\sigma$, except for CRISTAL-01, where the emission intensity is shown in logarithmic scale and the contours represent the [4, 6, 8, 10, 50, 150]$\sigma$ levels (the last two in red). The beam is shown in the lower left corner of each panel. } \label{family_portrait}
\end{figure*}

\begin{figure*}
\centering
   \includegraphics[scale=0.41]{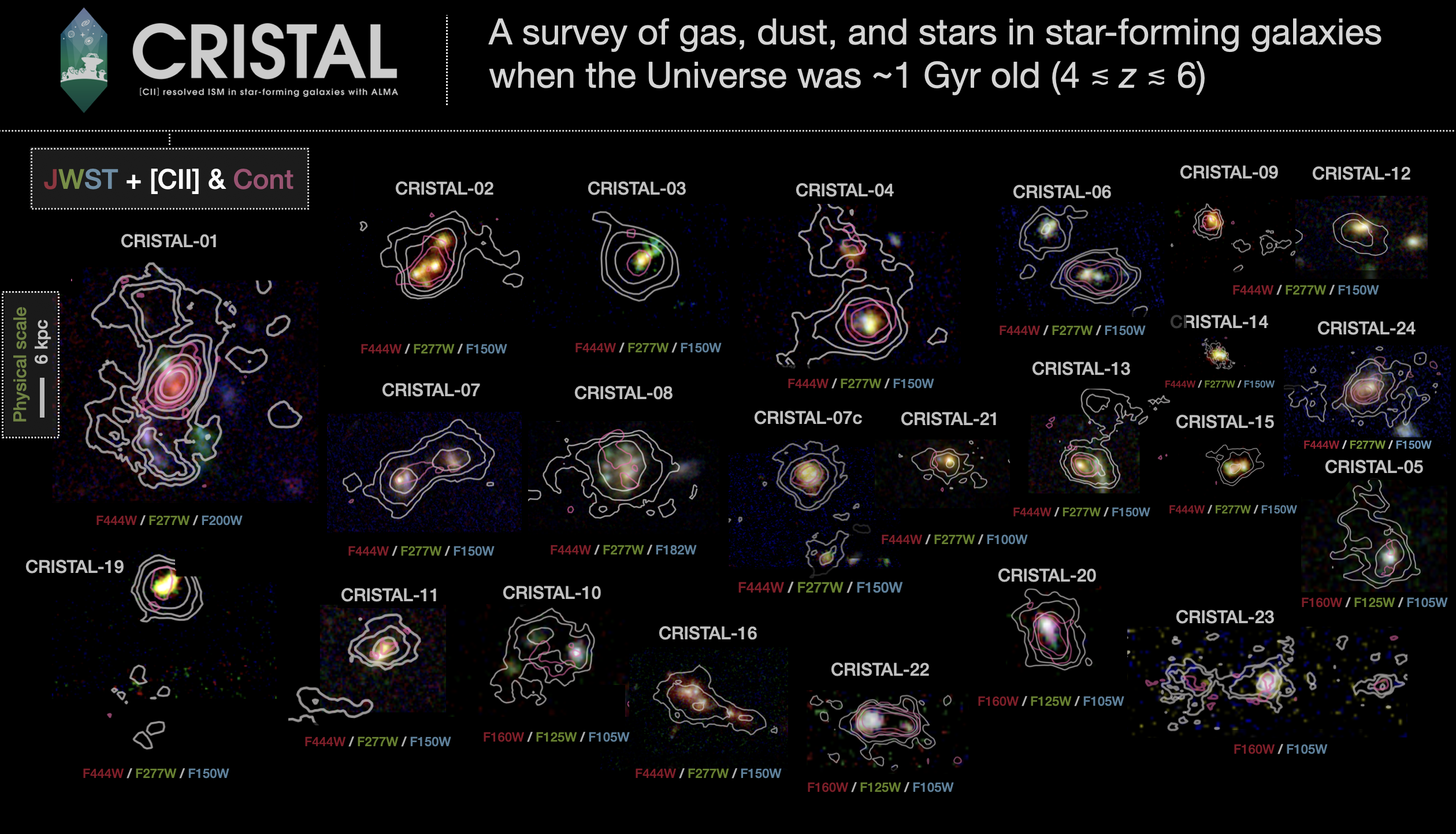}
      \caption{JWST composite images of CRISTAL galaxies, overlaid with \cii\ line emission (white contours) and dust continuum emission (pink contours). Contour levels correspond to [3, 5, 10]$\sigma$, except for CRISTAL-01, where the emission intensity is shown on a logarithmic scale, and contours represent [4, 6, 8, 20, 60, 100]$\sigma$ levels. All maps are scaled to the same physical size, with a 6 kpc scale bar shown as a white line on the left side of each panel. For galaxies CRISTAL-10, -20, -22, and -23 we show the HST composite images as there are no JWST/NIRCam observations available.} \label{family_portrait_jwst}
\end{figure*}

\section{Results} \label{sec:results}

\begin{table}
\caption{CRISTAL linewidths and fluxes}             
\label{table:fluxes}      
\centering                          
\begin{tabular}{l c c c}        
\hline\hline                
Name	&		 \cii\ FWHM				&			 \cii\ Flux			&			Band 7 Cont. 			\\
	&		km s$^{-1}$				&			Jy km s$^{-1}$			&			mJy			\\
\hline	&						&						&						\\
CRISTAL-01a	&	$	375.5	\pm	47	$	&	$	0.99	\pm	0.19	$	&	$	<0.03			$	\\
CRISTAL-01b	&	$	205.9	\pm	31	$	&	$	0.98 	\pm 0.13	$	&	$	0.47	\pm	0.08	$	\\
CRISTAL-01c	&	$	117.3	\pm	28	$	&	$	0.37 	\pm 0.07	$	&	$	<0.03			$	\\
CRISTAL-02	&	$	326.2	\pm	12	$	&	$	2.58	\pm	0.21	$	&	$	0.29	\pm	0.04	$	\\
CRISTAL-03	&	$	290.9	\pm	16	$	&	$	0.43	\pm	0.04	$	&	$	0.06	\pm	0.02	$	\\
CRISTAL-04a	&	$	210.8	\pm	12	$	&	$	0.86	\pm	0.08	$	&	$	0.18	\pm	0.04	$	\\
CRISTAL-04b 	&	$	120.8	\pm	19	$	&	$	0.43	\pm	0.07	$	&	$	0.09	\pm	0.05	$	\\
CRISTAL-05	&	$	333.0	\pm	24	$	&	$	0.96	\pm	0.10	$	&	$	0.18	\pm	0.06	$	\\
CRISTAL-06a	&	$	183.3	\pm	7	$	&	$	1.85	\pm	0.14	$	&	$	0.26	\pm	0.04	$	\\
CRISTAL-06b	&	$	186.1	\pm	12	$	&	$	0.55	\pm	0.06	$	&	$	<0.03			$	\\
CRISTAL-07a	&	$	199.5	\pm	14	$	&	$	0.62	\pm	0.06	$	&	$	0.08	\pm	0.02	$	\\
CRISTAL-07b	&	$	214.1	\pm	16	$	&	$	0.64	\pm	0.08	$	&	$	0.06	\pm	0.01	$	\\
CRISTAL-07c	&	$	258.0	\pm	16	$	&	$	0.77	\pm	0.08	$	&	$	0.08	\pm	0.02	$	\\
CRISTAL-07d	&	$	374.9	\pm	73	$	&	$	0.73	\pm	0.07	$	&	$	<0.03			$	\\
CRISTAL-08	&	$	247.0	\pm	12	$	&	$	1.61	\pm	0.14	$	&	$	0.25	\pm	0.07	$	\\
CRISTAL-09a	&	$	305.0	\pm	31	$	&	$	0.54	\pm	0.05	$	&	$	0.32	\pm	0.08	$	\\
CRISTAL-09b	&	$	147.1	\pm	31	$	&	$	0.33	\pm	0.09	$	&	$	<0.03			$	\\
CRISTAL-10a	&	$	324.3	\pm	31	$	&	$	0.50	\pm	0.10	$	&	$	0.28	\pm	0.06	$	\\
CRISTAL-10b	&	$	251.7	\pm	63	$	&	$	0.60	\pm	0.17	$	&	$	<0.03			$	\\
CRISTAL-11	&	$	306.0	\pm	24	$	&	$	0.97	\pm	0.08	$	&	$	0.23	\pm	0.07	$	\\
CRISTAL-12	&	$	124.3	\pm	24	$	&	$	0.12	\pm	0.02	$	&	$	<0.03			$	\\
CRISTAL-13a	&	$	234.5	\pm	16	$	&	$	1.12	\pm	0.10	$	&	$	0.12	\pm	0.03	$	\\
CRISTAL-13b	&	$	88.0	\pm	29	$	&	$	0.41	\pm	0.07	$	&	$	<0.05			$	\\
CRISTAL-14	&	$	307.4	\pm	33	$	&	$	0.33	\pm	0.08	$	&	$	<0.03			$	\\
CRISTAL-15 	&	$	316.8	\pm	35	$	&	$	0.52	\pm	0.06	$	&	$	<0.05			$	\\
CRISTAL-16	&	$	255.7	\pm	31	$	&	$	0.34	\pm	0.04	$	&	$	0.10	\pm	0.04	$	\\
CRISTAL-17 	&	$	154.9	\pm	69	$	&	$	0.11	\pm	0.03	$	&	$	<0.03			$	\\
CRISTAL-18	&	$				$	&	$	<0.03			$	&	$	<0.03			$	\\
CRISTAL-19	&	$	277.3	\pm	14	$	&	$	0.68	\pm	0.03	$	&	$	0.09	\pm	0.02	$	\\
CRISTAL-20	&	$	192.9	\pm	7	$	&	$	1.00	\pm	0.05	$	&	$	0.15	\pm	0.02	$	\\
CRISTAL-21	&	$	255.7	\pm	26	$	&	$	0.91	\pm	0.14	$	&	$	0.41	\pm	0.15	$	\\
CRISTAL-22a	&	$	588.2	\pm	24	$	&	$	4.08	\pm	0.23	$	&	$	1.04	\pm	0.09	$	\\
CRISTAL-22b	&	$	348.5	\pm	19	$	&	$	2.75	\pm	0.18	$	&	$	0.73	\pm	0.06	$	\\
CRISTAL-23a	&	$	281.8	\pm	16	$	&	$	4.48	\pm	0.46	$	&	$	0.77	\pm	0.21	$	\\
 CRISTAL-23b	&	$	177.9	\pm	14	$	&	$	2.37	\pm	0.37	$	&	$	0.37	\pm	0.10	$	\\
CRISTAL-23c	&	$	240.6	\pm	24	$	&	$	1.50	\pm	0.24	$	&	$	0.23	\pm	0.05	$	\\
CRISTAL-24	&	$	754.1	\pm	35	$	&	$	5.70	\pm	0.42	$	&	$	0.74	\pm	0.09	$	\\
CRISTAL-25	&	$	274.7	\pm	19	$	&	$	3.35	\pm	0.43	$	&	$	0.44	\pm	0.11	$	\\
\hline 
\end{tabular}
\end{table}

\subsection{Linewidths and fluxes}

Figure~\ref{spec_all} shows the \cii\ line spectra extracted for all CRISTAL galaxies, plus the two sub-millimeter galaxies J1000+0234 and CRLE located in CRISTAL fields (see Table~\ref{table:sample} for details). The global \cii\ spectra of all CRISTAL galaxies can be well fitted by a single Gaussian component, except four: CRISTAL-02, -04, -06 and -22. For CRISTAL-02, there is evidence that the need for a broad, second Gaussian component could be associated with widespread outflow activity. This is discussed in detail in \citep{rhc_davies25}. For CRISTAL-04, -06 and -22, the most likely explanation for the need of a second Gaussian component is the interacting nature of these systems. This is discussed in more detailed in the kinematic analysis of the CRISTAL galaxies presented in Lee et al. (in prep). 

Based on a single Gaussian fit, we measure the FWHM of the \cii\ line. The results can be found in Table~\ref{table:fluxes}. The linewidths range from $\sim$250-350~km~s$^{-1}$, for the most massive systems, to $\sim$120~km~s$^{-1}$ for the small companions detected around some of the main CRISTAL systems (e.g., CRISTAL-01c, CRISTAL-04b).
  
To measure the \cii\ and continuum fluxes, we use the Natural-weighted integrated intensity maps (see Section~\ref{sec:moments}) and three different methods. 

\begin{enumerate}
    \item {\it Aperture photometry:} Using the {\it astropy photutils} package, we used circular apertures of increasing size to measure the integrated flux and construct a flux curve-of-growth. We identified the radius where the curve-of-growth flattens out or where an increase in flux due to the presence of a companion galaxy is detected. This radius was used to measure the total flux. This method can be useful to disentangle the emission from interacting systems, as the presence of a companion system can be identified by a second increase in the enclosed emission after the initial convergence of the flux curve. \\
    
    \item  {\it Two-dimensional Gaussian fit:}  Using the CASA tool {\it imfit}, we measured the flux fitting a two-dimensional elliptical Gaussian choosing a fitting box with a variable size depending on the source size. For interacting systems, we defined the fitting box to minimize any additional flux contribution from the companion galaxy. The error in the flux measurement was calculated following the formalism described by \cite{rhc_condon97}.  \\
    
    \item  {\it Flux above a $S/N$ threshold:}  We measured the flux by integrating the emission above a specified $S/N$ threshold, analogous to the isophotal magnitude method used in optical astronomy. This technique can be particularly useful for measuring flux in sources with complex structures. However, it encounters difficulties in the presence of interacting systems, especially when galaxies are connected by diffuse emission.  To address this, we restricted our flux measurements to the area within the fitting box defined for the two-dimensional Gaussian fit method. For our measurements, we used a $S/N$ threshold of 2$\sigma$ on the integrated map.
    
\end{enumerate}

The \cii\ and continuum flux measurements obtained using the three methods are consistent within 10\%, with the exception of CRISTAL-04b, -07a, -07b, and -09b. In these cases, the flux measured using the $S/N$ threshold method is approximately 20\% higher than the flux obtained from the other methods. This discrepancy is expected, as these four galaxies are part of interacting systems connected by diffuse emission. 

Table~ \ref{table:fluxes} lists the \cii\ and dust continuum fluxes for the CRISTAL galaxies based on the aperture photometry method. The CRISTAL fluxes are consistent with those measured from the low-angular resolution ALPINE data reported by \cite{rhc_bethermin20}, with discrepancies within $\sim15\%$, as illustrated in Fig.~\ref{flux_comp}. The exceptions are four galaxies where we observe differences of around 50\%. In these cases, however, the integrated \cii\ line emission was detected with $S/N \lesssim 5$ in the ALPINE data, which could be the cause for the discrepancy. 

The \cii\ and continuum fluxes in Table~\ref{table:fluxes} agree on average within a $\sim10\%$ and $\sim20\%$ with the fluxes measured in \cite{rhc_ikeda25} and \cite{rhc_mitsuhashi24}. The latter fluxes were measured directly in the visibility plane, making them independent of imaging parameter choices.

\subsection{CRISTAL Family Portrait}

Figure~\ref{family_portrait} shows the integrated \cii\ 158~$\mu$m line emission maps for all CRISTAL galaxies, providing a comprehensive view of the gas distribution across the sample. Complementing this, Figure~\ref{family_portrait_jwst} displays composite JWST/NIRCam images of the same galaxies, overlaid with contours of \cii\ line and dust continuum emission. To ensure consistent comparison, all maps are scaled to the same physical size.

These visualizations highlight the diversity in sizes, morphologies, and structural features among CRISTAL galaxies, offering valuable insights into their assembly histories and evolutionary pathways during the first $\sim1$~Gyr of the Universe. A significant number exhibit evidence of morphological disturbances, likely driven by dynamic processes in their environments. For example, systems such as CRISTAL-02, -03, -09, -12, and -13 feature extended emission structures reminiscent of tidal tails, suggesting interactions with minor companions. Meanwhile, galaxies like CRISTAL-01, -04, -06, and -07 appear to be undergoing merger events, with their morphologies indicating complex dynamical states. The properties of individual CRISTAL systems are discussed in detail in the following section.

\begin{figure*}[h!]
\centering
   \includegraphics[width=\hsize]{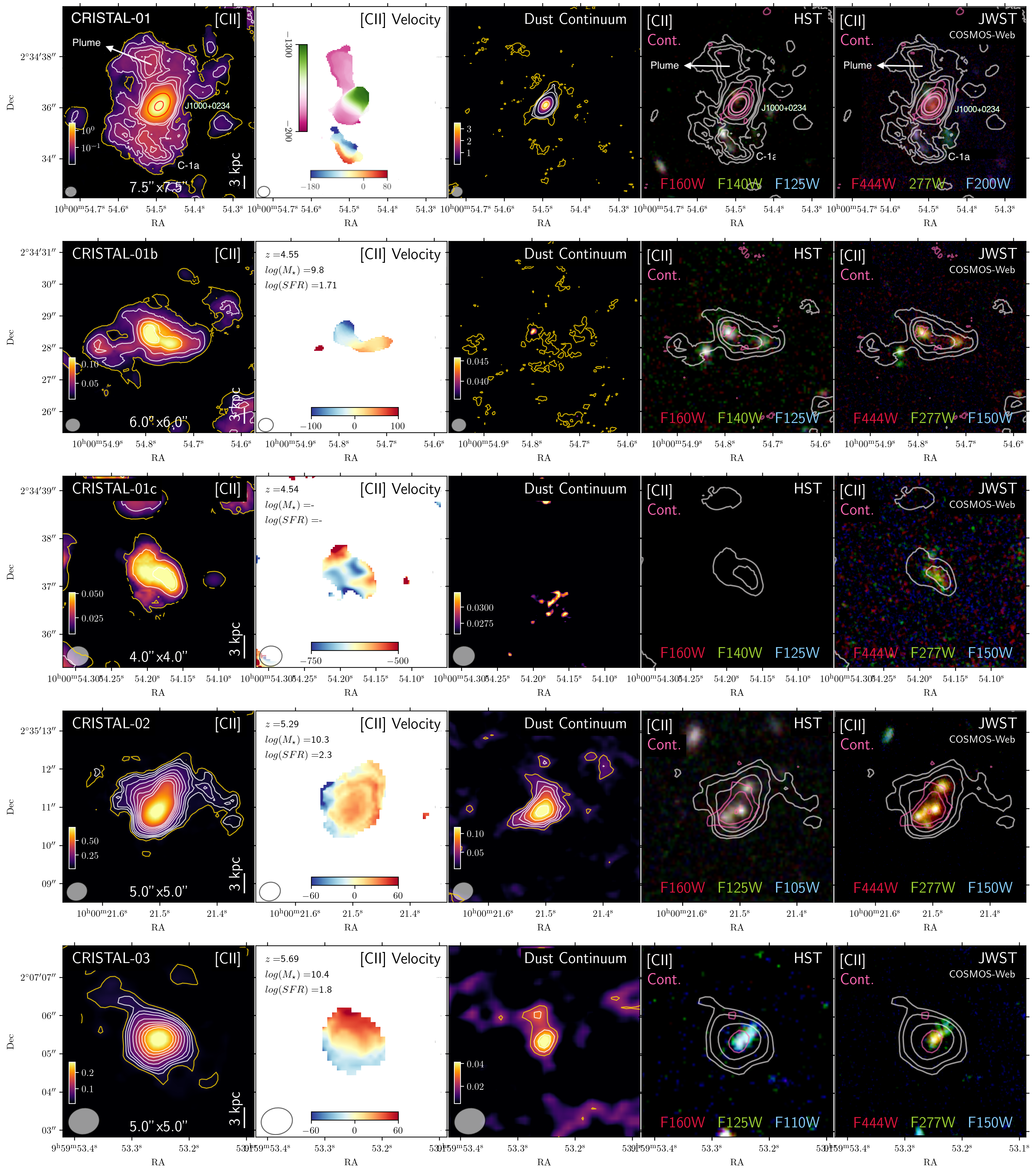}
      \caption{Multi-wavelength view of the CRISTAL galaxies including from left to right: integrated \cii\ line emission, \cii-based velocity field, dust continuum emission, \cii\ and dust continuum emission overlaid on a composite image based on HST/WFC3 and JWST/NIRCam observations. The redshift, stellar mass, and star formation rate are listed in the top left corner of the second panel. S/N contours correspond to 3, 4, and 5$\sigma$ and then increase in steps of 2$\sigma$, except for CRISTAL-01, where the emission intensity is shown in logarithmic scale and the contours represent the [4, 6, 8, 10, 50, 150]$\sigma$ levels (the last two in red). Additional figures showing the multi-wavelength view of the remaining CRISTAL galaxies are included in Appendix~\ref{appendix:view}.} \label{comb_panel1}
\end{figure*}

\subsection{Multi-wavelength view of individual CRISTAL galaxies}\label{multi-view}

\begin{figure*}
\centering
   \includegraphics[width=\hsize]{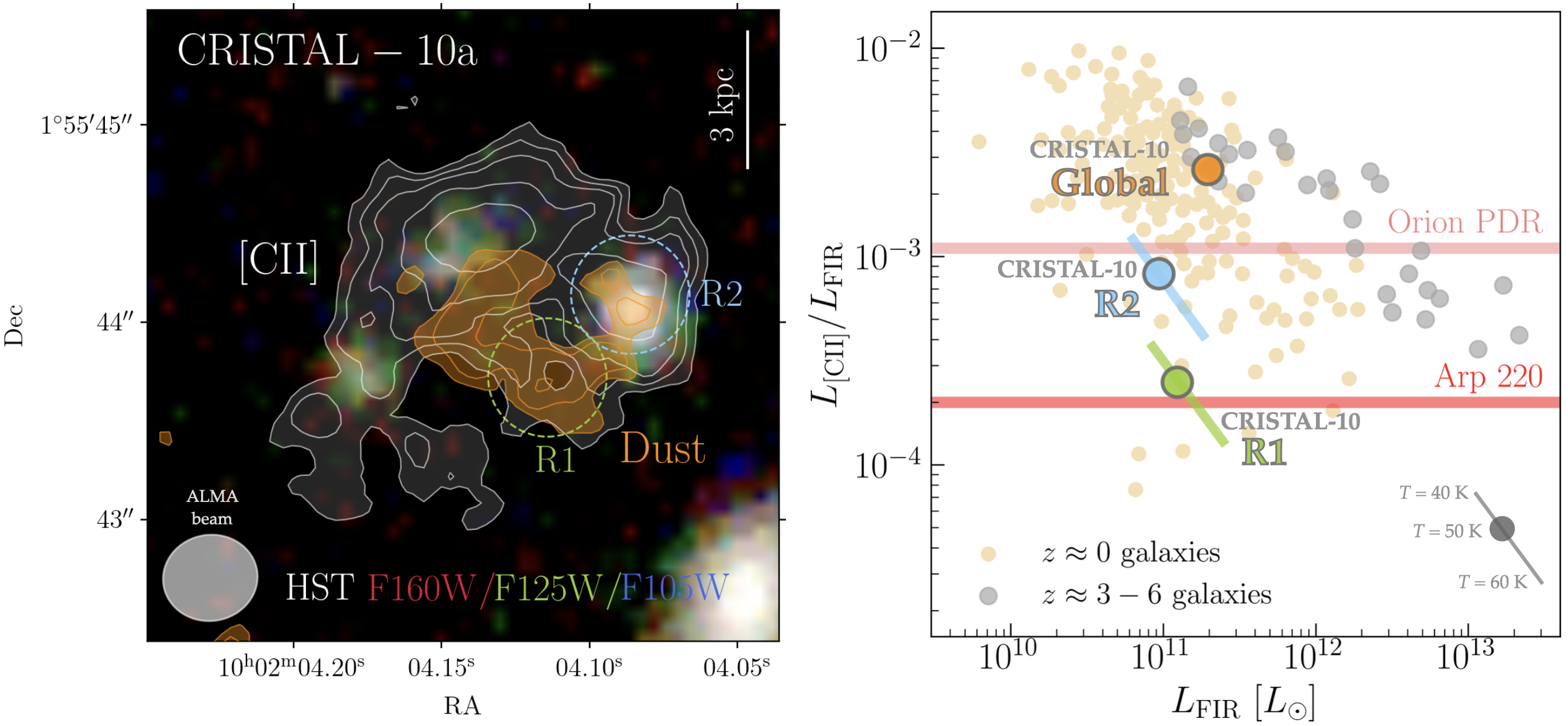}
      \caption{{\it (Left)} Multi-wavelength view of CRISTAL-10, a main-sequence star-forming galaxy at $z=5.67$. The background image is a composite of three HST/WFC3 filters, highlighting the rest-frame UV stellar light emitted by young, massive stars. Overlaid on this image are white contours showing the integrated \cii\ line emission (3, 4, 5, 7 and 10$\sigma$ levels), and orange contours showing the dust continuum emission at rest-frame 158~$\mu$m (2.5, 3, 4 and 5$\sigma$ levels). The dust continuum reveals two distinct peaks, designated as Region 1 and Region 2. In Region 1, the dust continuum emission lacks a corresponding rest-frame UV light counterpart, and the \cii\ line emission is faint. Region 2, on the other hand, correspond to a star-forming region where the rest-frame UV, dust, and \cii\ line emission peak. {\it (Right)} \cii/FIR luminosity ratio as a function of the FIR luminosity observed in nearby star-forming galaxies and starbursts \citep[beige points; ][]{rhc_lutz16,rhc_rhc18a} and high-$z$ star-forming galaxies \citep[gray points; ][ and references therein]{rhc_spilker16}. The \cii/FIR luminosity ratio measured in the star-forming Region~2 (light blue circle) is comparable to that measured in the dense PDR of the  Orion complex, while the \cii/FIR luminosity ratio of Region~1 (yellow) is significantly low, comparable only to extreme local systems such as Arp 220 \citep[e.g., ][]{rhc_luhman98,rhc_luhman03}, characterized by deeply embedded star formation. The FIR luminosities where measured based on the Band 7 continuum using the conversion factor by \cite{rhc_bethermin20} and assuming a dust temperature of 50~K. The diagonal line represents how much the measurements of Regions 1 and 2 change if we assume a dust temperature of 40~K and 60~K. We also show the global value (orange circle) measured using the $L_{\rm FIR}$ in \cite{rhc_mitsuhashi24} and the integrated \cii\ flux from Table~\ref{table:fluxes}.} \label{C10}
\end{figure*}

\begin{figure*}
\centering
   \includegraphics[width=\hsize]{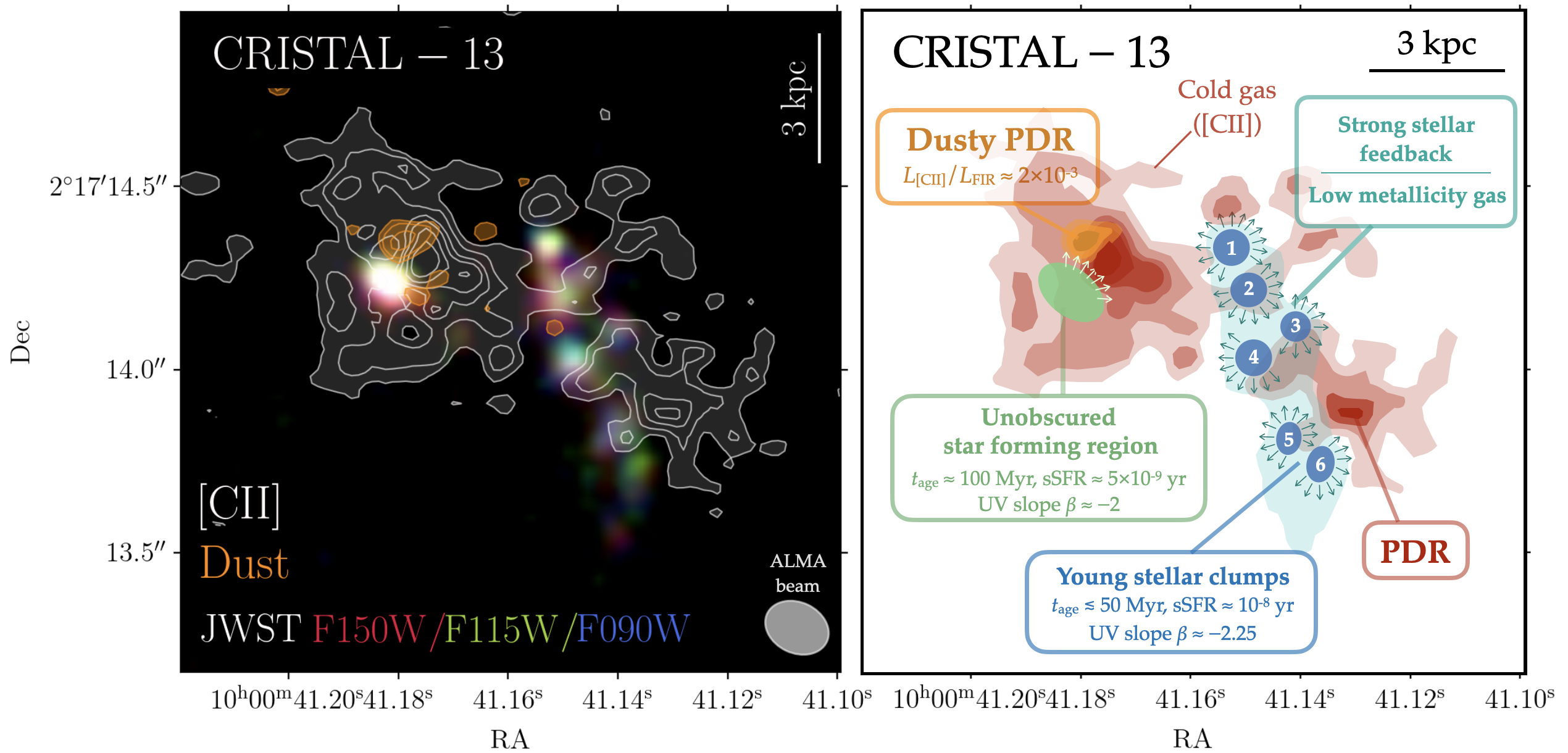}
      \caption{{\it (Left)} Composite image based on three JWST/NIRCam blue filters (F090W, F115W, F150W) that show the stellar light from multiple young stellar clumps in CRISTAL-13, a main-sequence star-forming galaxy at $z=4.58$. Overlaid are the integrated \cii\ line emission (white contours; ) and the dust continuum (orange; ). The ALMA beam is shown in the lower right corner. {\it (Right)} Cartoon representation of the distribution of the stellar clumps traced by JWST (blue and green), the cold gas traced by the \cii\ line (red), and the dust continuum (orange). The properties of the stellar clumps are based on the SED modeling presented in \cite{rhc_lines24}.} \label{C13}
\end{figure*}

One of the advantages of the CRISTAL survey is the availability of multi-wavelength data for a large fraction of the systems. This includes observations of the rest-frame UV and optical stellar light with HST/WFC3 and JWST/NIRCam, as well as more recent observations of the ionized gas with JWST/NIRSpec IFU (e.g., PID  1217, 3073, 4265, 5974).  Figures~\ref{comb_panel1} to \ref{comb_panel5} show, for each system detected in the CRISTAL survey at $z\approx4-6$, a series of panels that include the integrated \cii\ line emission, the \cii-based velocity field, the dust continuum, and the \cii\ line and dust continuum emission overlaid on composites images of the stellar light derived from HST/WFC3 and JWST/NIRCam observations. Based on this multi-wavelength view of the CRISTAL galaxies, we provide a brief description of their main characteristics: \\

\noindent  {\bf CRISTAL-01 (DC-842313):} This galaxy is among the most massive in the CRISTAL survey. While initial stellar mass estimates from ALPINE suggested \(M_{\star} = 10^{10.84}~M_\odot\) \citep{rhc_faisst20}, updated photometry by \cite{rhc_mitsuhashi24} revised this to \(M_{\star} = 10^{10.65}~M_\odot\). CRISTAL-01a is undergoing a major interaction with its companion, the SMG J1000+0234 \citep{rhc_fraternali21}, and this explains its very disturbed morphology. Despite its significant mass, CRISTAL-01a was not detected in dust continuum emission.

The depth of the \cii\ map from the combination of the CRISTAL and the ancillary data reveals, for the first time in this system,  an elongated structure (or plume) of \cii\ line emission connected to the SMG and extending north for about a projected distance of $\sim$15~kpc. The origin of this extended component is discussed in detail in \cite{rhc_solimano24a}, who consider four potential scenarios: a conical outflow, a cold accretion stream, ram pressure stripping, and gravitational interactions. New JWST/NIRSpec observations presented in detail in \cite{rhc_solimano25_jwst}, reveal a compact component at the base of the \cii\ plume exhibiting broad [OIII] line emission, consistent with the presence of an ionized outflow. The source for this broad line emission could be a compact AGN or represent the escape path for an outflow from the center of SMG J1000+0234. 

CRISTAL-01 resides in a protocluster environment, where multiple Ly$\alpha$ emitters and Lyman-break galaxies have been detected in its vicinity \citep{rhc_jimenez-andrade23}. In our ALMA-CRISTAL observations, we identify two additional \cii\ line emitters, CRISTAL-01b and CRISTAL-01c, at similar redshifts ($|\Delta z|\lesssim0.02$) and located approximately $\sim$7\arcsec\ ($\sim45$~kpc) and $\sim$5\arcsec\ ($\sim32$~kpc) from the CRISTAL-01a/J1000+0234 interacting pair, respectively. While CRISTAL-01c is not visible in the HST data, it is detected in the JWST/NIRCam imaging. \\

\noindent {\bf CRISTAL-02 (DC-848185, HZ6, LBG-1):} This galaxy is located in the field of the galaxy protocluster associated with the SMG AzTEC-3 at $z=5.3$, at a projected distance of $\sim15$\arcsec. CRISTAL-02 was first detected in \cii\ line emission  by \cite{rhc_riechers14}, and then re-observed  under the name HZ6 by \cite{rhc_capak15}, detecting for the first time the dust continuum emission. The connection between the \cii\ line emission and the distribution of the neutral gas as traced by the Ly$\alpha$ line emission is discussed in \cite{rhc_guaita22}.

The rest-frame UV and optical stellar emission in this galaxy is dominated by four stellar clumps of approximately kiloparsec size ($\sim0\farcs2$), three in the central part of the galaxy and one in the north. The dust continuum emission is strong and peaks in the central region, coincident with the three central stellar clumps. The \cii\  line emission traces the gas in the extended disk and shows interesting asymmetries perpendicular to the major morphological axis: a lump of  emission extending from the center to the east, and a tail of \cii\ emission to the west extending for about $\sim$10~kpc.  The \cii\ line profile in CRISTAL-02 is best modeled by a double Gaussian fit including a broad and a narrow component (see Fig.~\ref{spec_all}). The interpretation of the broad component as an outflow is discussed in detail in \cite{rhc_davies25}. \\

\noindent {\bf CRISTAL-03 (DEIMOS\_COSMOS\_536534, HZ1):} This system shows rest-frame UV and stellar light emission that is extended approximately in the same direction as the \cii\ kinematic axis. Similar to CRISTAL-02,  there is a tail of \cii\ line emission extending in the north-east direction for about $\sim$10~kpc. The dust continuum peaks in two locations: at the center, where reddened optical emission is observed in the NIRCam data, and to the north of the center, at the base of the \cii\ tail, where no stellar light counterpart is detected. \\

\noindent {\bf CRISTAL-04 (vuds\_cosmos\_5100822662):} This system is experiencing a merger event with a stellar mass ratio between CRISTAL-04a and CRISTAL-04b of approximately 17:1. Both galaxies are detected in the rest-frame UV, optical, \cii\ line, and dust continuum emission (although the dust continuum detection in CRISTAL-04b is marginal). It is important to note that the spatial distribution of the \cii\ line emission is very extended ($\sim$30~kpc long), disturbed, and connects both galaxies \citep{rhc_ikeda25}. This is expected from a tracer of the cold neutral gas, and resembles the perturbed and extended atomic HI emission observed in merging systems in the nearby universe \citep[e.g.,][]{rhc_hibbard96}.  \\

\noindent {\bf CRISTAL-05 (DEIMOS\_COSMOS\_683613, HZ3):} This galaxy, also shown in Fig.~\ref{C05} and analyzed in detail by \cite{rhc_posses24}, exhibits a complex structure. It comprises a close pair of interacting galaxies surrounded by an extended gas component traced by \cii\ line emission. As described by \cite{rhc_posses24}, this component extends to about four times the size of the star-forming disk traced by the rest-frame UV emission, and accounts for about 40\% of the total \cii\ emission. The $L_{\rm [CII]}/L_{\rm FIR}$ ratio upper-limits in this extended component are consistent with values found in shocked regions of nearby merging systems \cite[e.g.,][]{rhc_appleton13,rhc_peterson18}.\\

\noindent {\bf CRISTAL-06 (vuds\_cosmos\_5100541407):} Similar to CRISTAL-04, this system represents a major merger involving at least two galaxies, CRISTAL-06a and CRISTAL-06b, with a stellar mass ratio of 8:1. Both galaxies are detected in rest-frame UV and optical emission and are connected by a disturbed, extended gas component traced by the \cii\ line emission.

A closer inspection of CRISTAL-06a reveals two distinct components: one to the west, faint in rest-frame UV and optical data but bright in dust continuum and \cii\ line emission; and one to the east, bright in rest-frame UV and optical emission but less prominent in dust continuum and \cii\ line emission. Notably, CRISTAL-06a is the only system in the CRISTAL survey classified as ''multiple-\cii'' by \cite{rhc_ikeda25}, indicating the presence of multiple \cii\ line emission peaks associated with a single UV component. The kinematic analysis of CRISTAL-06a suggests two systems counter-rotating, implying that CRISTAL-06a itself may represent an advanced-stage merger of two galaxies.  \\

\noindent {\bf CRISTAL-07 (DEIMOS\_COSMOS\_873321, HZ8):} This system is composed of two galaxies, CRISTAL-07a and CRISTAL-07b, likely in an advanced stage of a major merger (1.5:1), separated by a projected distance of approximately $\sim$9~kpc. This was already hinted by the low-angular resolution \cii\ observations analyzed in \cite{rhc_capak15}. CRISTAL-07a and CRISTAL-07b have comparable \cii\ luminosities, although only CRISTAL-07a is detected in the dust continuum. Additionally, another system of galaxies, CRISTAL-07c and CRISTAL-07d, is detected about  $\sim$75~kpc west of CRISTAL-07ab. CRISTAL-07c is more massive and brighter in \cii\ line emission than CRISTAL-07a and CRISTAL-07b, and it shows disturbed, extended emission to the south, which is connected to a smaller companion in size, CRISTAL-07d,  located at a projected distance of approximately  $\sim$15~kpc. CRISTAL-07d is nearly as bright in \cii\ line emission as CRISTAL-07c, but has a wider line profile (${\rm FWHM}_{\rm [CII]}=375\pm73$~km~s$^{-1}$). \\

\noindent {\bf CRISTAL-08 (vuds\_efdcs\_530029038):} The rest-frame UV and optical emission from HST and JWST reveal at least eight $\sim$kiloparsec-sized stellar clumps with a range of colors. SED modeling of these clumps following the same method as \cite{rhc_li24} indicates stellar masses ranging from $\sim2-8\times10^{8}$~$M_{\odot}$ and ages between $\sim70-200$~Myr (Herrera-Camus et al. (in prep.)). The \cii\ line emission is extended and the peaks is offset from the stellar clumps, with prominent regions in the northwest and southwest. In contrast, the \cii\ line emission is faintest in the east, where the largest concentration of stellar clumps is located. The dust continuum peaks near the \cii\ line emission maximum in the northwest.  

Despite its clumpy structure, CRISTAL-08 exhibits evidence of smooth, ordered rotation (see Lee et al. in prep. for a detailed kinematic analysis). This behavior is reminiscent of main-sequence star-forming galaxies at cosmic noon, where regular rotating disks are observed in $\approx$70-80\% of cases, even when stellar light reveals significant clumpiness \citep[e.g.,][]{rhc_wisnioski15,rhc_nfs20}.  

Notably, there is evidence of outflows traced by the \cii\ line emission, potentially emerging from some of the giant star-forming clumps. The outflow velocities are consistent with those observed in other CRISTAL systems \citep[e.g.,][; Birkin et al. in prep.]{rhc_rhc21,rhc_davies25}. These outflows, along with their properties, will be analyzed in detail in Herrera-Camus et al. (in prep.). \\

\noindent  {\bf CRISTAL-09 (DEIMOS\_COSMOS\_519281):} This galaxy exemplifies how ALMA \cii\ observations can provide a complementary and more revealing view of the complexity of $z\approx4-6$ galaxies compared to HST and JWST observations. The rest-frame UV and optical emission of CRISTAL-09a is compact ($R_{\rm eff}\approx0.7$~kpc) relative to the \cii\ line emission that extends for about $\sim$12~kpc to the west and connect CRISTAL-09a with a potential minor companion, CRISTAL-09b \citep{rhc_ikeda25}. Notably, CRISTAL-09b has no counterpart in HST or JWST observations.  \\

\noindent {\bf CRISTAL-10 (DEIMOS\_COSMOS\_417567, HZ2):} This system is particularly intriguing due to the complexity of its gas, dust, and stellar morphology. The stellar emission reveals three clumps, with the brightest located to the west. The cold neutral gas, traced by the \cii\ line emission, covers all three clumps but peaks at the central stellar clump. Interestingly, the stellar and \cii\ emissions form a semi-ring around the dust continuum emission, which is detected in an offset position. This complex morphology and the implications of the observed $L_{\rm [CII]}/L_{\rm FIR}$ ratio are discussed in more detail in Figure\ref{C10} and Section~\ref{zoom-C10}.

 We detect an additional galaxy in the field, designated CRISTAL-10b, located at a projected distance of approximately $\sim$47~kpc to the northeast of CRISTAL-10a. The \cii\ morphology of this system is highly disturbed, with no corresponding rest-frame UV or optical emission detected.  \\

\noindent {\bf CRISTAL-11 (DEIMOS\_COSMOS\_630594):} Similar to CRISTAL-09, the rest-frame and optical emission in this galaxy is compact ($R_{\rm eff}\approx0.8$~kpc) relative to the extent of the \cii\ line emission.  The \cii\ line emission extends significantly to the east, forming a tail with a disturbed morphology that stretches for a projected distance of approximately $\sim$36~kpc. There is no corresponding stellar light emission detected in the HST or JWST data for this extended component \citep{rhc_ikeda25}. The dust continuum is detected at the 3$\sigma$ level, peaking in the northern part of the stellar distribution. \cite{rhc_lines24} model the SED of CRISTAL-11 based on NIRCam observations from the PRIMER survey. They find that the northern region shows higher dust attenuation and an older stellar population ($\approx$200~Myr) compared to the southern region, which exhibits minimal dust attenuation and a young stellar population ($\approx$10~Myr).  \\

\noindent {\bf CRISTAL-12 (CANDELS\_GOODSS\_21):} Similar to CRISTAL-01a, the updated SED modeling by \cite{rhc_mitsuhashi24} scaled down the stellar mass and SFR of this system by a factor of $\sim3$ relative to the values reported by the ALPINE survey \citep{rhc_faisst20}, \textbf{although both values remain consistent within a 1$\sigma$ uncertainty. This difference could be due to the use of updated fluxes from the ASTRODEEP-GS43 catalogue \citep{rhc_merlin21} in \cite{rhc_mitsuhashi24}.} CRISTAL-12 is detected in rest-frame UV emission, and the \cii\ line emission peaks at the position of the stellar light, and then extends to the west. There is a hint of \cii\ line extended emission towards the northwest, resembling the tails of \cii\ line emission observed in other CRISTAL systems (e.g., CRISTAL-09, -11). \\

\noindent {\bf CRISTAL-13 (vuds\_cosmos\_5100994794):} This galaxy exhibits a complex structure, with rest-frame UV and optical stellar emission revealing a main component to the east and a tail of stellar emission extending approximately $\sim10$~kpc  to the northwest. \cite{rhc_lines24} use the NIRCam data from the PRIMER survey to model the SED and find that this extended tail is composed of at least 5 stellar clumps with high specific SFR (${\rm sSFR}\approx10^{-8}$~yr$^{-1}$), young age ($t_{\rm age}\approx50$~Myr), and blue rest-frame UV slope ($\beta_{\rm UV}\approx-2.25$). The stellar mass of the system is dominated by the stellar component on the east, which has an age of $\approx$100~Myr. This complex system is discussed in more detail in Section~\ref{zoom-C13} and Figure~\ref{C13}. \\

\noindent {\bf CRISTAL-14 (DEIMOS\_COSMOS\_709575):} This galaxy has one of the smallest rest-frame UV sizes in the CRISTAL sample, therefore our planned ALMA observations aimed at higher angular resolution than the average of the other CRISTAL targets to spatially resolve the source. The resulting beam size, for natural weighting, was  $0\farcs11 \times 0\farcs12$, corresponding to a physical scale of approximately $\approx700$~pc at the redshift of the source. The \cii\ line emission is more extended than the stellar component traced by NIRCam observations, with a hint of extended \cii\ line emission towards the northwest, which our ALMA observations appear to tentatively detect. \\

\noindent {\bf CRISTAL-15 (vuds\_cosmos\_5101244930):} The composite HST and JWST imaging reveal three main clumps of rest-frame UV and optical light emission. However, detailed analysis of the blue filters of JWST/NIRCam show that there are at least five separate star-forming clumps that are only a few Myr old \cite{rhc_lines24}. The \cii\ line emission is elongated and connect these three clumps displaying a well defined velocity gradient from northwest to southeast. Notably, the \cii\ line emission also extends towards the northeast, resembling the tail-like \cii\ line emission observed in other CRISTAL systems such as CRISTAL-02, CRISTAL-03, and CRISTAL-09. \\

\noindent {\bf CRISTAL-16 (CANDELS\_GOODSS\_38):} This galaxy appears to have two distinct components. The first is the main component, which shows a clear counterpart in HST rest-frame UV emission and exhibits a velocity gradient oriented from north to south. Additionally, there is a tentative detection of dust continuum emission in this region. The second component, located to the west, initially seems to be an extension of the disk. However, its distinct kinematic properties suggest that it is, in fact, a minor companion. We refer to this system as CRISTAL-16b. \\

\noindent {\bf CRISTAL-17 (DEIMOS\_COSMOS\_742174):} This system is only weakly detected in rest-frame UV emission with HST. The  JWST NIRCam image from the PRIMER survey is also low signal-to-noise but it is still possible to identify two main stellar clumps and a faint tail of emission extending from the eastern component \citep{rhc_lines24}. The \cii\ line emission is detected, but only with a $S/N\approx4$. \\

\noindent {\bf CRISTAL-18 (vuds\_cosmos\_5101288969):} This system was selected to be included in the ALPINE survey based on spectroscopic data from the VIMOS Ultra-Deep Survey \citep{rhc_lefevre15}. The spectrum contained a single emission line, which, if identified as Ly-$\alpha$, corresponds to a redshift of $z_{\rm Ly\alpha}=5.6982$. Given the presence of only one line in the spectrum and the observed continuum, this redshift has an estimated probability of $\sim$80\% to be correct \citep{rhc_lefevre15}. The ALPINE ALMA follow-up observations tentatively detected the \cii\ line emission with a $S/N\approx4$ at a redshift $z_{\rm [CII]}=5.7209$, differing by only $|\Delta{z}|=0.002$ from the Ly$\alpha$ redshift. 

As described in Section~\ref{sec:obs}, we placed two of the four spectral windows next to each other with a small overlap centered at the frequency of the line, based on the ALPINE \cii\ line tentative detection. Despite a deep integration, we did not detect the \cii\ line in any of the four spectral windows. If the Ly$\alpha$ detection is real and the ALPINE \cii\ line detection spurious, this would imply a record velocity offset of the \cii\ line relative to the Ly$\alpha$ line of $\gtrsim1000$~km~s$^{-1}$ \citep[e.g., ][]{rhc_hashimoto19,rhc_baier-soto22}.

At the sensitivity of our observations ($S_{\rm [CII],Nat}=0.1$~mJy~beam$^{-1}$), and if we assume a linewidth of 250~km~s$^{-1}$, the 3$\sigma$ upper limit for the integrated \cii\ line flux is 0.03~Jy~km~s$^{-1}$. This translates to a 3$\sigma$ upper limit \cii\ line luminosity of $\approx10^{6.7}$~$L_{\odot}$.
\\

\noindent {\bf CRISTAL-19 (DEIMOS\_COSMOS\_494763):} The \cii\ line and dust continuum emission in this system are significantly detected and are co-spatial with the rest-frame UV and optical emission. The velocity gradient observed in the \cii\ line emission is also aligned with the major morphological axis of the stellar light. 
As seen in other systems like CRISTAL-09 and CRISTAL-10, CRISTAL-19 exhibits extended \cii\ line emission to the south, spanning approximately 23 kpc. \\

This concludes the discussion of the 19 fields observed as part of the CRISTAL ALMA Large Program. The following six systems discussed here have been included in the final CRISTAL sample given that they fulfill all relevant sample criteria outlined in Section~\ref{sample}. \\

\noindent {\bf CRISTAL-20 (DEIMOS\_COSMOS\_494057, HZ4):} This galaxy has been extensively discussed in \cite{rhc_rhc21,rhc_rhc22}. The \cii\ line emission extends well beyond the star-forming disk as traced by the rest-frame UV emission and the dust continuum. There is also evidence for a neutral gas outflow traced by a broad component of the \cii\ line emission that is aligned with the minor morphological and kinematic axis, similar to the case of nearby starbursts such as M~82 or NGC~253 \citep[for a review see][]{rhc_veilleux20}. The kinematic analysis of the \cii\ line emission suggest that CRISTAL-20 (or HZ4) has a regular rotating disk ($V_{\rm rot}/\sigma_{0}\approx2$) with a high intrinsic velocity dispersion ($\sigma_{0}\approx65$~km~s$^{-1}$), but the analysis of JWST/NIRSpec data suggest that the galaxy is undergoing a merger \citep{rhc_parlanti25}. \\

\noindent {\bf CRISTAL-21 (HZ7):} A detailed analysis of the \cii\ line emission in this galaxy is presented in \cite{rhc_lambert23}. The analysis reveals that the system exhibits a complex \cii\ line morphology and kinematics, with the \cii\ line emission being approximately twice as extended as the rest-frame UV emission. This evidence strongly suggests that the galaxy is an interacting system. \\

\noindent {\bf CRISTAL-22 (HZ10):} This system is one of the brightest in the CRISTAL sample, exhibiting a broad \cii\ line profile that requires fitting with a double Gaussian profile. The linewidths of the two components are ${\rm FWHM_1}=201\pm24$ km~s$^{-1}$ and ${\rm FWHM_2}=590\pm27$ km~s$^{-1}$. The HST rest-frame UV emission shows two main young stellar components, with the one in the west significantly fainter. Both stellar components are detected in the dust continuum under the \cii\ line, and analysis of new Band~9 observations reveal the rest-frame UV emission from the West component is strongly attenuated by the dust \citep{rhc_villanueva24}. The \cii\ line emission connects both components, and exhibit complex kinematics which will be discussed in detail in \cite{rhc_telikova24}. The complex nature of the system has been confirmed by \cite{rhc_jones24} based NIRSpec IFU observations of the main nebular lines, which also discovered a minor companion to the east of the main stellar component.

At about a projected distance of $\sim77$~kpc south of CRISTAL-22 (HZ10), our ALMA \cii\ line and dust continuum observations also include in the field the starbursting galaxy CRLE \citep{rhc_pavesi18}, at a similar redshift than CRISTAL-22  ($|\Delta z|=0.014$).  \\

\noindent {\bf CRISTAL-23 (DEIMOS\_COSMOS\_818760):} The \cii\ line emission in this system has been extensively analyzed in \cite{rhc_jones21} and \cite{rhc_devereaux24}. Their analysis concludes that the system comprises three components: two primary components involved in a major merger  \citep[stellar mass ratio almost 1:1; ][see also Table~\ref{table:sample}]{rhc_mitsuhashi24}, and a third component located approximately $\sim18$~kpc to the west, likely representing an upcoming minor merger. \\

\noindent {\bf CRISTAL-24 (DEIMOS\_COSMOS\_873756):} This galaxy is the brightest in \cii\ line emission within the ALPINE survey, comparable in our sample only to CRISTAL-22. The integrated \cii\ line emission reveals a complex structure, which \cite{rhc_devereaux24} attribute to the presence of multiple merging components or  clumps of star formation within the system. The dust continuum peaks co-spatially with the \cii\ in the main system but also extends toward the northwest, following the \cii\ line emission. JWST/NIRCam observations reveal two peaks of stellar light, with the western peak being brighter. The dust continuum emission peaks offset from the stellar light peaks by approximately $\sim$1~kpc to the south. Additionally, a secondary 3$\sigma$ dust continuum peak is detected at a projected distance of $\sim$10 kpc. There is no HST or JWST counterpart for this outer \cii\ or dust continuum emission. \\

\noindent {\bf CRISTAL-25 (vuds\_cosmos\_5101218326):} The \cii\ line and dust continuum emission of this system are analyzed in detail in \cite{rhc_devereaux24}. The integrated \cii\ line emission shows two distinct peaks. The primary peak, located in the southwest, spatially coincides with the main dust continuum peak. Interestingly, in between the two \cii\ peaks we observe a second and fainter dust continuum peak, spatially coincident with the bulk of the stellar light traced by the HST and JWST images.

\section{Case studies: CRISTAL-10 and CRISTAL-13} \label{case-studies}

The multi-wavelength observations on $\sim$kiloparsec scales available for the CRISTAL galaxies provide a unique opportunity to conduct comprehensive analyses of their gas, dust, and stellar properties. Previous studies have focused on individual CRISTAL sources, such as CRISTAL-01 \citep{rhc_solimano24a}, CRISTAL-05 \citep{rhc_posses24}, and CRISTAL-22 \citep{rhc_villanueva24}, and in this section we present two additional cases where the CRISTAL data offer valuable insights into the interstellar medium properties of high-redshift galaxies. 

\subsection{CRISTAL-10: Arp~220 conditions at $z\approx5$?}\label{zoom-C10}

As initially described in Section~\ref{multi-view}, CRISTAL-10a presents an intriguing case due to the spatial distribution of the dust continuum relative to the \cii\ and stellar light emission. The left panel of Figure~\ref{C10} shows that there are three main stellar clumps detected in rest-frame UV emission (unfortunately there are no JWST/NIRCam observations available). The two primary clumps, located in the north and northwest, align with peaks in the \cii\ line emission. Conversely, the bulk of the dust continuum emission is concentrated in a region offset to the south of the \cii\ and stellar peaks, where the \cii\ line emission is faint and the stellar light is undetected.

To investigate the nature of this dusty component of CRISTAL-10a, we measure the \cii/FIR ratio placing two circular apertures of 0\farcs5 ($\sim$3~kpc) radius centered in regions where the dust continuum is detected. Region~1 is the region where the dust continuum is strong, the \cii\ line emission is faint, and there is no detectable rest-frame UV light. In contrast, Region~2 is the star-forming region where the \cii, dust, and stellar light emission are spatially coincident. Assuming a dust temperature of $T_{\rm dust}=50$~K \cite[following $T_{\rm dust}$ measured in the West component of CRISTAL-22 (HZ10) that also shows a high level of dust obscuration; ][]{rhc_faisst20b, rhc_villanueva24}, we measure \cii/FIR ratios of $2.5\times10^{-4}$ and $8.3\times10^{-4}$ for Region~1 and Region~2, respectively. For comparison, the \cii/FIR ratio measured globally is higher ($L_{\rm [CII]}/L_{\rm FIR}=2.6\times10^{-3}$), driven in large part because the \cii\ line emission is much more extended than the dust continuum, peaking in the other two main star forming regions traced by the rest-frame UV emission. 

The right panel of Figure~\ref{C10} shows the \cii/FIR ratio measured in Region~1 (green) and Region~2 (blue) of CRISTAL-10a in context with \cii/FIR ratios measured in nearby and high-$z$ systems (similar to the left panel of \ref{CII2FIR}). The diagonal bar shows how much the \cii/FIR ratio and $L_{\rm FIR}$ change if we assume a dust temperature $\pm10$~K around $T_{\rm dust}=50$~K. However, given the characteristics of this source, the dust temperature could potentially be higher.

The \cii/FIR ratio measured in the star-forming Region~2 is comparable to that found in typical dense PDRs like that in the Orion nebula, where the FUV radiation field and neutral gas densities are high \citep[e.g., $G_{0}\gtrsim10^4$, $n_{\rm H}\gtrsim10^5$~cm$^{-3}$; ][]{rhc_goicoechea15}. In contrast, the \cii/FIR ratio measured in Region~1 is comparable to that measured in Arp~220, a highly dust obscured ULIRG with deeply embedded star forming formation and potential AGN activity \citep[e.g.,][]{rhc_luhman98,rhc_luhman03,rhc_gonzalez-alfonso04,rhc_barcos18,rhc_perna24}. Finally, the global \cii/FIR ratio is a factor of $\sim10$ higher than that measured in Region~1, highlighting that spatially-resolved data are important to properly interpret source-integrated measurements.

The notably low \cii/FIR ratio observed in Arp~220 relative to other star-forming galaxies and (U)LIRGs can be attributed to different factors, including:  a) significant non-PDR contributions to the FIR emission due to a high ionization parameter $U$ \citep[e.g., ][]{rhc_gonzalez-alfonso04,rhc_gracia-carpio11} and/or strong AGN activity \citep[e.g.,][]{rhc_spoon04}, and 2) optically thick or self-absorbed \cii\ line emission due to the high FIR optical depth \cite[$\tau\sim5$ at 100~$\mu$m and $\tau\sim1$ at 240~$\mu$m; ][]{rhc_rangwala11}, although this scenario is likely not dominant \citep[e.g., ][]{rhc_luhman03}. Despite the extensive dataset available for Arp220, the primary cause of the observed \cii\ deficit remains unclear, complicating the interpretation of the similarly low \cii/FIR ratio observed in Region~1 of CRISTAL-10a, which has far less ancillary data available.

The discovery of a comparably low \cii/FIR ratio in CRISTAL-10a at $z=5.67$ as observed in the extreme nearby system Arp~220 is particularly intriguing. This finding highlights the need for additional observations to further understand the power source and properties of the interstellar medium in Region~1 of CRISTAL-10a, such as observations of additional FIR lines.

\subsection{CRISTAL-13: Burst of star formation and offset with \cii\ line emission}\label{zoom-C13}

To gain deeper insight into the spatial distribution of cold gas, dust, and stellar emission within this complex system, we analyzed high-resolution Briggs-weighted \cii\ and dust continuum maps. These maps provide an angular resolution of approximately \(\sim1\) kpc (\(\theta_{\rm beam} \approx 0.16''\)). The left panel of Figure~\ref{C13} illustrates the distribution of cold gas (traced by the \cii\ line) and dust (traced by the continuum) on kiloparsec scales, compared to the young stellar light traced by combining three NIRCam blue filters (F090W, F115W, F150W). The NIRCam imaging reveals at least seven star-forming clumps: a primary clump in the east and six clumps in the west, distributed from north to south. In the main eastern component, the peak of stellar light is offset by approximately $\sim1$~kpc from the peak of the \cii\ line and dust continuum emission. One possibility is that this offset is caused by dust obscuration, although stellar light from this position is also visible in the redder NIRCam filters, including F444W. In contrast, the western component, dominated by young stellar clumps, shows only faint \cii\ line emission, with \cii\ primarily tracing neutral gas surrounding the clumps.

The right panel of Figure~\ref{C13} shows a cartoon representation of the main components of CRISTAL-13. The physical properties of the stellar clumps are derived based on the modeling of the stellar light SED presented in \cite{rhc_lines24}. Starting with the east component, there is a stellar clump with a stellar population of age $\approx100$~Myr that contains about half of the total stellar mass of the system. The stellar mass and SFR of this clump (${\rm sSFR}\approx0.5\times10^{-8}$~yr$^{-1}$) place it on the main-sequence of star-forming galaxies at $z\approx5$ \citep[e.g., ][]{rhc_speagle14}. Right next to this massive star-forming clump there is a peak in the \cii\ and dust continuum that can be interpreted as a massive and dusty PDR illuminated by the stellar light from the massive stellar clump. The \cii/FIR ratio measured in this region is $L_{\rm [CII]}/L_{\rm FIR}=2\times10^{-3}$ \citep[assuming $T_{\rm dust}=40$~K; ][]{rhc_faisst20b,rhc_villanueva24}, comparable to that measured in the dense PDR in the Orion nebula.

The situation in the western component of CRISTAL-13 appears to be more interesting due to the offset between the six giant stellar clumps, which are aligned from north to south, and the neutral gas traced by the \cii\ line emission. In the Briggs-weighted map, the \cii\ emission peaks west of these giant clumps. In contrast, the natural-weighted map (Fig.~\ref{comb_panel4}) reveals some diffuse \cii\ emission overlapping with the stellar clumps, though it is significantly fainter than the stronger \cii\ emission surrounding them. The 6 stellar clumps are young ($t_{\rm age}\lesssim50$~Myr), blue ($\beta\approx-2.25$), and exhibit a sSFR that places them at least a factor 10 above the main-sequence relation of star-forming galaxies. The more concentrated \cii\ line emission is distributed around the stellar clumps, with a peak in the south resembling a PDR illuminated mainly by the emission from clumps 5 and 6. This spatial anti-correlation between young stellar clusters and the peak \cii\ line emission is expected and observed in \hii\ region-PDR complexes such as M~17, the Orion nebula and 30~Doradus \citep[e.g., ][]{rhc_goicoechea15,rhc_pellegrini17,rhc_okada19,rhc_pabst21}. The key difference in CRISTAL-13 is the physical scale, as we are discussing giant stellar clumps and PDRs on $\sim$kiloparsec scales, where much of the complexity is hidden by our angular resolution.

Simulations of star-forming galaxies have investigated the impact of stellar feedback on the interstellar medium  under conditions similar to those observed in high-redshift galaxies like CRISTAL-13. For example, \cite{rhc_vallini17} study the photo-evaporation timescale of molecular clouds as a function of metallicity and the FUV radiation field intensity, $G_0$. In models with gas metallicity $Z=0.2Z_\odot$, expected for the gas in CRISTAL-13, the photo-evaporation timescales are approximately $\sim10$~Myr for $G_0$ values ranging from $\approx1-10^3$, and then decrease from $\sim10$~Myr to $\sim1$~Myr as $G_0$ increases from $\sim10^3$ to $\sim10^5$. This is consistent with the scenario where strong stellar feedback from the six young ($\lesssim50$~Myr) stellar clusters in the west component of CRISTAL-13 are responsible for clearing or photo-evaporating the surrounding cold gas. 

\begin{figure*}
\centering
   \includegraphics[width=\hsize]{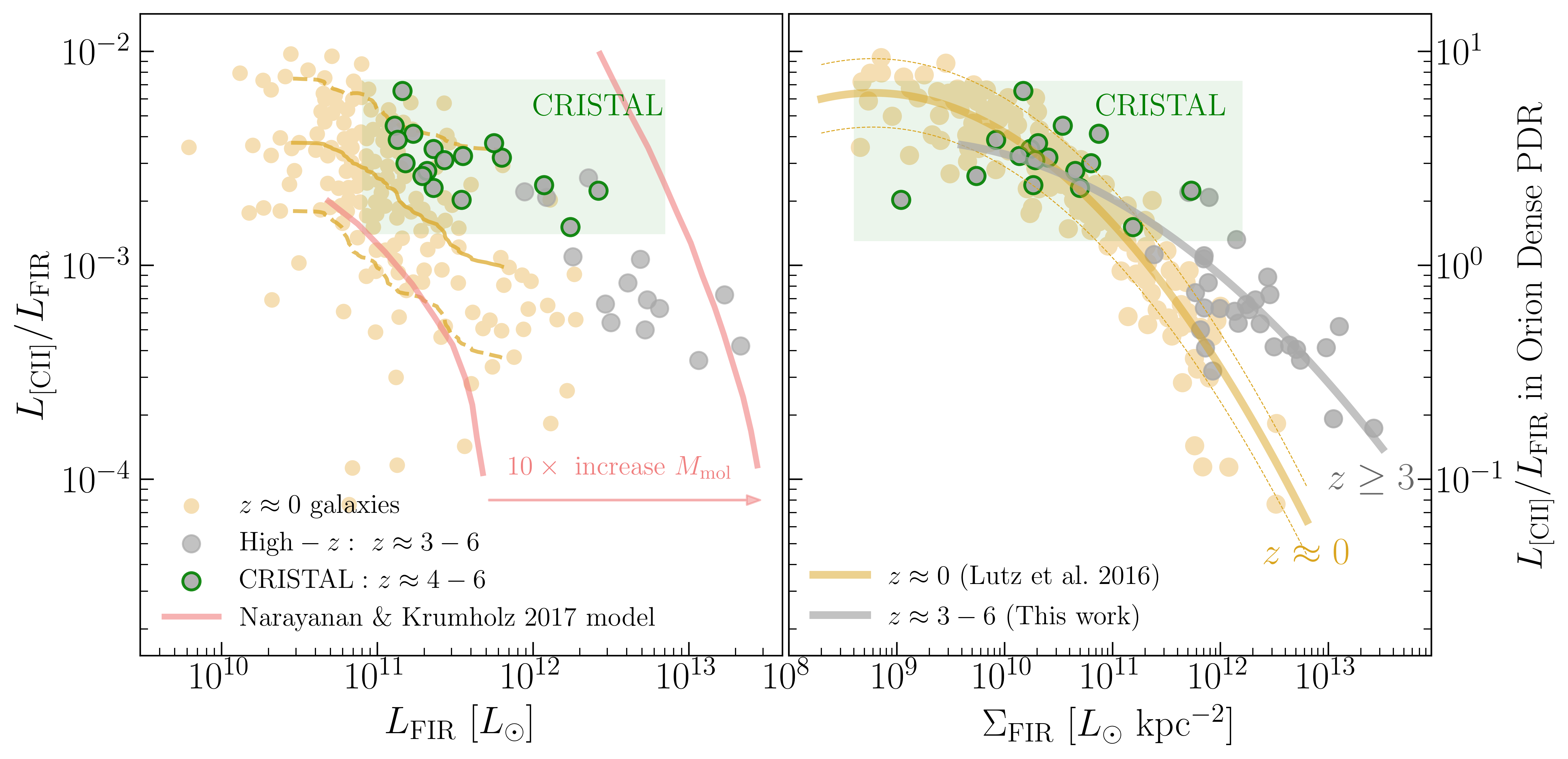}
      \caption{{\it (Left)} \cii/FIR luminosity ratio as a function of the FIR luminosity observed in nearby star-forming galaxies and starbursts \citep[beige points; ][]{rhc_lutz16,rhc_rhc18a} and high-$z$ star-forming galaxies \citep[gray points; ][ and references therein]{rhc_spilker16}. CRISTAL galaxies are shown as gray points with green borders. As a reference, the y-axis on the right shows the \cii/FIR luminosity ratio normalized to the value measured in the dense PDR of the Orion nebula \citep[$L_{\rm [CII]}/L_{\rm FIR}=1.1\times10^{-3}; $][]{rhc_goicoechea15}.  The predictions from the models by \cite{rhc_narayanan16} of how the relation between $L_{\rm [CII]}/L_{\rm FIR}$ and $L_{\rm FIR}$ changes if the molecular gas content increase by a factor of $\times$10 are  shown as light-blue lines. The solid golden line represents the running median for $z\approx0$ galaxies, while the dashed lines indicate the 10th and 90th percentiles. {\it (Right)} Similar to the left panel, but this time we plot the \cii/FIR luminosity ratio as a function of the FIR surface brightness ($\Sigma_{\rm FIR}$). The best fit to the data at $z\approx0$ \citep[][]{rhc_lutz16,rhc_rhc18a} and $z\approx3-6$ (this work; Eq.~\ref{eq_cristal}) are shown as beige and gray lines, respectively.} \label{fig_CII2FIR}
\end{figure*}

Similar results are found in the synthetic observations of smoothed particle hydrodynamics simulations of a dwarf galaxy merger ($Z=0.1Z_{\odot}$), where supernova feedback from the stellar clusters disperses the dense gas, leading to a decrease in the \cii\ luminosity \citep{rhc_bisbas22}. Furthermore, \cite{rhc_katz17} use cosmological simulations of star-forming galaxies at $z\approx6$ to investigate the spatial distribution of star formation activity (traced by UV light), dust, and \cii\ line emission. They find that in regions with low gas metallicity and dust content, UV and optical light from young stars escapes without significant reprocessing, and the \cii\ is suppressed. This scenario is also consistent with our observations of the western component of CRISTAL-13. Conversely, in the central regions of galaxies where metallicity and dust content are higher, the \cii\ emission produced in dense PDRs pass through the dust, while UV light is obscured, resulting in a spatial offset similar to that observed in the main component of CRISTAL-13. 

\section{The \cii/FIR ratio in CRISTAL galaxies and systems across redshift}\label{CII2FIR}

The photoelectric heating efficiency (\(\epsilon_{\rm ph}\))—defined as the ratio of the energy transferred to the gas to the energy absorbed by dust grains from UV radiation—plays a key role in regulating the thermal balance of the ISM. This efficiency typically ranges from 0.1\% to 1\% \citep[e.g.,][]{rhc_malhotra01,rhc_smith17,rhc_rhc18a}. A key factor influencing \(\epsilon_{\rm ph}\) is the ratio of the UV radiation field intensity to the neutral gas density, \(G_0/n_{\rm H}\), which affects the charge state of dust grains and, consequently, their heating efficiency \citep[e.g.,][]{rhc_rubin09,rhc_diaz-santos17}. Observationally, $\epsilon_{\rm ph}$ is often estimated through the ratio of \cii\ to FIR luminosities, based on the assumption that the photoelectric effect on dust grains is the dominant heating mechanism, and that the \cii\ emission is the main coolant of the cold neutral gas. However, it is important to note that there may be non-negligible contributions from other cooling lines, such as the \oi\ 63~$\mu$m transition \cite[e.g.,][]{rhc_rosenberg15}.

One of the key advantages of the CRISTAL survey is its ability to extend the study of the \(L_{\rm [CII]}/L_{\rm FIR}\) ratio, and consequently \(\epsilon_{\rm ph}\), to star-forming galaxies at \(z\approx4-6\), including lower-mass galaxies than previously explored at high-redshift \citep[e.g.,][]{rhc_spilker16}. Figure~\ref{fig_CII2FIR} shows the \(L_{\rm [CII]}/L_{\rm FIR}\) ratio measured in nearby star-forming galaxies and starbursts (gold), and in high-redshift galaxies (grey), including the CRISTAL galaxies (green contours), as a function of FIR luminosity (\(L_{\rm FIR}\); left panel) and surface density (\(\Sigma_{\rm FIR}\); right panel). FIR luminosities of CRISTAL galaxies are drawn from \cite{rhc_mitsuhashi24}. The vertical axis on the right shows the \(L_{\rm [CII]}/L_{\rm FIR}\) ratio normalized to the value observed in the dense PDR of Orion. Notably, all global values for the CRISTAL galaxies exceed the Orion PDR ratio, although, as we discuss in the case of CRISTAL-10, this may not hold on kiloparsec scales. 

It is well established that the \(L_{\rm [CII]}/L_{\rm FIR}\) ratio decreases with increasing \(L_{\rm FIR}\) in both nearby and high-redshift galaxies, a phenomenon known as the ``\cii\ deficit'' \citep[e.g.,][]{rhc_malhotra97,rhc_malhotra01,rhc_luhman03,rhc_diaz-santos13,rhc_diaz-santos17,rhc_rhc18a,rhc_rhc18b}. The left panel of Figure~\ref{fig_CII2FIR} illustrates this trend, showing that while both local and high-redshift galaxies follow this pattern, high-redshift galaxies are shifted towards higher FIR luminosities \cite[e.g.,][]{rhc_maiolino09,rhc_stacey10,rhc_gracia-carpio11}. 

\citet{rhc_narayanan16} propose a theoretical explanation for the ``\cii\ deficit'', based on models that consider clouds composed of atomic and molecular hydrogen, coupled with chemical equilibrium networks and radiative transfer models. According to their framework, the decline in the \cii/FIR ratio with increasing \(L_{\rm FIR}\) is driven by the relationship between star formation rate (and thus \(L_{\rm FIR}\)) and the surface density of gas clouds (\(\Sigma_{\rm gas}\)). As \(\Sigma_{\rm gas}\) increases, \(L_{\rm FIR}\) rises, and the carbon in molecular gas is principally in the form of CO, leading to a decrease in C$^+$ abundance, and consequently, a lower \cii/FIR ratio.

Their model also accounts for the scatter observed around the declining \cii/FIR trend with \(L_{\rm FIR}\). Since star formation rate, and therefore \(L_{\rm FIR}\), scales with both gas mass (\(M_{\rm gas}\)) and \(\Sigma_{\rm gas}\), any increase in \(M_{\rm gas}\), while keeping \(L_{\rm FIR}\) fixed, must result in a lower \(\Sigma_{\rm gas}\). As \(\Sigma_{\rm gas}\) decreases, C$^+$ abundance increases, raising the \cii/FIR ratio. This effect is illustrated by the model curves in the left panel of Figure~\ref{fig_CII2FIR}, which shows the impact of increasing the gas mass by a factor of 10. Under this interpretation, the offset in the ``\cii\ deficit'' observed between local and high-$z$ galaxies arises because high-$z$ galaxies are generally richer in molecular gas \cite[e.g.,][]{rhc_tacconi20,rhc_dessauges-zavadsky24,rhc_aravena24}.  The CRISTAL galaxies appear to occupy the upper envelope of nearby galaxies with similar FIR luminosities and overlap with high-$z$ systems at comparable FIR luminosities, bridging the characteristics of these two populations.

The right panel of Figure~\ref{fig_CII2FIR} presents the ``\cii\ deficit'' as a function of FIR surface brightness, \(\Sigma_{\rm FIR}\). For the CRISTAL galaxies, \(\Sigma_{\rm FIR}\) is calculated using the dust continuum sizes measured by \citet{rhc_mitsuhashi24}. The relationship between the \cii/FIR ratio and \(\Sigma_{\rm FIR}\) is significantly tighter than when plotted against \(L_{\rm FIR}\), and the offset between nearby and high-$z$ galaxy populations becomes less pronounced. This tighter correlation between the \cii/FIR ratio and \(\Sigma_{\rm FIR}\) has been reported in previous studies \citep[e.g.,][]{rhc_lutz16,rhc_spilker16,rhc_diaz-santos17,rhc_rhc18a}, and aligns with other findings that show a stronger relationship between \(\Sigma_{\rm [CII]}\) and \(\Sigma_{\rm SFR}\) compared to \(L_{\rm [CII]}\) and \(L_{\rm SFR}\) \citep[e.g.,][]{rhc_rhc15,rhc_rhc18b,rhc_delooze14}. This likely originates from the closer connection between \(G_0/n_{\rm H}\), the primary parameter controlling the physical and thermal structure of PDRs, and \(\Sigma_{\rm FIR}\), compared to \(L_{\rm FIR}\).

With the CRISTAL survey, we can now extend the study of the \cii/FIR–\(\Sigma_{\rm FIR}\) relation in high-redshift galaxies to values of \(\Sigma_{\rm FIR}\lesssim10^{11}\,L_{\odot}\,{\rm kpc}^{-2}\), covering a range similar to that of previous studies of nearby galaxies. Following \citet{rhc_lutz16}, the relation between the \cii/FIR ratio and \(\Sigma_{\rm FIR}\) for star-forming galaxies at \(z\approx0\) is parameterized as:

\begin{multline}
{\rm log}_{10}(L_{\rm [CII]}/L_{\rm FIR})~[z\approx0] = -11.7044 + 2.1676\times {\rm log}_{10}(\Sigma_{\rm FIR}) \\ - 0.1235\times ({\rm log}_{10}(\Sigma_{\rm FIR}))^2,
\end{multline}

\noindent shown as the golden solid line in the right panel of Figure~\ref{CII2FIR}.

Using a similar approach for star-forming galaxies at \(z\approx3-6\), including the CRISTAL galaxies covering the range \(10^{9}\lesssim \Sigma_{\rm FIR} \lesssim10^{11}\,L_{\odot}\,{\rm kpc}^{-2}\), and the galaxies from \citet{rhc_spilker16} (and references therein) covering \(10^{11}\lesssim \Sigma_{\rm FIR} \lesssim10^{14}\,L_{\odot}\,{\rm kpc}^{-2}\), we derive the following relation:

\begin{multline}\label{eq_cristal}
{\rm log}_{10}(L_{\rm [CII]}/L_{\rm FIR})~[z\approx3-6] = -8.4163 + 1.3245\times {\rm log}_{10}(\Sigma_{\rm FIR}) \\ - 0.0731\times ({\rm log}_{10}(\Sigma_{\rm FIR}))^2,
\end{multline}

\noindent represented by the gray solid line in the right panel of Figure~\ref{fig_CII2FIR}.

Notably, in the range \(10^{11.5} \lesssim \Sigma_{\rm FIR} \lesssim 10^{13.5}\,L_{\odot}\,{\rm kpc}^{-2}\), star-forming galaxies at \(z \approx 3-6\) exhibit \cii/FIR ratios that are higher by factors of \(3\) to \(10\), compared to galaxies at \(z \approx 0\), for a given value of \(\Sigma_{\rm FIR}\).  

One alternative to explain the physical origin driving the elevated \cii/FIR ratios observed in high-redshift galaxies compared to nearby star-forming systems could be metallicity. Metallicity plays an important role in determining the cooling efficiency of the ISM. In low-metallicity environments, the reduced dust content allows a larger fraction of FUV photons to penetrate deeper into molecular clouds, leading to more extended PDRs. Elevated \cii/FIR ratios have been observed in gas with metallicities around $\sim1/2-1/5$ of solar abundance—similar to what could be expected for massive star-forming galaxies at \(z\gtrsim4-6\) like CRISTAL systems \cite[e.g.,][]{rhc_curti24,rhc_nakajima23}. Examples include the Small Magellanic Cloud \citep{rhc_israel11} and low-metallicity, kiloparsec-scale regions of nearby galaxies \citep{rhc_smith17}. Consequently, high-$z$ galaxies with lower metal content could exhibit enhanced \cii\ emission relative to their FIR luminosity, contributing to the elevated \cii/FIR ratios observed in these systems.

A second factor, not as significant as metallicity but complementary, could be the presence of diffuse, extra-planar \cii\ gas, which extends beyond the star-forming regions of CRISTAL galaxies. In this scenario, the \cii\ emission may originate not only from PDRs but also from lower-density, more diffuse gas in the CGM. Evidence for \cii\ line emission significantly extending beyond the star-forming disk as traced by the UV/optical/infrared emission has been found in CRISTAL galaxies \citep{rhc_ikeda25} and other star-forming high-\(z\) galaxies \citep{rhc_fujimoto19,rhc_fujimoto20,rhc_ginolfi20}, including massive, intensely star-forming galaxies at \(z\sim3\) \citep[e.g.,][]{rhc_rybak19}. In the particular case of interacting galaxies, which represent at least one-third of the CRISTAL galaxies, extra-planar, low-velocity shocked gas can also contribute to the \cii\ line emission \cite[e.g.,][]{rhc_appleton13,rhc_peterson18}. However, it is important to note that at \(z\sim5\), CMB suppression of the \cii\ emission from diffuse gas becomes significant and works against this interpretation \citep{rhc_kohandel19}. In all these cases, the diffuse gas adds to the total \cii\ flux without a corresponding increase in FIR, potentially contributing to the elevated \cii/FIR ratios observed at high redshift relative to systems with comparable \(\Sigma_{\rm FIR}\) at lower redshift. 

In the future, JWST/NIRSpec observations of the main nebular lines in CRISTAL galaxies will provide valuable insights into the role of metal content in determining the \cii/FIR ratio on $\sim$kiloparsec scales (Herrera-Camus et al. in prep.). This will enable direct comparisons with similar spatial scale observations of nearby galaxies,  where a clear dependence of the \cii/FIR ratio with metallicity is observed \citep[e.g.,][]{rhc_smith17}. 

\section{Summary} \label{conclusions}

We present the CRISTAL survey, an ALMA Cycle 8 Large Program designed to explore the physical properties of star-forming galaxies in the early Universe through spatially resolved, multi-wavelength observations. CRISTAL focuses on main-sequence star-forming galaxies at redshifts \(4 \lesssim z \lesssim 6\), selected from the ALPINE survey \citep{rhc_lefevre20, rhc_bethermin20, rhc_faisst20}. Based on ALMA Band 7 data to observe the \cii\ line emission and dust continuum, and complemented by HST observations of rest-frame UV emission and JWST/NIRCam imaging of stellar light, CRISTAL offers a comprehensive view of the gas, dust, and stars on $\sim$kiloparsec scales at the end of the era of reionization.

\medskip

The main results of our study can be summarized as follows:

\begin{enumerate}

\item {\bf Sample size:} The initial CRISTAL sample consist of 19 galaxies; however, the depth and angular resolution of the observations allowed us to detect seven additional galaxies within CRISTAL fields and to spatially resolve four interacting systems into distinct pairs. Additionally, we incorporated three galaxies from pilot programs (HZ4, HZ7, HZ10) and included three galaxies from the ALMA archive that meet the CRISTAL selection criteria. The sample also includes two SMGs within CRISTAL fields: J1000+0234 and CRLE.  In total, the extended CRISTAL sample consists of 39 galaxies, 24 of which are detected in the dust continuum. The main properties of these systems are detailed in Table~\ref{table:sample} and Table~\ref{table:fluxes}.

\medskip

\item {\bf Diverse nature of CRISTAL galaxies:} The CRISTAL galaxies exhibit a wide range of morphologies and kinematic properties. This includes evidence for rotating disks (e.g., CRISTAL-11, -15, -20, -23c), mergers (e.g., CRISTAL-01, -04, -06, -07, -16, -22) and galaxies with \cii\ emission tails likely indicative of interactions with minor companions (e.g., CRISTAL-02, -03, -09, -11, -12, -19). We also identify systems displaying extended \cii\ emission without a corresponding stellar counterpart in HST or JWST/NIRCam images (e.g., CRISTAL-02, -05, -13). Some galaxies exhibit clumpy stellar structures with spatially offset \cii\ emission (e.g., CRISTAL-13), while others show a smoother \cii\ spatial distribution and kinematics (e.g., CRISTAL-08). Overall, the \cii\ line emission reveals the complex nature of star-forming galaxies at \(4 \lesssim z \lesssim 6\), with \cii\ emission often extending beyond the stellar light captured by HST and JWST, and displaying in many cases a disturbed morphology.

\end{enumerate}

The multi-wavelength, spatially resolved nature of the CRISTAL survey enables in-depth studies of four main aspects of galaxies: kinematics, outflows, morphology, and the physical conditions of the ISM \citep[see, for example, CRISTAL papers by][]{rhc_li24, rhc_lines24, rhc_mitsuhashi24, rhc_villanueva24, rhc_posses24, rhc_ikeda25, rhc_solimano24a}. In this overview paper, we highlight two case studies --CRISTAL--10 and -13-- that exemplify these primary research areas.

\begin{enumerate}

\setcounter{enumi}{2}

\medskip

\item {\bf CRISTAL-10: Arp 220-like conditions in the ISM of a $z\sim5$ galaxy?} CRISTAL-10a, a main-sequence star-forming galaxy at \(z = 5.67\), presents an intriguing scenario: the peaks of the dust continuum and \cii\ line emission are significantly offset, separated by over \(\sim3\) kpc. This pronounced offset points to an extreme \cii\ deficit. In the region where the dust continuum is most intense, the \cii/FIR ratio drops to \(2.5 \times 10^{-4}\), comparable to the extreme low value observed in the highly dust-obscured ultra-luminous infrared galaxies (ULIRG) Arp 220 \citep{rhc_luhman98}. Notably, the global \cii/FIR ratio for CRISTAL-10a is nearly an order of magnitude higher, emphasizing the localized nature of the deficit and the importance of spatially-resolved observations to characterize the ISM properties of high-$z$ systems.

\medskip

\item {\bf CRISTAL-13: Burst of star formation and \cii\ emission offset.} CRISTAL-13, a star-forming galaxy at \(z=4.579\), showcases a complex morphology characterized by at least seven giant star-forming clumps identified through multiband JWST/NIRCam imaging. The most massive clump, situated in the eastern region, coincides spatially with the \cii\ and dust continuum peaks, and has a \cii/FIR ratio comparable to that found in dense PDRs like Orion. In contrast, the western region hosts a group of at least six young (\(\lesssim50\) Myr) giant star-forming clumps \citep{rhc_lines24}. We observe a strong spatial anti-correlation between these young stellar clusters and \cii\ line emission, likely due to strong feedback effects from the star-forming clumps.

\end{enumerate}

\begin{enumerate}

\setcounter{enumi}{5} 

\item {\bf The \cii/FIR ratio in CRISTAL galaxies and its comparison across cosmic time.} CRISTAL galaxies have global \cii/FIR ratios ranging from \(1.5 \times 10^{-3}\) to \(8 \times 10^{-3}\), following a similar trend of decreasing \cii/FIR with increasing FIR luminosity as observed in nearby galaxies, but shifted toward higher FIR luminosities. One likely explanation is that CRISTAL galaxies at \(z \sim 4-6\) are significantly more molecular gas rich compared to typical nearby star-forming systems \citep[e.g.,][]{rhc_gracia-carpio11, rhc_narayanan16}. When considering FIR surface brightness—a proxy for the \(G_0/n_{\rm H}\) ratio—CRISTAL galaxies span a unique range of \(\Sigma_{\rm FIR}\) values between \(10^9\) and \(10^{11}\) \(L_{\odot} \, \text{kpc}^{-2}\), previously unexplored at high-$z$. Complemented by observations of bright $z\gtrsim3$ sources like SMGs, we find that the \cii/FIR ratio tend to be higher than star-forming galaxies at $z\sim0$. This behavior could be attributed to lower metallicities in high-$z$ galaxies or the presence of significant extraplanar gas.

\end{enumerate}

\medskip

Through the CRISTAL ALMA Large Program, we have advanced our understanding of the complex processes driving galaxy evolution during the first Gyr of the Universe. Future JWST/NIRSpec programs (e.g., PID 3073, PI Faisst; PID 5974, ORCHIDS, PI Aravena) will build on this progress, expanding the CRISTAL survey by incorporating detailed observations of the physical conditions of ionized gas and metals. This addition will provide a more complete picture of the interplay between the neutral and ionized gas, dust, stars, and metals in the early Universe.

\begin{acknowledgements}
RH-C thanks the Max Planck Society for support under the Partner Group project "The Baryon Cycle in Galaxies" between the Max Planck for Extraterrestrial Physics and the Universidad de Concepción. R.H-C. also gratefully acknowledge financial support from ANID - MILENIO - NCN2024\_112 and ANID BASAL FB210003. RJA was supported by FONDECYT grant number 1231718. RH-C, MA and RJA were supported by the ANID BASAL project FB210003. N.M.F.S. acknowledges funding by the European Union (ERC Advanced Grant GALPHYS, 101055023). Views and opinions expressed are, however, those of the author(s) only and do not necessarily reflect those of the European Union or the European Research Council. Neither the European Union nor the granting authority can be held responsible for them.  J. G-L., acknowledge support from ANID BASAL project FB210003 and Programa de Inserción Académica 2024 Vicerrectoría Académica y Prorrectoría Pontificia Universidad Católica de Chile. RB acknowledges support from an STFC Ernest Rutherford Fellowship (grant number ST/T003596/1). HÜ acknowledges support through the ERC Starting Grant 101164796 “APEX”. TDS acknowledges the research project was supported by the Hellenic Foundation for Research and Innovation (HFRI) under the "2nd Call for HFRI Research Projects to support Faculty Members \& Researchers" (Project Number: 03382). K.T. acknowledges support from JSPS KAKENHI Grant No. 23K03466. MR acknowledges support from project PID2023-150178NB-I00 (and PID2020-114414GB-I00), financed by MCIU/AEI/10.13039/501100011033, and by FEDER, UE. This paper makes use of the following ALMA data: ADS/JAO.ALMA\#2021.1.00280.L, 2017.1.00428.L, 2012.1.00523.S, 2018.1.01359.S, 2019.1.01075.S. ALMA is a partnership of ESO (representing its member states), NSF (USA) and NINS (Japan), together with NRC (Canada), NSTC and ASIAA (Taiwan), and KASI (Republic of Korea), in cooperation with the Republic of Chile. The Joint ALMA Observatory is operated by ESO, AUI/NRAO and NAOJ. This work is based in part on observations made with the NASA/ESA/CSA James Webb Space Telescope and NASA/ESA Hubble Space Telescope. The data were obtained from the Mikulski Archive for Space Telescopes at the Space Telescope Science Institute, which is operated by the Association of Universities for Research in Astronomy, Inc., under NASA contract NAS 5-03127 for JWST and NAS 1432 5–26555 for HST. Some of the data products presented in this work were retrieved from the DJA. DJA is an initiative of the Cosmic Dawn Center, which is funded by the Danish National Research Foundation under grant No. 140. We acknowledge that the research in this paper was conducted on Noongar land and pay our respects to its traditional custodians and elders, past and present. T.N. acknowledges support from the Deutsche Forschungsgemeinschaft (DFG, German Research Foundation) under Germany’s Excellence Strategy - EXC-2094 - 390783311 from the DFG Cluster of Excellence "ORIGINS". MA acknowledges support from ANID Basal Project FB210003 and and ANID MILENIO NCN2024\_112.

\end{acknowledgements}

%
%

\bibliographystyle{aa}
\bibliography{references.bib}

\begin{appendix} 

\section{Main properties of the CRISTAL galaxy sample}

Table~\ref{table:sample} summarizes the main properties of the CRISTAL galaxies, including names, coordinates, \cii-based redshift, and stellar masses and star formation rates derived by \cite{rhc_mitsuhashi24}. 

\section{Summary of ALMA observations}

Table~\ref{table:obs} provides a summary of the ALMA-CRISTAL observations, detailing the array configuration, number of antennas, flux calibrator, synthesized beam size, as well as the cube and continuum noise levels achieved with both Natural and Briggs weighting.

\begin{sidewaystable*}
\caption{Summary of ALMA observations}             
\label{table:obs}   
\centering                          
\begin{tabular}{l c c c | c c c | c c c}        
\hline\hline               
& & & & & Natural & & & Briggs & \\   
\hline
Name &  Array Config. & \# Antennas & Flux Calibrator & Beam Size & Cube Noise$^{a}$ & Cont. Noise & Beam Size & Cube Noise$^{a}$ & Cont. Noise \\ 
 & & {\it mean} & & & mJy beam$^{-1}$ & $\mu$Jy beam$^{-1}$  &  & mJy beam$^{-1}$ &  $\mu$Jy beam$^{-1}$ \\ 
\hline		
CRISTAL-01	&	C43-5+C43-2	&	45	&	J1058+0133	&	$	0.42	\arcsec	\times	0.47	\arcsec	$	&	0.13	&	10.2	&	$	0.22	\arcsec	\times	0.24	\arcsec	$	&	0.13	&	10.3	\\
CRISTAL-02	&	C43-4+C43-1	&	43	&	J1058+0133, J0854+2006	&	$	0.45	\arcsec	\times	0.55	\arcsec	$	&	0.13	&	12.7	&	$	0.29	\arcsec	\times	0.36	\arcsec	$	&	0.16	&	13.5	\\
CRISTAL-03	&	C43-4+C43-1	&	43	&	J1058+0133	&	$	0.68	\arcsec	\times	0.82	\arcsec	$	&	0.15	&	9.5	&	$	0.44	\arcsec	\times	0.54	\arcsec	$	&	0.18	&	10.1	\\
CRISTAL-04	&	C43-4+C43-1	&	44	&	J1058+0133	&	$	0.57	\arcsec	\times	0.75	\arcsec	$	&	0.18	&	16.6	&	$	0.36	\arcsec	\times	0.43	\arcsec	$	&	0.23	&	17.6	\\
CRISTAL-05	&	C43-5+C43-2	&	46	&	J1058+0133	&	$	0.31	\arcsec	\times	0.38	\arcsec	$	&	0.16	&	9.0	&	$	0.19	\arcsec	\times	0.22	\arcsec	$	&	0.21	&	9.7	\\
CRISTAL-06	&	C43-4+C43-1	&	45	&	J1058+0133	&	$	0.45	\arcsec	\times	0.54	\arcsec	$	&	0.13	&	9.1	&	$	0.30	\arcsec	\times	0.38	\arcsec	$	&	0.17	&	10.9	\\
CRISTAL-07	&	C43-5+C43-2	&	42	&	J1058+0133	&	$	0.49	\arcsec	\times	0.74	\arcsec	$	&	0.18	&	10.3	&	$	0.30	\arcsec	\times	0.46	\arcsec	$	&	0.22	&	11.5	\\
CRISTAL-08	&	C43-3+C43-1	&	44	&	J0519-4546	&	$	0.54	\arcsec	\times	0.80	\arcsec	$	&	0.17	&	15.3	&	$	0.36	\arcsec	\times	0.43	\arcsec	$	&	0.20	&	21.5	\\
CRISTAL-09	&	C43-5+C43-2	&	46	&	J1058+0133	&	$	0.33	\arcsec	\times	0.36	\arcsec	$	&	0.16	&	10.7	&	$	0.18	\arcsec	\times	0.19	\arcsec	$	&	0.19	&	11.6	\\
CRISTAL-10	&	C43-5+C43-2	&	43	&	J1058+0133	&	$	0.44	\arcsec	\times	0.48	\arcsec	$	&	0.09	&	8.6	&	$	0.27	\arcsec	\times	0.30	\arcsec	$	&	0.13	&	9.2	\\
CRISTAL-11	&	C43-5+C43-2	&	43	&	J1037-2934, J1058+0133	&	$	0.38	\arcsec	\times	0.47	\arcsec	$	&	0.22	&	20.0	&	$	0.21	\arcsec	\times	0.25	\arcsec	$	&	0.29	&	21.0	\\
CRISTAL-12	&	C43-4+C43-1	&	45	&	J0334-4008, J0519-4546	&	$	0.42	\arcsec	\times	0.58	\arcsec	$	&	0.12	&	7.1	&	$	0.24	\arcsec	\times	0.34	\arcsec	$	&	0.15	&	7.6	\\
	&		&		&	J0238+1636	&	$						$	&		&		&	$						$	&		&		\\
CRISTAL-13	&	C43-5+C43-2	&	47	&	J1058+0133	&	$	0.44	\arcsec	\times	0.52	\arcsec	$	&	0.18	&	15.7	&	$	0.14	\arcsec	\times	0.18	\arcsec	$	&	0.24	&	17.3	\\
CRISTAL-14	&	C43-6+C43-3	&	49	&	J1058+0133	&	$	0.11	\arcsec	\times	0.12	\arcsec	$	&	0.15	&	11.1	&	$	0.08	\arcsec	\times	0.08	\arcsec	$	&	0.17	&	12.1	\\
CRISTAL-15 	&	C43-5+C43-2	&	43	&	J1058+0133	&	$	0.36	\arcsec	\times	0.42	\arcsec	$	&	0.17	&	10.7	&	$	0.22	\arcsec	\times	0.24	\arcsec	$	&	0.18	&	12.2	\\
CRISTAL-16	&	C43-4+C43-1	&	43	&	J0519-4546, J0334-4008	&	$	0.42	\arcsec	\times	0.58	\arcsec	$	&	0.14	&	7.7	&	$	0.26	\arcsec	\times	0.30	\arcsec	$	&	0.17	&	8.9	\\
CRISTAL-17 	&	C43-4+C43-1	&	43	&	J1058+0133, J0854+2006	&	$	0.91	\arcsec	\times	1.30	\arcsec	$	&	0.12	&	7.5	&	$	0.81	\arcsec	\times	1.00	\arcsec	$	&	0.14	&	8.2	\\
	&		&		&	J1037-2934 	&	$						$	&		&		&	$						$	&		&		\\
CRISTAL-18	&	C43-5+C43-2	&	45	&	J1058+0133	&	$	0.45	\arcsec	\times	0.40	\arcsec	$	&	0.10	&	6.6	&	$	0.23	\arcsec	\times	0.21	\arcsec	$	&	0.12	&	7.0	\\
CRISTAL-19	&	C43-4+C43-1	&	44	&	J1037-2934, J1058+0133	&	$	0.54	\arcsec	\times	0.67	\arcsec	$	&	0.10	&	9.4	&	$	0.31	\arcsec	\times	0.40	\arcsec	$	&	0.15	&	10.2	\\
	&		&		&	J0854+2006	&	$						$	&		&		&	$						$	&		&		\\
CRISTAL-20	&		&		&		&	$	0.41	\arcsec	\times	0.45	\arcsec	$	&	0.06	&	6.2	&	$	0.29	\arcsec	\times	0.33	\arcsec	$	&	0.08	&	7.4	\\
CRISTAL-21	&		&		&		&	$	0.32	\arcsec	\times	0.35	\arcsec	$	&	0.22	&	13.6	&	$	0.22	\arcsec	\times	0.30	\arcsec	$	&	0.25	&	15.1	\\
CRISTAL-22	&		&		&		&	$	0.27	\arcsec	\times	0.34	\arcsec	$	&	0.22	&	14.9	&	$	0.23	\arcsec	\times	0.26	\arcsec	$	&	0.27	&	15.7	\\
 CRISTAL-23	&		&		&		&	$	0.25	\arcsec	\times	0.30	\arcsec	$	&	0.35	&	27.3	&	$	0.16	\arcsec	\times	0.23	\arcsec	$	&	0.44	&	32.6	\\
CRISTAL-24	&		&		&		&	$	0.26	\arcsec	\times	0.31	\arcsec	$	&	0.46	&	29.2	&	$	0.16	\arcsec	\times	0.23	\arcsec	$	&	0.54	&	34.7	\\
CRISTAL-25	&		&		&		&	$	0.25	\arcsec	\times	0.30	\arcsec	$	&	0.38	&	26.7	&	$	0.15	\arcsec	\times	0.23	\arcsec	$	&	0.45	&	30.9	\\
\hline 
\end{tabular}
\tablefoot{$^{a}$ Cube noises are measured in 20 km s$^{-1}$ channels}
\end{sidewaystable*}

\section{Summary of ALMA observing programs combined in the data processing}

Table~\ref{table:ALMA_ID} lists the IDs of the ALMA observing programs that were combined to produce the final CRISTAL data products.

\begin{table}
\caption{IDs of ALMA program used to produce final CRISTAL products}         
\label{table:ALMA_ID}      
\centering                          
\begin{tabular}{l c c c c}       
\hline\hline                 
Name	&		&	ALMA Program ID	&		&		\\
\hline  									\\
CRISTAL-01	&	2017.1.00428.L	&	2021.1.00280.L	&	2019.1.01587.S	&		\\
CRISTAL-02	&	2012.1.00523.S	&	2017.1.00428.L	&	2021.1.00280.L	&	2011.0.00064.S	\\
CRISTAL-03	&	2012.1.00523.S	&	2017.1.00428.L	&	2021.1.00280.L	&		\\
CRISTAL-04	&	2017.1.00428.L	&	2021.1.00280.L	&		&		\\
CRISTAL-05	&	2012.1.00523.S	&	2017.1.00428.L	&	2018.1.01359.S	&	2021.1.00280.L	\\
CRISTAL-06	&	2017.1.00428.L	&	2021.1.00280.L	&		&		\\
CRISTAL-07	&	2017.1.00428.L	&	2021.1.00280.L	&		&		\\
CRISTAL-08	&	2017.1.00428.L	&	2021.1.00280.L	&		&		\\
CRISTAL-09	&	2017.1.00428.L	&	2021.1.00280.L	&		&		\\
CRISTAL-10	&	2012.1.00523.S	&	2017.1.00428.L	&	2021.1.00280.L	&		\\
CRISTAL-11	&	2017.1.00428.L	&	2021.1.00280.L	&		&		\\
CRISTAL-12	&	2017.1.00428.L	&	2021.1.00280.L	&		&		\\
CRISTAL-13	&	2017.1.00428.L	&	2021.1.00280.L	&		&		\\
CRISTAL-14	&	2017.1.00428.L	&	2021.1.00280.L	&		&		\\
CRISTAL-15	&	2017.1.00428.L	&	2021.1.00280.L	&		&		\\
CRISTAL-16	&	2017.1.00428.L	&	2021.1.00280.L	&		&		\\
CRISTAL-17	&	2017.1.00428.L	&	2021.1.00280.L	&		&		\\
CRISTAL-18	&	2017.1.00428.L	&	2021.1.00280.L	&		&		\\
CRISTAL-19	&	2017.1.00428.L	&	2021.1.00280.L	&		&		\\
CRISTAL-20	&	2018.1.01605.S	&		&		&		\\
CRISTAL-21	&	2018.1.01359.S	&		&		&		\\
CRISTAL-22	&	2019.1.01075.S	&		&		&		\\
CRISTAL-23	&	2019.1.00226.S	&		&		&		\\
CRISTAL-24	&	2019.1.00226.S	&		&		&		\\
CRISTAL-25	&	2019.1.00226.S	&		&		&		\\
\hline 
\end{tabular}
\end{table}

\section{Summary of HST and JWST data available for CRISTAL galaxies}

Table~\ref{table:ancillary} provides a summary of the HST/WFC3 and JWST/NIRCam observations available for the CRISTAL survey. All CRISTAL galaxies have HST/WFC3 F160W observations available, and at least $\sim75\%$ of them have JWST/NIRCam simultaneous observations in the F115W, F150W, F277W and F444W filters.

\begin{sidewaystable*}
\caption{Summary of HST/WFC3 and JWST/NIRCam data available for CRISTAL galaxies}            
\label{table:ancillary}      
\centering                         
\begin{tabular}{l | c c c c c | c c c c c c c c c c}        
\hline\hline                 
	&		&		&	HST/WFC3	&		&		&		&		&		&		&		&	JWST/NIRCam	&		&		&		&		\\
 \hline
ID	&	F105W	&	F110W	&	F125W	&	F140W	&	F160W	&	F070W	&	F090W	&	F115W	&	F150W	&	F200W	&	F277W	&	F356W	&	F410M	&	F444W	&	Program ID	\\
\hline
1	&	\checkmark	&		&	\checkmark	&	\checkmark	&	\checkmark	&		&		&	\checkmark	&	\checkmark	&		&	\checkmark	&		&		&	\checkmark	&	1727,4265	\\
2	&	\checkmark	&		&	\checkmark	&		&	\checkmark	&		&		&	\checkmark	&	\checkmark	&		&	\checkmark	&		&		&	\checkmark	&	1727	\\
3	&	\checkmark	&	\checkmark	&	\checkmark	&		&	\checkmark	&		&		&	\checkmark	&	\checkmark	&		&	\checkmark	&		&		&	\checkmark	&	1727	\\
4	&		&	\checkmark	&		&		&	\checkmark	&		&		&	\checkmark	&	\checkmark	&		&	\checkmark	&		&		&	\checkmark	&	1727	\\
5	&	\checkmark	&		&	\checkmark	&		&	\checkmark	&		&		&	\checkmark	&	\checkmark	&		&	\checkmark	&		&		&	\checkmark	&	1727	\\
6	&		&	\checkmark	&		&	\checkmark	&	\checkmark	&		&		&	\checkmark	&	\checkmark	&		&	\checkmark	&		&		&	\checkmark	&	1727	\\
7	&	\checkmark	&		&	\checkmark	&		&	\checkmark	&		&		&	\checkmark	&	\checkmark	&		&	\checkmark	&		&		&	\checkmark	&	1727	\\
8	&	\checkmark	&		&	\checkmark	&	\checkmark	&	\checkmark	&		&	\checkmark	&	\checkmark	&	\checkmark	&	\checkmark	&	\checkmark	&	\checkmark	&	\checkmark	&	\checkmark	&	1180	\\	
9	&		&	\checkmark	&		&		&	\checkmark	&		&		&	\checkmark	&	\checkmark	&		&	\checkmark	&		&		&	\checkmark	&	1727	\\
10	&	\checkmark	&		&	\checkmark	&		&	\checkmark	&		&		&		&		&		&		&		&		&		&		\\
11	&		&		&	\checkmark	&	\checkmark	&	\checkmark	&		&	\checkmark	&	\checkmark	&	\checkmark	&	\checkmark	&	\checkmark	&	\checkmark	&	\checkmark	&	\checkmark	&	1727,1837	\\
12	&	\checkmark	&		&	\checkmark	&	\checkmark	&	\checkmark	&	\checkmark	&	\checkmark	&	\checkmark	&	\checkmark	&	\checkmark	&		&		&		&		&	1286	\\
13	&		&		&	\checkmark	&	\checkmark	&	\checkmark	&		&	\checkmark	&	\checkmark	&	\checkmark	&	\checkmark	&	\checkmark	&	\checkmark	&	\checkmark	&	\checkmark	&	1727,1837	\\
14	&		&		&		&		&	\checkmark	&		&		&	\checkmark	&	\checkmark	&		&	\checkmark	&		&		&	\checkmark	&	1727	\\
15	&		&		&	\checkmark	&	\checkmark	&	\checkmark	&		&	\checkmark	&	\checkmark	&	\checkmark	&	\checkmark	&	\checkmark	&	\checkmark	&	\checkmark	&	\checkmark	&	1727,1837	\\
16	&	\checkmark	&		&	\checkmark	&		&	\checkmark	&	\checkmark	&	\checkmark	&	\checkmark	&	\checkmark	&	\checkmark	&		&		&		&		&	1286	\\
17	&		&		&	\checkmark	&	\checkmark	&	\checkmark	&		&	\checkmark	&	\checkmark	&	\checkmark	&	\checkmark	&	\checkmark	&	\checkmark	&	\checkmark	&	\checkmark	&	1727,1837	\\
18	&		&		&		&	\checkmark	&	\checkmark	&		&		&		&		&		&		&		&		&		&		\\
19	&		&		&		&		&	\checkmark	&		&		&	\checkmark	&	\checkmark	&		&	\checkmark	&		&		&	\checkmark	&	1727	\\
20	&	\checkmark	&		&	\checkmark	&	\checkmark	&	\checkmark	&		&		&		&		&		&		&		&		&		&		\\
21	&	\checkmark	&		&	\checkmark	&		&	\checkmark	&		&		&	\checkmark	&	\checkmark	&		&	\checkmark	&		&		&	\checkmark	&	1727	\\
22	&	\checkmark	&		&	\checkmark	&		&	\checkmark	&		&		&		&		&		&		&		&		&		&		\\
23	&	\checkmark	&		&		&	\checkmark	&	\checkmark	&		&		&		&		&		&		&		&		&		&		\\
24	&	\checkmark	&	\checkmark	&	\checkmark	&	\checkmark	&	\checkmark	&		&		&	\checkmark	&	\checkmark	&		&	\checkmark	&		&		&	\checkmark	&	1727	\\
25	&	\checkmark	&		&		&		&	\checkmark	&		&		&	\checkmark	&	\checkmark	&		&	\checkmark	&		&		&	\checkmark	&	1727	\\
\hline 
\end{tabular}
\end{sidewaystable*}

\section{ALMA, HST, and JWST View of the CRISTAL Galaxies}\label{appendix:view}

\begin{figure*}[h!]
\centering
   \includegraphics[width=\hsize]{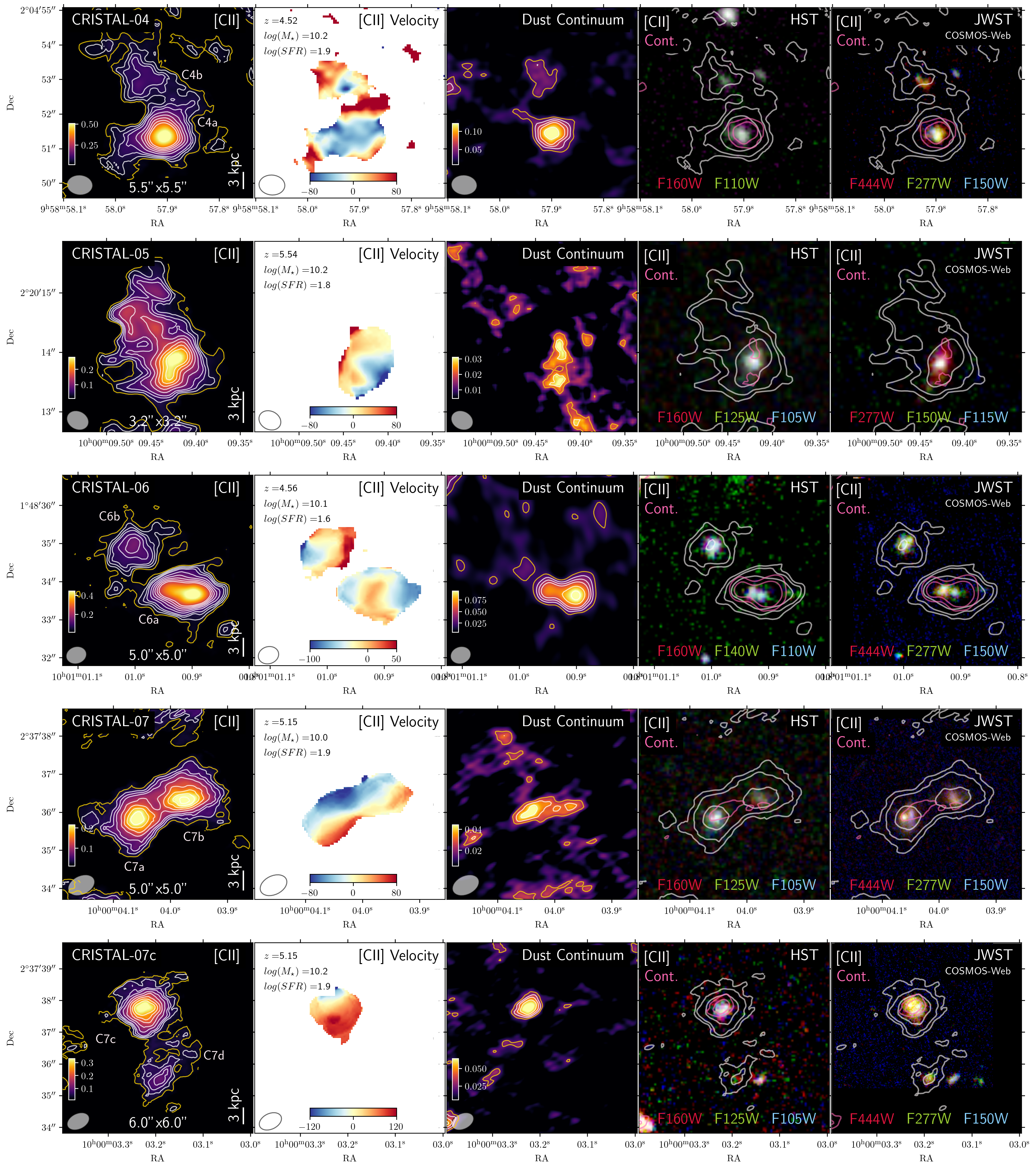}
      \caption{Multi-wavelength view of the CRISTAL galaxies including from left to right: integrated \cii\ line emission, \cii-based velocity field, dust continuum emission, \cii\ and dust continuum emission overlaid on a composite image based on HST/WFC3 and JWST/NIRCam observations. The redshift, stellar mass, and star formation rate are listed in the top left corner of the second panel. S/N contours correspond to 3, 4, and 5$\sigma$ and then increase in steps of 2$\sigma$.} \label{comb_panel2}
\end{figure*}

\begin{figure*}[h!]
\centering
   \includegraphics[width=\hsize]{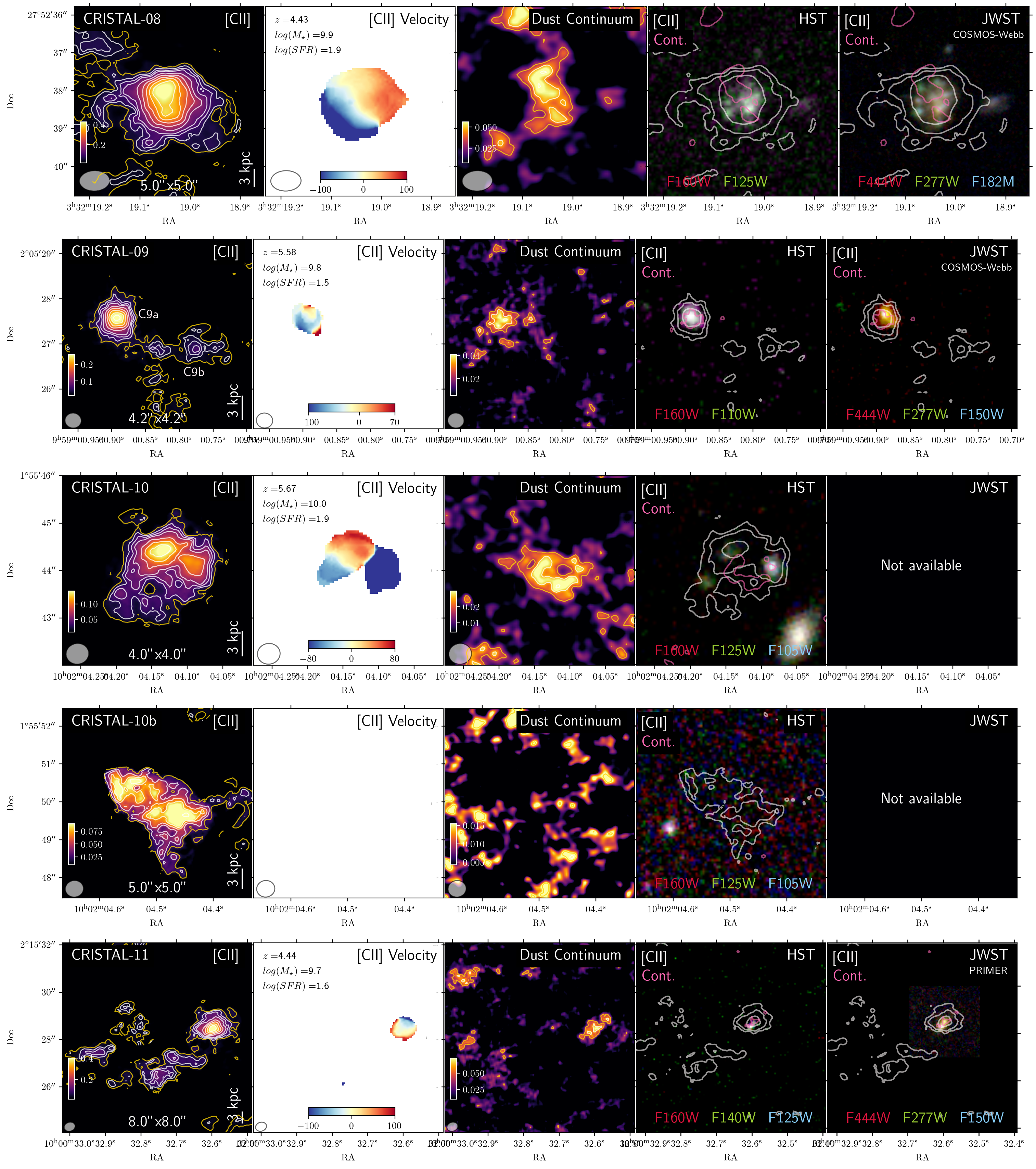}
      \caption{Multi-wavelength view of the CRISTAL galaxies including from left to right: integrated \cii\ line emission, \cii-based velocity field, dust continuum emission, \cii\ and dust continuum emission overlaid on a composite image based on HST/WFC3 and JWST/NIRCam observations. The redshift, stellar mass, and star formation rate are listed in the top left corner of the second panel. S/N contours correspond to 3, 4, and 5$\sigma$ and then increase in steps of 2$\sigma$.} \label{comb_panel3}
\end{figure*}

\begin{figure*}[h!]
\centering
   \includegraphics[width=\hsize]{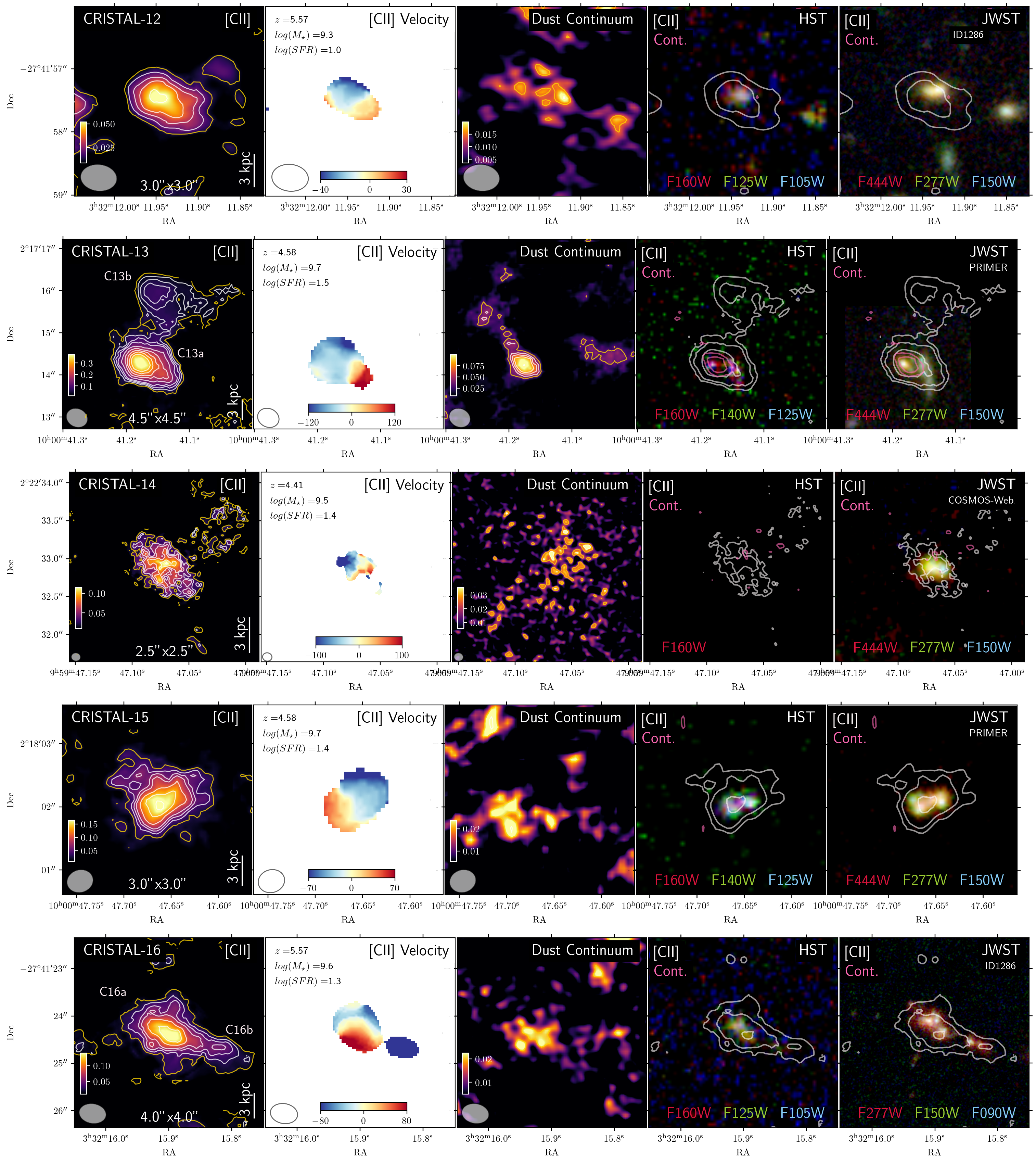}
      \caption{Multi-wavelength view of the CRISTAL galaxies including from left to right: integrated \cii\ line emission, \cii-based velocity field, dust continuum emission, \cii\ and dust continuum emission overlaid on a composite image based on HST/WFC3 and JWST/NIRCam observations. The redshift, stellar mass, and star formation rate are listed in the top left corner of the second panel. S/N contours correspond to 3, 4, and 5$\sigma$ and then increase in steps of 2$\sigma$.} \label{comb_panel4}
\end{figure*}

\begin{figure*}[h!]
\centering
   \includegraphics[width=\hsize]{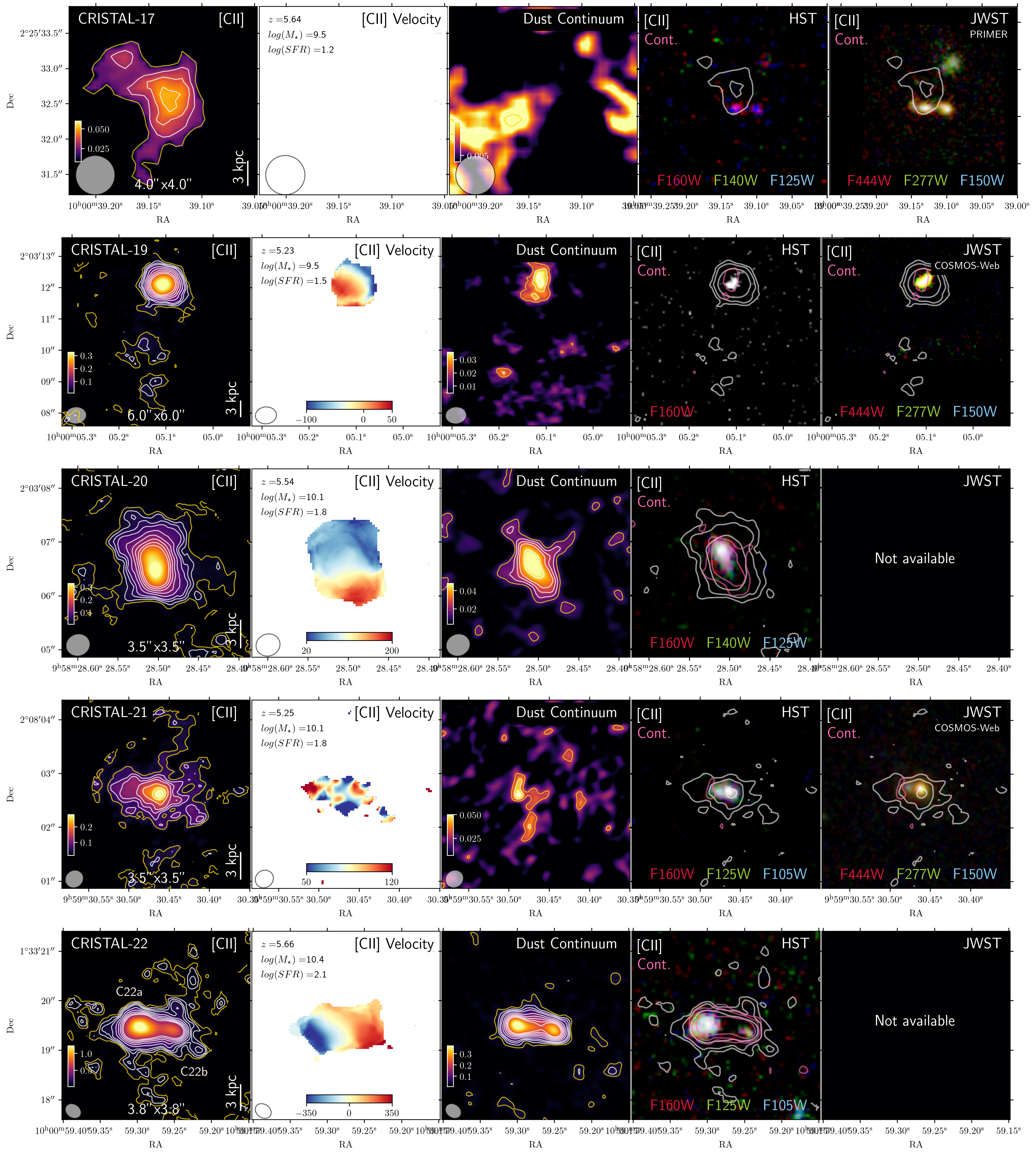}
      \caption{Multi-wavelength view of the CRISTAL galaxies including from left to right: integrated \cii\ line emission, \cii-based velocity field, dust continuum emission, \cii\ and dust continuum emission overlaid on a composite image based on HST/WFC3 and JWST/NIRCam observations. The redshift, stellar mass, and star formation rate are listed in the top left corner of the second panel. S/N contours correspond to 3, 4, and 5$\sigma$ and then increase in steps of 2$\sigma$.} \label{comb_panel5}
\end{figure*}

\begin{figure*}[h!]
\centering
   \includegraphics[width=\hsize]{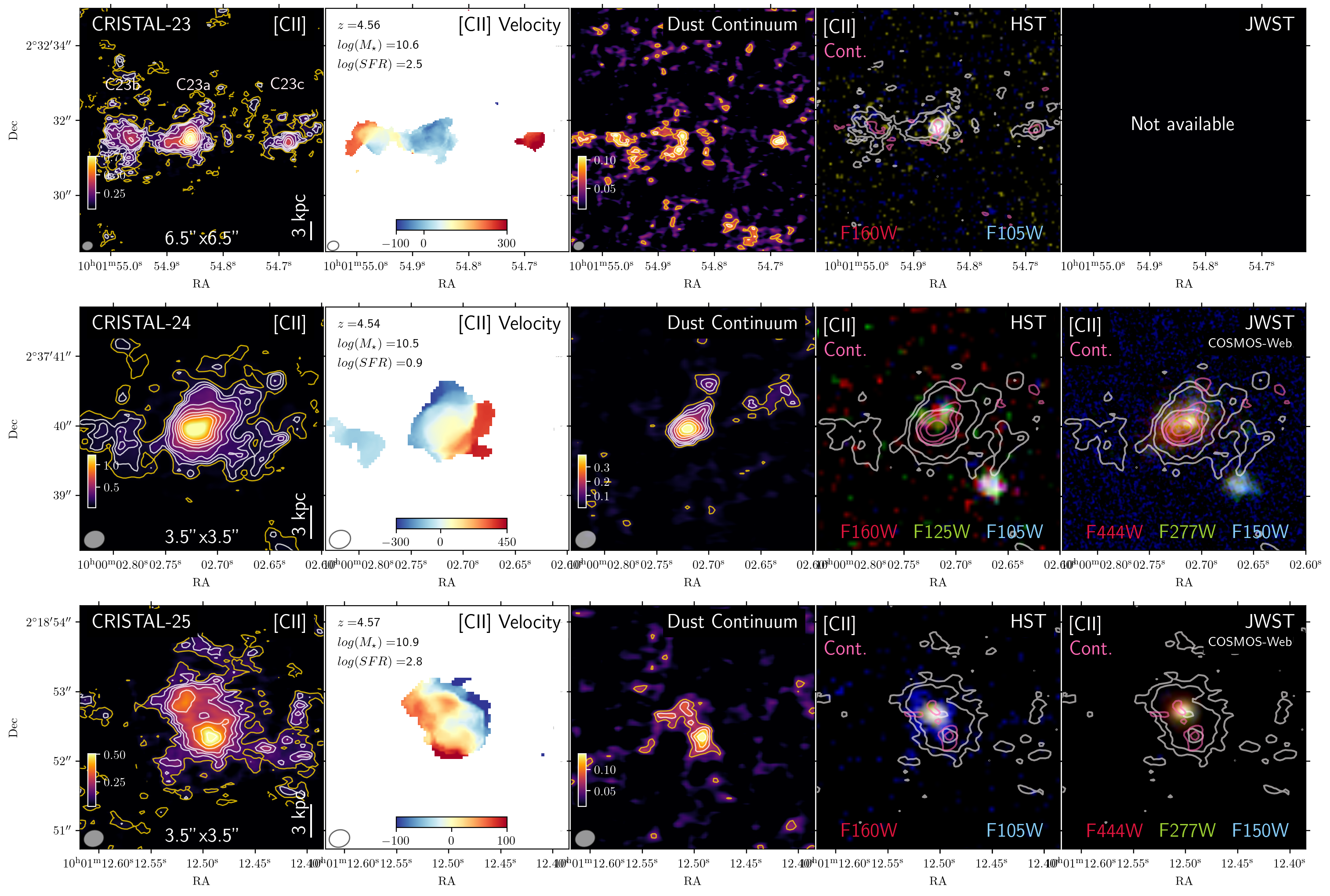}
      \caption{Multi-wavelength view of the CRISTAL galaxies including from left to right: integrated \cii\ line emission, \cii-based velocity field, dust continuum emission, \cii\ and dust continuum emission overlaid on a composite image based on HST/WFC3 and JWST/NIRCam observations. The redshift, stellar mass, and star formation rate are listed in the top left corner of the second panel. S/N contours correspond to 3, 4, and 5$\sigma$ and then increase in steps of 2$\sigma$.} \label{comb_panel6}
\end{figure*}

Figure~\ref{comb_panel1} in the main text, along with Figures~\ref{comb_panel2}, \ref{comb_panel3}, \ref{comb_panel4}, \ref{comb_panel5}, and \ref{comb_panel6}, presents a multi-wavelength view of the CRISTAL galaxies. From left to right, each panel shows maps of: \cii\ integrated intensity (moment 0), \cii\ velocity field (moment 1), ALMA Band 7 dust continuum, composite HST/WFC3 imaging, and composite JWST/NIRCam imaging of the stellar light. The filters used in the composite images are listed at the bottom of each panel.

\end{appendix}

\end{document}